\documentclass[prb,aps,twocolumn,superscriptaddress]{revtex4}

\usepackage{graphicx}
\usepackage{dcolumn}

\begin{document}


\title{
Thermodynamic Properties of the Dipolar Spin Ice Model
}

\author{Roger G. Melko}
\affiliation{Department of Physics, University of Waterloo, Waterloo,
Ontario, N2L 3G1, Canada}
\affiliation{Department of Physics, University of California, Santa
Barbara, California 93106}

\author{Matthew Enjalran}
\affiliation{Department of Physics, University of Waterloo, Waterloo,
Ontario, N2L 3G1, Canada}

\author{Byron C. den Hertog}
\affiliation{Department of Physics, University of Waterloo, Waterloo,
Ontario, N2L 3G1, Canada}

\author{Michel J. P. Gingras}
\affiliation{Department of Physics, University of Waterloo, Waterloo,
Ontario, N2L 3G1, Canada}
\affiliation{Canadian Institute for Advanced Research, 180 Dundas Street
West, Toronto, Ontario, M5G 1Z8, Canada}

\date{\today}

\begin{abstract}
We present a detailed theoretical overview of the thermodynamic properties 
of the dipolar spin ice model, which
has been shown to be an excellent quantitative descriptor of the Ising 
pyrochlore materials   
Dy$_{\rm 2}$Ti$_{\rm 2}$O$_{\rm 7}$ and Ho$_{\rm 2}$Ti$_{\rm 2}$O$_{\rm 7}$.
We show that the dipolar spin ice model can reproduce an effective 
{\it quasi} macroscopically degenerate ground state and spin-ice behavior 
of these materials when the long-range nature of dipole-dipole
interaction is handled carefully using Ewald summation techniques.  This 
degeneracy is, however, ultimately lifted at low temperature.  The long-range 
ordered state is identified via 
mean field theory and Monte Carlo simulation techniques. 
 Finally, we investigate the behavior of the dipolar spin ice model in 
an applied magnetic field, and compare our predictions with 
experimental results.  We find that a number of different long-range ordered 
states are favored by the model depending on field direction.
\end{abstract}

\maketitle





\section{Introduction}
\label{sec-intro}

\subsection{Water Ice and Spin Ice}

Frustrated or competing interactions are a common feature of many condensed 
matter systems.\cite{Toulouse}
In magnetic materials, frustration arises when the system 
cannot minimize its total classical ground state energy by minimizing the energy
of each spin-spin interaction individually.\cite{Villain,Diep,HFM2000}
When competing interactions cannot be simultaneously 
satisfied as a consequence of the arrangement of spins on a geometrical unit,
such as a triangle or a tetrahedron, a system made of an assembly of such units
is said to be geometrically frustrated. 
Geometric frustration has been studied extensively in 
recent years, with the discovery of classical systems that do not display any 
ordering or dynamical phase transitions down to the lowest temperatures (for recent
reviews see
Refs.~\onlinecite{Ramirez_review}-\onlinecite{BSS_rev}).
Furthermore, much current research effort is being deployed to investigate
the exotic behavior of quantum frustrated 
systems.\cite{Misguich,Misguich_qudimer,Lhuillier_hfm,Sindzingre}
In highly frustrated systems, weak 
quantum fluctuations may work to select a 
unique ground state that is not stabilized at the classical level, while
strong quantum fluctuations (e.g. small spin number value, $S$) can give rise
to novel quantum disordered states.\cite{quantum_pyro}
Real material \cite{LiViO4_buntgenn,LiV2O4_lacroix,Ballou} and model systems with
strongly correlated electrons in the presence of strong	magnetic frustration 
display interesting exotic properties. 

\begin{figure}[ht]
\begin{center}
\includegraphics[height=6cm]{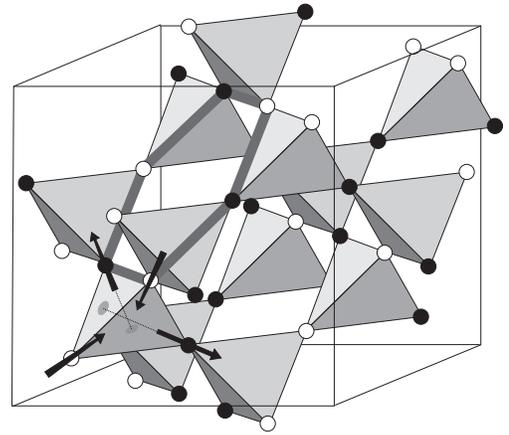}
\caption{The $\left<{111}\right>$ Ising pyrochlore lattice.  
The lower left ``downward'' tetrahedron of the pyrochlore 
lattice shows Ising spins as arrows.					
Each spin axis is along the local $\langle  111 \rangle $ 
quantization axis, which goes from one site
to the middle of the opposing triangular face (as shown by the disks)
and meets with the three other $\langle 111 \rangle $  axes in the middle
of the tetrahedron. For clarity,  black and white circles on the lattice points
denote other spins. White represents a spin pointing into a downward
tetrahedron while black is the opposite. The entire  lattice is shown in
an ice-rules state (two black and two white sites for every tetrahedron).
The hexagon (thick gray line) shows a minimal size loop move, which corresponds to
reversing all colors (spins) on the loop to produce a new ice-rules state.}
\label{Pyro}
\end{center}
\end{figure}

While geometric frustration most commonly arises between spins interacting
antiferromagnetically (AF), Harris and collaborators \cite{Harris_prl1,Harris_jpc}
showed that the pyrochlore lattice of corner sharing tetrahedra with
Ising spins pointing along a local cubic $\left<{111}\right>$ axis constitutes
a new class of geometrical frustration when nearest neighbor interactions
are ferromagnetic (FM) (See Fig.~\ref{Pyro}).\cite{Anderson,frust_ferro}
As a consequence of the frustration on this lattice, the Ising pyrochlore
ferromagnet has a lowest energy ground state configuration that is very closely 
analogous to an entirely different yet very common frustrated
condensed matter system $-$ namely water ice.\cite{Bramwell_science,BSS_rev}
\begin{figure}[ht]
\begin{center}
\includegraphics[height=3cm]{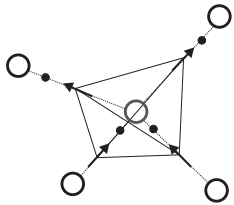}
\end{center}
\caption{The local proton arrangement in ice, showing oxygen atoms (large white
circles) and hydrogen atoms (small black circles) arranged to obey the ice
rules.  The displacement of the hydrogen atoms from the mid-points of the
oxygen-oxygen bonds are represented as arrows, which translate into spins on
the pyrochlore lattice in Fig.~\ref{Pyro}.} 
\label{H2Otet}
\end{figure}
In the low temperature $-$ low pressure
phase of water ice (the so-called ``hexagonal ice'', phase $I_{h}$), the 
oxygen atoms are arranged on a hexagonal lattice, each oxygen having four nearest 
neighbors.  Bernal and Fowler \cite{Bernal} and Pauling \cite{Pauling} were the 
first to propose that the hydrogen atoms (protons) within the H$_{\rm 2}$O lattice 
are not arranged periodically, but are disordered.  
These hydrogen atoms  on the O$-$O bonds are 
not positioned at the mid  point between both oxygen atoms, but rather each
proton is (covalently) bonded ``near'' one oxygen and (hydrogen-bonded)
``far'' from the other such that the water
solid consists of hydrogen-bonded H$_{\rm 2}$O molecules (see Fig.~\ref{H2Otet}).
In the Pauling 
model,  ice $I_{h}$ is established when the whole system is arranged according 
to the two {\it ice rules}:
\begin{enumerate}
\item
 Precisely one hydrogen atom is on each proton bond that links two nearest 
neighbor oxygen atoms.
\item
 Precisely two hydrogen atoms are near each oxygen atom (spin in) and two
 are far (spin out) (see Fig.~\ref{H2Otet}).
\end{enumerate}
A consequence of this structure, and the subsequent ice rules,
is that there is no single unique lowest energy state.  Indeed, there 
exists an infinitely large number of degenerate low energy
states that fulfill
the ice rules and, if the degeneracy was truly exact, would
manifest itself as a residual entropy at zero temperature (called 
zero point entropy).  Linus Pauling 
\cite{Pauling2} estimated theoretically the residual entropy, 
$S(T \rightarrow 0)$, of ice as
\begin{equation}
S(T \rightarrow 0) \approx  \frac{R}{2} \ln \frac{3}{2}
\;\;\; ,				
\label{PaulingENT}
\end{equation}
where $R \approx 8.31$ J mole$^{-1}$K$^{-1}$ is the molar gas constant. 
Pauling's result is not exact, but is accurate to within a 
few percent compared to experiments.\cite{giauque}

Returning to the magnetic Ising pyrochlores, the analogy to water 
ice arises if the spins are chosen to represent hydrogen displacements from 
the mid-points of the O$-$O bonds (Fig.~\ref{H2Otet}).  
The ice rules of {\it two protons close, two 
protons further away} corresponds to the {\it two spins in $-$ two
spins out} configuration of 
each tetrahedron on the pyrochlore lattice.   Because of this direct analogy 
between water ice and the Ising pyrochlores, 
Harris {\it et al.} \cite{Harris_prl1,Harris_jpc} 
called the latter {\it spin ice}.\cite{Bramwell_science,BSS_rev,Anderson,frust_ferro}
We note, howerver, that common water ice at atmospheric pressure, ice I$_h$, 
has a hexagonal
structure while here, the magnetic lattice has cubic symmetry.
Strictly speaking, the Ising pyrochlore problem is equivalent to
cubic ice, and not the hexagonal phase. Yet, this
does not modify the ``ice-rule'' analogy (or mapping) or the connection between
the statistical mechanics of the local proton coordination in water ice
and the low temperature spin structure of the spin ice materials.    

An important point must be emphasized here. 
In both ice water and spin ice, the microscopic
origin of the residual zero point entropy arises from the ``simplicity'' 
and ``under-constraints'' in the problem. Indeed, the constraints (rules) to
construct a minimum energy ground state, which arise from the
underlying microscopic Hamiltonian,
are so ``simple'' that an infinite
number of configurations of the dynamical variables at stake 
(proton position in ice, and spin direction in spin ice) 
can be used to make a minimum energy state from which the extensive residual
ground state entropy $S(T\rightarrow 0)$ results.

\subsection{Dipolar Spin Ice}
\label{sec-into3}

\begin{figure}[ht]
\begin{center}
\includegraphics[height=8cm]{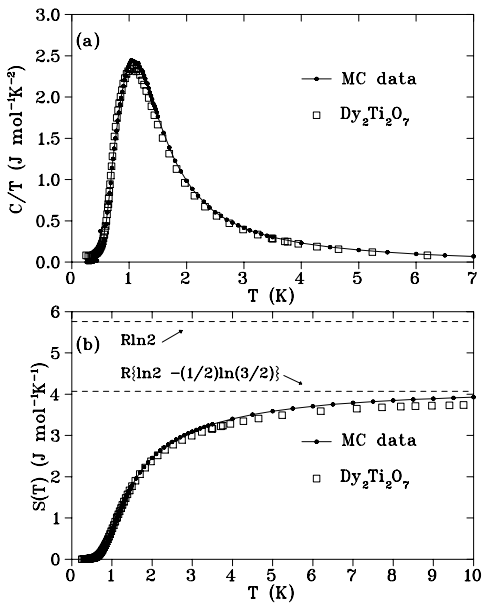}
\caption{(a) Specific heat and (b) entropy data for
Dy$_{\rm 2}$Ti$_{\rm 2}$O$_{\rm 7}$
from Ref.~\onlinecite{Ramirez}, compared with Monte Carlo simulation
results for the dipolar spin ice model, with $J_{\rm nn}=-1.24$K and
$D_{\rm nn}=2.35$K.}
\label{ResEnt}
\end{center}
\end{figure}

Experimentally, it is known that the single
ion ground states of the rare earth
ions Dy$^{\rm 3+}$ and Ho$^{\rm 3+}$ in the pyrochlore structure are
described by an effective classical Ising doublet.\cite{Harris_prl1,Rosenkranz}
Specific heat measurements by Ramirez 
\cite{Ramirez} on the compound Dy$_{\rm 2}$Ti$_{\rm 2}$O$_{\rm 7}$ have
shown
that the ``missing'' magnetic entropy not recovered 
upon warming the system from $T\approx 0.4$ K to 10 K, 
agrees reasonably well with Pauling's entropy calculation above,
$S\approx S(T\rightarrow 0)$, 
thereby providing compelling thermodynamic evidence that 
Dy$_{\rm 2}$Ti$_{\rm 2}$O$_{\rm 7}$ is a spin ice 
material \cite{Harris_nature} (see Fig.~\ref{ResEnt}).  
While early neutron scattering and magnetization measurements
first suggested that Ho$_{\rm 2}$Ti$_{\rm 2}$O$_{\rm 7}$ 
was a spin ice material,\cite{Harris_prl1}
some subsequent specific heat measurements and numerical simulations by
Siddharthan and co-workers were interpreted 
as evidence for a freezing transition to a partially
ordered state as opposed to spin ice behavior in that material.
\cite{BSS_rev,Siddharthan_prl,Siddharthan_prb}
However, more recent specific heat,\cite{Bramwell_prl,Cornelius}
magnetization \cite{Cornelius,newPetrenko} and
neutron scattering experiments,\cite{Bramwell_prl} supported
by Monte Carlo (MC) simulations,\cite{Bramwell_prl} appear to confirm
the initial proposal\cite{Harris_prl1} that Ho$_{\rm 2}$Ti$_{\rm 2}$O$_{\rm 7}$ 
is indeed a spin ice material akin to Dy$_{\rm 2}$Ti$_{\rm 2}$O$_{\rm 7}$.
Other magnetization measurements have recently been reported that
also argue for spin ice behavior in the closely related 
Ho$_{\rm 2}$Sn$_{\rm 2}$O$_{\rm 7}$,\cite{R2Sn2O7,Ho2Sn2O7_Kadowaki}
Dy$_{\rm 2}$Sn$_{\rm 2}$O$_{\rm 7}$, \cite{Dy2Sn2O7} and 
Ho$_{\rm 2}$Ru$_{\rm 2}$O$_{\rm 7}$ \cite{Ho2Ru2O7} materials.
The dynamical properties of these materials at 
the spin ice freezing point 
appear somewhat puzzling and are the subject of an increasing 
number of studies.\cite{freeze_Ho,freeze_Dy1,freeze_Dy2,Ehlers}

Following the initial spin ice proposal in 1997 by Harris and 
co-workers,\cite{Harris_prl1,Harris_jpc}
it appeared that the spin ice materials obeyed the simple ferromagnetic 
nearest neighbor model mentioned above.  This model intuitively gives rise to a 
degenerate spin ice ground state because of the equivalent energies of the 
six different tetrahedron configurations that make up the ground state of this 
geometrically frustrated unit.  However, the nearest 
neighbor spin ice model is too simple to accurately describe
the physical properties of real materials composed of the rare-earth ions Ho$^{3+}$ and Dy$^{3+}$
(see Ref.~\onlinecite{Siddharthan_prl}).
Firstly, the magnetic cations Ho$^{3+}$ and Dy$^{3+}$ in
Ho$_2$Ti$_2$O$_7$ and Dy$_2$Ti$_2$O$_7$ carry a large magnetic
moment,\cite{Harris_prl1,Rosenkranz}
$\mu$, of approximately 10$\mu_{\rm B}$. This entails strong
magnetic dipole-dipole interactions in these materials. Indeed, the 
strength of the dipolar interaction at nearest neighbor distances, $D_{\rm nn}$,
is of order 2 K, which is of the
same order of magnitude as the overall magnetic interaction energy scale
in these materials as estimated by the Curie-Weis temperature, 
$\theta_{\rm CW} \sim 1$ K, extracted from DC magnetization measurements.
Secondly, rare-earth ions possess very small exchange energies,
 which is roughly
the same order of magnitude as $\theta_{\rm CW}$ and $D_{\rm nn}$.
Consequently, dipole-dipole interactions 
in 						
Ho$_{\rm 2}$M$_{\rm 2}$O$_{\rm 7}$ 
and 
Dy$_{\rm 2}$M$_{\rm 2}$O$_{\rm 7}$ 
(M=Ti, Sn) 
are very significant and constitute an order one energy scale in the problem.
This is the reverse of what is observed in transition metal
compounds, where the exchange interaction predominates and the 
dipolar interaction can be treated as a very weak perturbation.
Finally, the nearest neighbor exchange interaction
in Ho$_{\rm 2}$Ti$_{\rm 2}$O$_{\rm 7}$ and
 Dy$_{\rm 2}$Ti$_{\rm 2}$O$_{\rm 7}$ is actually antiferromagnetic, which
would by itself cause a phase transition to a N\'eel 
long-range ordered $q=0$ state \cite{Harris_jpc,frust_ferro} (see Fig.~\ref{PhaseD}).
Consequently, we consider the 
simplest model of $\langle 111 \rangle$ Ising pyrochlore magnets with both
nearest-neighbor exchange and long-range magnetic dipole-dipole interactions
with the Hamiltonian:
\begin{eqnarray}
\label{DSPhamiltonian}
H&=&-J\sum_{\langle (i,a),(j,b)\rangle}{\bf S}_{i}^{a}\cdot{\bf S}_{j}^{b}  \\
&+&  Dr_{{\rm nn}}^{3}\sum_{ \begin{array}{c} i>j \\ a,b \end{array}    }
\frac{{\bf S}_{i}^{a}\cdot{\bf
S}_{j}^{b}}{|{\bf R}_{ij}^{ab}|^{3}} - \frac{3({\bf S}_{i}^{a}\cdot{\bf 
R}_{ij}^{ab}) ({\bf S}_{j}^{b}\cdot{\bf R}_{ij}^{ab})}{|{\bf R}_{ij}^{ab}|^{5}} 
\nonumber .
\end{eqnarray}
Here the spin vector ${\bf S}_{i}^{a}=\sigma_{i}^{a}\hat{z}^a$ 
labels the Ising moment of magnitude 
$|{\bf S}_{i}^{a}|=1$ at FCC lattice site ${\bf R}_i$ and tetrahedral sub-lattice site 
coordinate ${\bf r}^a$, 
where the local Ising axis is denoted by ${\hat z}^{a}$ and the Ising variable is $\sigma_{i}^{a} = \pm 1$.   
The vector ${\bf R}_{ij}^{ab}={\bf R}_{ij}+{\bf r}^{ab}$ connects spins ${\bf S}_{i}^{a}$
and ${\bf S}_{j}^{b}$.
$J$ represents the exchange energy and $D$ the dipolar energy scale
($J>0$ and $D=0$ in the spin ice model originally proposed by Harris {\it et al.}
\cite{Harris_jpc} which we refer to as the ``near neighbor spin ice model''). 
Because of the relative local $\left< {111} \right>$
Ising orientations, the nearest neighbor exchange energy between two spins is 
$J_{\rm nn}\equiv J/3$.  The dipole interaction is calculated from 
\begin{equation}
D=\frac{\mu_{0}}{4\pi}\;\frac{\mu^{2}}{r_{\rm nn}^{3}}.
\label{Dstrength}
\end{equation}
Experimentally, from magnetization measurements\cite{Harris_jpc} and analysis
of the crystal-field levels via inelastic neutron scattering,\cite{Rosenkranz}
it is known that the moments of the Dy$^{3+}$ and Ho$^{3+}$
rare-earth ions in the 
pyrochlore lattice are $\mu \approx 10 \mu_{\rm{B}}$, and the nearest neighbor 
distance $r_{\rm nn}$ is approximately 3.54 Angstroms.  
From Eq.~(\ref{DSPhamiltonian}), we get the dipole-dipole interaction at
nearest neighbor distances to be $D_{\rm nn} \equiv 5D/3$, since 
$\hat{z}^{a} \cdot \hat{z}^{b} = -1/3$ and $(\hat{z}^{a} 
\cdot {\bf R}_{ij}^{ab})({\bf R}_{ij}^{ab} \cdot \hat{z}^{b}) = -2/3$ 
in Eq.~(\ref{DSPhamiltonian}). For both 
Ho$_{\rm 2} $Ti$_{\rm 2}$O$_{\rm 7}$ and 
Dy$_{\rm 2}$Ti$_{\rm 2}$O$_{\rm 7}$, $D_{\rm nn} 
\approx 2.35 \rm{ K}$.

In order to consider the combined role of exchange and dipole-dipole
interactions, it is useful to define an effective nearest neighbor
energy scale, $J_{\rm eff}$, for
$\langle 111 \rangle $ Ising spins:
 \begin{equation}
 J_{\rm eff} \; \equiv \;        J_{\rm nn}+D_{\rm nn} \;\;\; ,
\end{equation}
\noindent
where $J_{\rm nn} \equiv J/3$ is the nearest neighbor exchange
energy between $\langle 111\rangle $ Ising moments. This simple
near-neighbor description of the system 
suggests that a $\langle 111\rangle $ Ising system could
display spin ice properties, even for antiferromagnetic nearest
neighbor exchange, $J_{\rm nn}<0$, so long as $J_{\rm eff}=J_{\rm
nn}+D_{\rm nn}>0$.  Fits to experimental data give 
$J_{\rm nn} \sim
-0.52$ K for Ho$_2$Ti$_2$O$_7$\hspace{0.15cm} \cite{Bramwell_prl} and $J_{\rm nn} \sim
-1.24$ K for Dy$_2$Ti$_2$O$_7$.\hspace{0.15cm}\cite{Hertog}  Thus, $J_{\rm
eff}$ is positive (using $D_{\rm nn}=2.35$K), hence ferromagnetic and
frustrated, for both Ho$_2$Ti$_2$O$_7$ ($J_{\rm eff} \sim 1.8$ K) and
Dy$_2$Ti$_2$O$_7$ ($J_{\rm eff} \sim 1.1$ K).  It would therefore
appear natural to ascribe the spin ice behavior in both
Ho$_2$Ti$_2$O$_7$ and Dy$_2$Ti$_2$O$_7$ to the positive $J_{\rm eff}$
value as in the simple model of Bramwell and Harris.\cite{Harris_jpc}
However, the situation is more complex than it appears.
					
Dipole-dipole interactions are ``complicated'': (i) they are
strongly anisotropic since they couple the spin, ${\bf S}_{i}^{a}$, 
and space, ${\bf R}_{ij}^{ab}$, directions, and (ii) they are also
very long-ranged ($\propto |{\bf R}_{ij}^{ab}|^{-3}$). For example, the second
nearest neighbor distance is $\sqrt 3$ times larger than the nearest
neighbor distance, which means that the second nearest neighbor dipolar energy is 
$D_{\rm nnn} \sim 0.2D_{\rm nn}$. This implies an important perturbation
compared to $J_{\rm eff}=J_{\rm nn}+D_{\rm nn} < D_{\rm nn}$, especially 
for antiferromagnetic (negative) $J_{\rm nn}$.  Specifically, for
Dy$_2$Ti$_2$O$_7$, the second nearest neighbor energy scale is about 40\%
of the effective nearest neighbor energy scale, $J_{\rm eff}$,
a large proportion! Therefore, one might have expected 
that the dipolar interactions beyond nearest neighbor 
would cause the different ice-rules states to have
different energies, hence possibly breaking the degeneracy
of the spin ice manifold, similar to what happens 
in the kagome  \cite{Kallin}			
and pyrochlore Heisenberg antiferromagnets \cite{Reimers} when exchange 
interactions beyond nearest-neighbor are considered. 
In Eq.(~\ref{DSPhamiltonian}), if the dipolar term is summed beyond nearest neighbor, 
one could expect a long-ranged 
N\'eel ordered state at a critical temperature $T_N \sim O(D_{\rm nn})$.
Thus, here arises one of the main puzzling and interesting 
problems posed by the dipolar spin ice
materials that can be summarized by two questions:

\begin{enumerate}
\item {Are the experimental observations of spin ice behavior
in real materials
consistent with dominant long-range dipolar interactions?}
\item {If so, why do long-range dipolar interactions fail to
destroy spin ice behavior and give rise to long-range order at a temperature
$T_N \sim O(D_{\rm nn})$ ? }
\end{enumerate}

Results from 
Monte Carlo simulations on the dipolar spin ice model attempting to answer
the first question above were first reported 
in Ref.~\onlinecite{Siddharthan_prl} and Ref.~\onlinecite{Siddharthan_prb}.  
In that work, the dipole-dipole interactions
were cut-off at a distance of five \cite{Siddharthan_prl} or ten and 
twelve nearest-neighbors.\cite{Siddharthan_prb} In those studies the thermodynamic
behavior was found to be consistent with spin ice behavior for a model
of Dy$_{\rm 2}$Ti$_{\rm 2}$O$_{\rm 7}$,  provided the exchange interaction
was made to extend far beyond nearest neighbor,\cite{greedan-open-pyro} but not 
for a model of Ho$_{\rm 2}$Ti$_{\rm 2}$O$_{\rm 7}$.
A subsequent work,\cite{Hertog} considered the Hamiltonian 
of Eq.~(\ref{DSPhamiltonian}) with only 
nearest-neighbor exchange and the value of $J$ as an adjustable parameter.
In that work, the long-range dipole-dipole interaction was
handled using the well-known Ewald method, which derives an effective 
dipole-dipole interaction between spins within the cubic simulation cell. 
The Monte Carlo simulations were carried out by slowly cooling the 
simulated lattice, subject to the usual Metropolis algorithm.  
Numerical  integration of the specific heat divided by temperature
 was performed to determine the entropy of the system.\cite{Hertog} 
For a parameter $J$  appropriate for
the Dy$_{\rm 2}$Ti$_{\rm 2}$O$_{\rm 7}$ spin ice material (see Section II below), 
the
dipolar spin ice model retained 
Pauling's entropy (Eq.~(\ref{PaulingENT})), in good agreement with 
experiments on Dy$_{\rm 2}$Ti$_{\rm 2}$O$_{\rm 7}$ (Fig.~\ref{ResEnt}).
Following the same approach as in Ref.~\onlinecite{Hertog}, 
recent Monte Carlo simulations 
have found good agreement between the dipolar spin ice model,
specific heat measurements and elastic neutron scattering, as well as
experiments on  Ho$_{\rm 2}$Ti$_{\rm 2}$O$_{\rm 7}$.\cite{Bramwell_prl}
Finally, mean-field theory calculations of the neutron scattering intensity,
valid in the (paramagnetic) temperature regime $T\gg \theta_{\rm CW}$
that consider large distance cut-off of the dipole-dipole interactions
have been found to be in good agreement with experiments
on Ho$_{\rm 2}$Sn$_{\rm 2}$O$_{\rm 7}$ \cite{Ho2Sn2O7_Kadowaki} and
Ho$_{\rm 2}$Ti$_{\rm 2}$O$_{\rm 7}$.\cite{Ho2Ti2O7_Kadowaki}
Consequently, there is now strong compelling evidence that
the long-range dipolar interaction is responsible for the
ice behavior and the subsequent retention of zero-point entropy
in rare-earth based insulating pyrochlore magnets.\cite{Bramwell_science}

\subsection{True Long Range Order at Low Temperature in the Dipolar Spin Ice Model}

Having answered question \#1 above in the affirmative, 
one is then faced with addressing question \#2. 
The Monte Carlo results 
mentioned above \cite{Hertog}
show that spin ice behavior arises
from the combination of nearest-neighbor exchange, $J_{\rm nn}$,
and dipole energies, $D_{\rm nn}$,
which create an {\it effective} ferromagnetic Ising model
$J_{\rm eff}$ at nearest neighbor as long as
$J_{\rm eff}\equiv J_{\rm nn}+D_{\rm nn} \gtrsim  0$,
akin to Harris and Bramwell's simple nearest-neighbor Ising
model.\cite{Harris_prl1,Harris_jpc,Harris_prl2,Champion_epl}
However, the long-range dipolar interaction does not appear to destroy
the spin ice degeneracy (and subsequent retention of zero point entropy)
created by this effective ferromagnetic nearest neighbor interaction.
In support of this picture, a mean-field
theory (MFT) calculation finds that remaining 
(beyond nearest neighbor) dipole-dipole  
interaction terms, which couple every spin in the system with varying strength 
depending on their separation distance, is ``screened'' to 
a large degree.\cite{Gingras_cjp,Enjalran}
This means that the degeneracy
between different ice-rules obeying states is almost exactly fulfilled 
by carefully including
the long distance dependence of the dipolar term in the Hamiltonian.  
However, and perhaps most interestingly for these Ising pyrochlore
systems, the same
mean-field calculation suggests that the screening of the
long-range terms is not perfect, and that the 
associated spin ice manifold is only {\it quasi}-degenerate, due to some small 
remaining effective energy scale, and that a unique ordering wavevector
is selected.\cite{Ho2Ti2O7_Kadowaki,Gingras_cjp,Enjalran}
This suggests that at some temperature 
below the onset temperature of spin ice correlations, 
the dipolar spin ice model should in principle favor the unique long-range 
ordered state selected by this remaining (``unscreened'') 
perturbative dipole-dipole energy.
\begin{figure}[ht]
\begin{center}
\includegraphics[height=9cm]{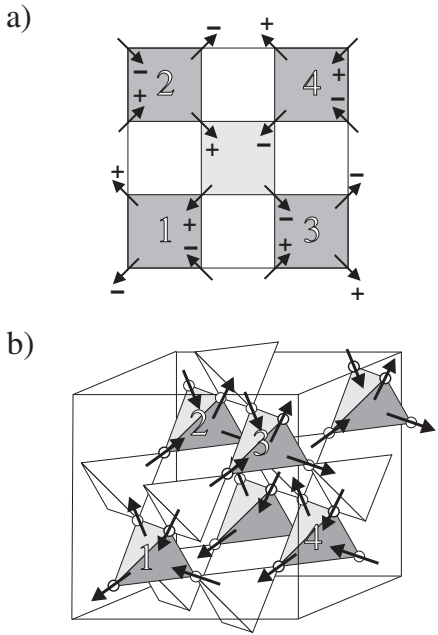}
\caption[The Long Range Ordered Dipolar Spin Ice Ground State.]
{The long-range ordered ${\bf q}=(0,0,2\pi/a)$ dipolar spin ice ground state. 
Projected down the z axis (a),
the four tetrahedra making up the cubic unit cell appear as dark gray
squares.  The light gray square in the middle does not represent 
a tetrahedron, but
its diagonally opposing spins are in the same lattice plane.  The component of
each spin parallel to the z axis is indicated by a + and - sign.  
In perspective (b), the four tetrahedra of the unit cell 
are numbered to enable comparison with (a).}
\label{GSsmall}
\end{center}
\end{figure}

One might naively expect that such an ordered state should 
be found in the MC
simulations.\cite{Bramwell_prl,Hertog,already_found}
However, this does not happen, as
measurements of the temperature dependent acceptance rate of the simulations 
make it apparent that the standard single (Ising) spin flip Metropolis algorithm 
experiences a dynamical ``freezing'' at a temperature $\approx 0.3$ K
for $J_{\rm nn}$ and $D_{\rm nn}$ parameters appropriate for
Dy$_{\rm 2}$Ti$_{\rm 2}$O$_{\rm 7}$ \cite{Hertog} and $T \approx 0.6$ K for 
Ho$_{\rm 2}$Ti$_{\rm 2}$O$_{\rm 7}$.\cite{Bramwell_prl}
If the dipolar interactions are
cut-off at some arbitrary distance, $R_c$, one can generate scenarios where,
depending on specific numerical values for
$J_{\rm nn}, D_{\rm nn}$ and $R_c$,  a
selected state is dynamically accessible before the spin-ice manifold
freeze-out, as was found in
simulations where dipole interactions are cut-off.
\cite{Siddharthan_prl,Siddharthan_prb}
Consequently, akin to the approaches used
in ice lattice models,\cite{Barkema,BarkNewBook} one must introduce 
non-local dynamics in the simulation to 
combat this freezing-out and maintain simulation equilibrium down
to lower temperatures.  The inclusion of non-local ``loop moves''
in the dipolar spin ice model promotes 
the development of a long-range ordered phase 
via a sharp first order phase transition 
at $T \approx 0.18$ K,\cite{Melko_prl,Melko_thesis} a much lower temperature than the
onset temperature for spin ice correlations at $T\sim 1.2$ K in 
Dy$_2$Ti$_2$O$_7$ \cite{Ramirez,Hertog}		
and $T\sim 1.9$ K in 
Ho$_2$Ti$_2$O$_7$.\cite{Bramwell_prl}
The ground state found in the loop MC simulations
has zero total (bulk) magnetization (recall that each tetrahedron individually
carries a net magnetic moment in each of the ice-rule obeying states). 
See Fig.~\ref{GSsmall} for the spin configurations in this ground state.
The pre-transitional specific heat and the latent heat associated with the
first order transition recovers all of Pauling's
missing entropy in the model. The ordered state that is found in the
loop MC simulations \cite{Melko_prl} corresponds to the ordered
state predicted by mean field theory.\cite{Gingras_cjp,Enjalran}
In other words, the dipolar spin ice model possesses on its own, without invoking
energetic perturbations and/or thermal and quantum fluctuations,
a unique (up to trivial global symmetry relations)
classical ground state with zero extensive entropy.

Also, using MC simulations and direct Ewald energy calculations, we
investigate the behavior of the dipolar spin ice model in an external 
magnetic field.  With application of a large field along three
different crystal symmetry directions, three different long-range ordered ground
states appear.  With large fields parallel to the [100] crystal direction, the 
ground
state is the ice-rules ${\bf q}=0$ structure identified by 
Harris.\cite{Harris_prl1}  For large fields parallel to [110], the ground 
state is the ice-rules ${\bf q}=X$ 
state,\cite{Harris_prl1} and for large fields along [111], the ice rules are 
broken and a three-spin in, one-spin out spin configuration becomes the lowest 
energy state.
The experimentally determined field dependence of the magnetization 
and specific heat for fields along the
[100], [110] and [111] directions in the Dy$_{\rm 2}$Ti$_{\rm 2}$O$_{\rm 7}$ 
spin ice material agree quantitatively well with the Monte Carlo results for 
the long-range dipolar spin ice model. 
\cite{Fukazawa,Higashinaka,hiroi,kagome-ice,kagomeMAT}

\subsection{Phases of Dipolar Spin Ice}

\begin{figure}[ht]
\begin{center}
\includegraphics[height=5.5cm]{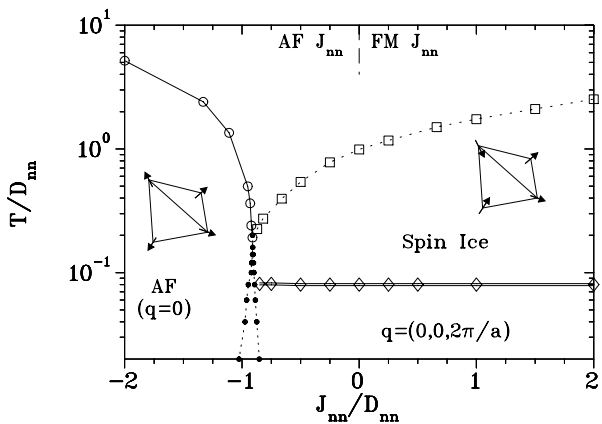} 
\end{center}
\caption{The phase diagram for the dipolar spin ice model.  The 
antiferromagnetic ground state is an all-spins-in or all-spins-out configuration
for each tetrahedron.  The spin ice configuration, which includes the
${\bf q}=(0,0,2\pi/a)$ ground state,  is a two spins in-two spins out configuration 
for each tetrahedron. The region encompassed between the quasi vertical dotted
lines displays hysteresis in the long-range ordered state selected
(${\bf q}=0$ vs. 
${\bf q}=(0,0,2\pi/a)$) as $J_{\rm nn}/D_{\rm nn}$ is varied at fixed temperature $T$.}
\label{PhaseD}
\end{figure}

Using Monte Carlo simulations, the phase diagram for the dipolar spin ice
model can be mapped out (Fig.~\ref{PhaseD}).  To summarize the results, 
spin ice correlations develop
for all cases where the effective nearest neighbor energy scale
$J_{\rm eff}/D_{\rm nn} > 0.095$ (ferromagnetic), and the temperature is below the 
broad peak in the specific heat, $T_{\rm peak}$.  For $T/D_{\rm nn} \leq 0.08$,
independent of the value of $J_{\rm nn}$ (as long as $J_{\rm eff}/D_{\rm
nn} > 0.095$), the system orders
into the long-range ordered state, with the help of the loop moves in the simulation.		
For $J_{\rm nn}/D_{\rm nn}$ less than -0.905 
($J_{\rm eff}/D_{\rm nn} < 0.095$), 
the system orders into an 
antiferromagnetic ${\bf q}=0$ N\'eel 
ground state, where every tetrahedron in the 
system has an all-in or all-out spin configuration at low temperatures.
\cite{Harris_jpc,frust_ferro,BGR,Moessner_q0}  The 
region around $J_{\rm nn}/D_{\rm nn} = -1$ shows hysteresis at low 
temperatures. Because of the close cancellation of energy scales, we imagine
that real materials which fall into this region, e.g., 
Tb$_{\rm 2}$Ti$_{\rm 2}$O$_{\rm 7}$,\cite{Gardner_tb_prl,Gingras_prb,Gardner_Tb_prb,Yasui_Tb}
will be particularly susceptible
to the influence of small perturbations 
(such as exchange beyond nearest-neighbor or 
finite, as opposed to infinite, Ising anisotropy \cite{Enjalran,Kao})
with the result of possible ordering into long-range 
ordered states \cite{Mirebeau} distinct from the two shown in Fig.~\ref{PhaseD}.

\subsection{Outline}

The rest of the paper is organized as follows.
In the next section we present results from conventional single spin flip
Monte Carlo simulations that show how spin ice behavior develops at
finite temperatures in the dipolar spin ice system whenever the
effective nearest-neighbor coupling is ferromagnetic  
($J_{\rm eff}\equiv J_{\rm nn}+D_{\rm nn}\gtrsim 0$). Results from
mean-field theory are
presented in Section III. There we show 
that there exists a weak selection of a unique ordering (critical or soft mode) 
at ${\bf q}=(0,0,2\pi/a)$. Motivated by our mean-field results,
we undertake a numerical search for a long-range ordered state in the model of 
Eq.~(\ref{DSPhamiltonian}). 
Section IV discusses the details of a loop Monte Carlo algorithm that avoids 
the freezing phenomenon observed in a MC simulation employing 
local single spin flip dynamics.
Section V presents the detailed results
from the loop Monte Carlo simulations. 
The results for the field dependence of the ground state energy and magnetization 
for fields along [100], [110] and [111] are presented in Section VI. 
We conclude the paper with a brief discussion in Section VII. 
We have included several appendixes.  
Appendix \ref{ApEwald} contains a discussion of the Ewald technique for 
dipolar interactions in real space (MC simulations) and in momentum 
space (MFT). In Appendix \ref{ApHTSE}, the ${\bf q}-$dependent susceptibility, to 
quadratic order, is derived via high temperature series expansion 
to demonstrate the connection of the mean-field $T_c$ to the onset of long-range
correlations. Appendix \ref{ApDemag} discusses some of the effects of
a finite demagnetization factor on specific heat results.


\section{The Dipolar Spin Ice Model: Conventional Metropolis Monte Carlo}
\label{dsim}

In this Section we present the results of Monte Carlo simulations of the
dipolar spin ice Hamiltonian \ref{DSPhamiltonian} using a standard single spin 
flip Metropolis algorithm.  To use Eq.~(\ref{DSPhamiltonian}) within a 
simulation, the dipole-dipole interaction must be handled with care.  
A lattice summation of such interactions
is conditionally convergent due to its $1/R^{3}$ nature.  In order to 
properly handle the long-range nature of this term, we implement the well 
known Ewald method in the simulations, which derives an effective 
dipole-dipole interaction between spins within the simulation cell.\cite{Ewald}
Unlike in dipolar fluid simulations,\cite{Camp} 
the pyrochlore lattice constrains the positions of	
the spins in the simulation, allowing the Ewald interactions to be 
calculated only once, after which a numerical simulation can proceed as normal.
Appendix \ref{ApEwald} contains a brief discussion of the Ewald method 
applied to MC real space simulations. 

\begin{figure}[ht]
\begin{center}
\includegraphics[height=7cm,angle=90]{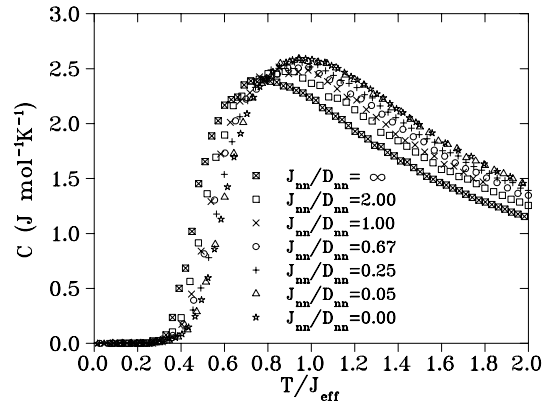}
\end{center}
\caption{Specific heat for system size $L=2$, with temperature, $T$, re-scaled into units of the 
effective nearest neighbor interaction $J_{\rm eff} \equiv J_{\rm nn}+D_{\rm nn}$.
$J_{\rm nn}/D_{\rm nn}=0$ corresponds to purely dipolar interactions, while
$J_{\rm nn}/D_{\rm nn}= \infty$ corresponds to nearest neighbor FM exchange only.
Simulation runs for $L=4$ were also performed, but revealed no important finite size 
effects.}
\label{AllCv}
\end{figure}

\begin{figure}[ht]
\begin{center}
\includegraphics[height=7cm,angle=90]{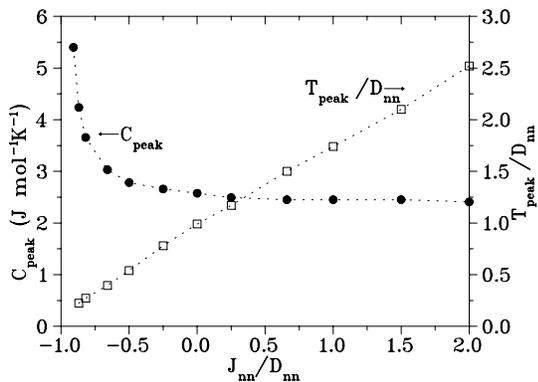}
\end{center}
\caption{Dependence of the simulated specific heat peak height $C_{\rm peak}$
and temperature location of $C_{\rm peak}$ and $T_{\rm peak}$ on exchange and 
dipole-dipole interaction parameters.  In this figure $D_{\rm nn}$ is set to 
2.35K.}
\label{Cdep}
\end{figure}

Simulations on the dipolar spin ice model are carried out using the standard
single spin flip Metropolis algorithm.  To mimic
the experimental conditions pertinent to real materials, the simulation 
sample is cooled slowly. At each temperature step, the system is equilibrated 
carefully, then thermodynamic quantities of interest are calculated.
Since $D_{\rm nn}$ can be determined once the crystal field 
structure of the magnetic ion is known, 
the nearest neighbor exchange $J_{\rm nn}$ 
is the only adjustable parameter in the model. 
The single spin flip Metropolis
algorithm is able to map out three different regions of the phase diagram 
shown in Fig.~\ref{PhaseD}.
Thermodynamic data indicates that when the nearest neighbor exchange is AF and 
sufficiently large compared to the dipolar interactions
($J_{\rm nn}<0$ and $\vert J_{\rm nn}\vert \gg D_{\rm nn}$),
the system undergoes a second order phase transition (in the
three dimensional Ising universality class)
to an all-in or all-out ${\bf q}=0$ ground state.  In the 
spin ice regime,
$J_{\rm eff} = J_{\rm nn}+D_{\rm nn}\gtrsim 0$,
each specific heat data set
for different $J_{\rm nn}$ shows qualitatively the same
broad peak as observed in the nearest neighbor FM exchange model,
\cite{Harris_prl2} which vanishes
at high and low temperatures  (see Fig.~\ref{AllCv}).  
The height, $C_{\rm peak}$, and peak temperature
position of this peak, $T_{\rm peak}$, show very little 
dependence on system size, for simulation cells of 
$L=2,3,4,5$ and $6$.  However, both $C_{\rm peak}$
and $T_{\rm peak}$ are found to depend strongly
on the ratio of $J_{\rm nn}/D_{\rm nn}$, as illustrated in Fig.~\ref{Cdep}.

By re-scaling the temperature scale for the specific heat corresponding to a number of different interaction
parameters $J_{\rm nn}$ and $D_{\rm nn}$,
 one can expose more clearly the dependence of the specific 
heat on the competition between the nearest neighbor exchange $J_{\rm nn}$ and
the dipole-dipole interactions.  In Fig.~\ref{AllCv}, this
dependence is illustrated in the regime $J_{\rm nn}/D_{\rm nn} >0$.  This figure shows
that in terms of an effective energy scale, $J_{\rm eff}$, 
the medium to long-range effects of 
the dipolar interactions are in some sense ``screened'' by the system, and one recovers 
qualitatively the short range physics of the nearest neighbor spin ice model.
As the nearest neighbor exchange interaction becomes AF 
(see Fig.~\ref{Cdep} for $J_{\rm nn}/D_{\rm nn}<0$), we find that the 
approximate collapse onto a single energy scale becomes less accurate, with
the specific heat becoming dependent on $J_{\rm nn}/D_{\rm nn}$.  
It is within this regime that we believe that both 
Ho$_{\rm 2} $Ti$_{\rm 2}$O$_{\rm 7}$ 
and Dy$_{\rm 2}$Ti$_{\rm 2}$O$_{\rm 7}$ are realized, as we now discuss.

\begin{figure}[ht]
\begin{center}
\includegraphics[height=7cm,angle=90]{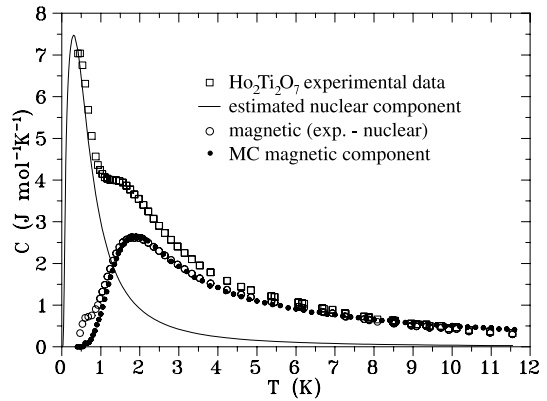}
\end{center}
\caption{The total experimental specific heat of 
Ho$_{\rm 2}$Ti$_{\rm 2}$O$_{\rm 7}$ is shown by the open squares.  The expected
nuclear contribution is indicated by the line, while the resulting
magnetic specific heat estimation is shown by the open circles.  Near 0.7 K the 
estimation is prone to a large error.  Dipolar spin ice simulation results are 
indicated by the filled circles.}
\label{HoCv}
\end{figure}

Since $D_{\rm nn}$ is calculated from Eq.~(\ref{Dstrength}), $J_{\rm nn}$
must be determined from experimental data.
By fitting either the height $C_{\rm peak}$ or the peak temperature $T_{\rm peak}$ of the
maximum of the specific heat curves of the Monte Carlo simulation to the 
experimental results \cite{Ramirez} (Fig.~\ref{ResEnt}a), 
we obtain a value of $J_{\rm nn} = -1.24 {\rm K}$ for 
Dy$_{\rm 2}$Ti$_{\rm 2}$O$_{\rm 7}$.  The results of this fitting are 
illustrated in the top panel of Fig.~\ref{ResEnt}.  
A fitting of the height or peak temperature of the experimental magnetic 
contribution to specific heat for 
Ho$_{\rm 2}$Ti$_{\rm 2}$O$_{\rm 7}$ 
gives $J_{\rm nn} = -0.52 {\rm K}$ for this material.\cite{Bramwell_prl}
Contrary to what is reported in Ref.~\onlinecite{Siddharthan_prb}, we,
therefore, conclude that 
Ho$_{\rm 2}$Ti$_{\rm 2}$O$_{\rm 7}$ is ``deeper'' 
($J_{\rm eff}$ more positive for Ho$_{\rm 2}$Ti$_{\rm 2}$O$_{\rm 7}$ than
for Dy$_{\rm 2}$Ti$_{\rm 2}$O$_{\rm 7}$) in the spin ice regime
(farther to the right in Fig.~\ref{PhaseD}) than 
Dy$_{\rm 2}$Ti$_{\rm 2}$O$_{\rm 7}$ .
As initially reported in Ref.~\onlinecite{cond-mat-Ho},
the temperature dependence of the specific heat
for Ho$_{\rm 2}$Ti$_{\rm 2}$O$_{\rm 7}$, is 
less straightforward to interpret than for 
Dy$_{\rm 2}$Ti$_{\rm 2}$O$_{\rm 7}$.\cite{Ramirez}
In Ho$_{\rm 2}$Ti$_{\rm 2}$O$_{\rm 7}$ the specific heat
possesses an important contribution from a nuclear component due to a large
 hyperfine splitting of the nuclear levels well known to occur for 
Ho$^{3+}$ cations, as discussed in
Ref.~\onlinecite{Blote} and Ref.~\onlinecite{Jensen}. This nuclear component was
estimated by Bl\"{o}te {\it et al.} \cite{Blote} for
Ho$_2$GaSbO$_7$. By subtracting it off from the
(total) experimental specific heat, we can uncover the underlying
magnetic contribution and compare to the theoretically calculated Monte Carlo 
specific heat data, from which $T_{\rm peak}$ or $C_{\rm peak}$ 
can be determined directly (Fig.~\ref{HoCv}).
We note here,
as recently observed in Ref.~\onlinecite{Hodges}, 
that for Dy$_{\rm 2}$Ti$_{\rm 2}$O$_{\rm 7}$ there should be
a hyperfine nuclear contribution to the specific heat manifesting itself at
a temperature below $T\lesssim 0.4$ K in the data of Ref.~\onlinecite{Ramirez} 
if one uses the typical hyperfine contact interaction expected for
a Dy$^{3+}$ insulating salt. The absence of the high temperature $1/T^2$ tail
of the nuclear specific heat (on the descending low temperature side of the
magnetic specific heat of Dy$_{\rm 2}$Ti$_{\rm 2}$O$_{\rm 7}$) below $0.4$ K
in Fig. 3a is, therefore, somewhat puzzling.\cite{Blote_note}

The shoulder-like feature in the estimated magnetic contribution
to the experimental specific heat data 
of Fig.~\ref{HoCv} (open circles) near 0.7 K can be entirely
eliminated by adjusting			
the nuclear hyperfine splitting by $\sim 2$\% 
percent around the
value estimated by Bl\"ote for Ho$_2$GaSbO$_7$, resulting in
an exceedingly good agreement with the Monte Carlo 
results down to $T=0.4$ K. 
Such a slight adjustment to account for any small deviations 
in the hyperfine parameters of $4f$ rare-earth ions (dependent
upon electric field gradients, chemical shift, etc.) would seem
reasonable.  
However, we do not do this in order 
to emphasize that the unbiased use of the estimated nuclear
specific heat contribution from the isostructural material Ho$_2$GaSbO$_7$
 \cite{Blote} already allows for a very good agreement with the
theoretical magnetic specific heat.

Having determined $J_{\rm nn}$ and $D_{\rm nn}$ for 
Ho$_{\rm 2}$Ti$_{\rm 2}$O$_{\rm 7}$ from specific heat measurements,
 we are able to compare
the experimental elastic neutron scattering against that
determined via the Monte Carlo simulations. 
The results, reported in Ref.~\onlinecite{Bramwell_prl}, show excellent
agreement between experiment and simulation. More recent neutron
scattering experiments on the Ho$_{\rm 2}$Sn$_{\rm 2}$O$_{\rm 7}$ show 
similar results.\cite{Ho2Sn2O7_Kadowaki}
Such comparison between theory and experiments for
Dy$_{\rm 2}$Ti$_{\rm 2}$O$_{\rm 7}$ is more difficult due to
the large neutron absorption cross section of naturally occurring
Dy isotopes.  Work in that direction using isotopically enriched 
samples with $^{162}$Dy isotope is in progress.\cite{Dy162}

Numerical integration of the specific heat 
divided by temperature can be performed to determine the entropy of 
both Ho$_{\rm 2}$Ti$_{\rm 2}$O$_{\rm 7}$  
and Dy$_{\rm 2}$Ti$_{\rm 2}$O$_{\rm 7}$.
Specifically, the entropy, $S(T)$, removed between temperature
$T_1$ and $T_2$,  $S(T_2)-S(T_1)$, can be calculated using
the thermodynamic relation:
\begin{equation}
S(T_2)-S(T_1)=\int_{T_1}^{T_2}\frac{C(T)}{T} \; dT.
\end{equation}
The results for Dy$_{\rm 2}$Ti$_{\rm 2}$O$_{\rm 7}$ are illustrated in 
Fig.~\ref{ResEnt}b.
The entropy recovered between $T=0.4$K, where
$C(T)$ is very small, up to a temperature $T=10$K, is
$S(T=10{\rm K})-S(T\approx 0) \approx 3.930$ J mol$^{-1}$ K$^{-1}$
As we can see in Fig.~\ref{ResEnt}b, 
the Monte Carlo data for $S(T)$ at $T=10$ K 
is slightly below the Pauling's value ${\rm R}\{\ln(2)-(1/2)\ln(3/2)\}$.
To perform the calculation of the recovered entropy between
$T=10$ K  up to $T=\infty$, 
we extrapolate the temperature specific heat $C(T)$ for $T>10$ K
by matching the Monte Carlo value of $C(T)$ at $T=10$ K with the 
$1/T^2$ high temperature paramagnetic temperature regime, 
$C(T) = C_\infty /T^2$ for $T > 10$K. 
This gives a value $C_\infty = 29.015$ J mol$^{-1}$ K, 
and an extra entropy of $S(T=\infty)-S(T=10) = 0.145$ J mol$^{-1}$ K$^{-1}$, 
hence a value $S(T=\infty)-S(T\approx 0) = 4.075$ J mol$^{-1}$ K$^{-1}$, 
in exceedingly close agreement with Pauling's value, 
$4.077$ J mol$^{-1}$ K$^{-1}$. Hence, we find that the simulation with 
the appropriate experimental parameters retains Pauling's entropy 
(Eq.~(\ref{PaulingENT})), similar to
what is found experimentally for Dy$_{\rm 2}$Ti$_{\rm 2}$O$_{\rm 7}$ 
(Fig.~\ref{ResEnt}b and in Ref.~\onlinecite{Ramirez}).  
A similar experimental procedure was done using the magnetic contribution of
the specific heat data of Ho$_{\rm 2}$Ti$_{\rm 2}$O$_{\rm 7}$, also giving
a residual entropy close to Pauling's entropy.\cite{Cornelius}

While the above conventional Monte Carlo
simulations of the model Hamiltonian 
for the spin ice compounds, Eq.~(\ref{DSPhamiltonian}), yields a 
reasonably successful quantitative theory of spin ice behavior in Ising
pyrochlore materials, there still remains the second question 
(\# 2, Section \ref{sec-into3}) as to why dipolar 
interactions, despite their anisotropic and long-range nature,
do not (appear to) lift the macroscopic degeneracy associated with
the ice rules, and select an ordered state. 
As a first attempt to address this, we investigate the 
spectrum of soft modes (i.e. critical modes or ordering wave vectors) 
accessible to the dipolar spin ice model within the context of mean-field
theory.

\section{Mean Field Theory}
\label{sect-mft}

In this section we present the main results of a mean-field theory
calculation aimed at determining the spectrum of soft modes 
in the dipolar spin ice model. The details
of the method can be found elsewhere.\cite{Reimers,Enjalran,ABHarris}

The MC simulation results presented in the previous section answer 
in the affirmative the question as to whether or not long-range dipole-dipole interactions
in real materials are consistent with the manifestation of spin ice behavior 
in a temperature range $0<T<\theta_{\rm CW}$. However, these results do not
address the question of whether or not a true ground state degeneracy and 
failure to order at any nonzero temperature
is an exact symmetry consequence of the long-range dipolar interactions for 
$\langle 111 \rangle$ Ising
spins on the pyrochlore lattice.  A direct way to address this question is to ask whether or
not there actually exists at the Gaussian level (i.e. MFT) a soft or critical
mode in the model at a well-defined ordering wave vector ${\bf q}$.
The results of this calculation for finite distance cut-off, $R_c$, 
of the dipole-dipole interactions have been reported in a conference 
proceedings.\cite{Gingras_cjp} We briefly review the essence 
of the calculation and extend it to untruncated (true long-range $1/R^3$) 
dipole-dipole interactions using the Ewald summation technique.
In MFT the Ewald technique is implemented in ${\bf q}$-space, in contrast
to real space for MC simulations. The approach is briefly discussed in
Appendix \ref{ApEwald}.
 
Our MF derivation begins with the Hamiltonian for 
$\langle 111 \rangle$ Ising spins, Eq.~(\ref{DSPhamiltonian}), but 
expressed in terms of the Ising variables, $\sigma_{i}^{a}$, and local quantization 
axes, $\hat{z}^a$,
\begin{equation}
H = - \frac{1}{2}\sum_{i,j} \sum_{a,b}
{\mathcal J}^{a b}(i,j) \sigma_{i}^{a} \sigma_{j}^{b} \; ,
\label{HIsing}
\end{equation}
where
\begin{eqnarray}
\label{J111}
{\mathcal J}^{a b}(i,j) & = & J (\hat{z}^{a} \cdot \hat{z}^{b}) 
\delta_{{\bf R}_{ij}^{ab},{\bf R}_{nn}} \\
&-& D_{dd}
\left ( \frac{\hat{z}^{a} \cdot \hat{z}^{b}}{|{\bf R}_{ij}^{ab}|^3}
- \frac{3(\hat{z}^{a} \cdot {\bf R}_{ij}^{ab})
(\hat{z}^{b} \cdot {\bf R}_{ij}^{ab})}{|{\bf R}_{ij}^{ab}|^5} \right ).
\nonumber
\end{eqnarray}
Recall that indices $a$ and $b$ denote the sub-lattice
and ${\bf R}_{ij}^{ab}$ is the vector that connects spins 
$\sigma_i^a$ and $\sigma_j^b$.
The pyrochlore lattice is a non-Bravais lattice which is described
in a rhombohedral basis with four atoms per unit cell located at
positions ${\bf r}^a$ given by 
$(0,0,0)$,
$(1/4,1/4,0)$,
$(1/4,0,1/4)$, and
$(0,1/4,1/4)$ 
in units of the conventional cubic unit cell of size $a=R_{\rm nn}\sqrt{8}$. 
Each of these four points define a fcc sub-lattice of cubic unit cell size $a$.
Using $H$ from Eq.~(\ref{HIsing}), we form the free energy of our
system,  
\begin{equation}
F = {\rm Tr} \{\rho H \} + T {\rm Tr}\{\rho \ln \rho \},
\label{free-energy}
\end{equation}
where $\rho$ is the many-body density matrix. 

The first step in the mean-field approximation entails replacing $\rho$ with the product of 
single-particle density matrices ($\rho(\{ \sigma_i^a \})=\prod_{i,a} \rho_i^a(\sigma_i^a)$).
Next, a variational free energy is obtained by treating the $\rho_i^a(\sigma_i^a)$ as
variational parameters subject to the constraints ${\rm Tr}\{ \rho_i^a\}=1$ and 
${\rm Tr}\{ \rho_i^a \sigma_i^a\}=m_i^a$, where $m_i^a$ is the local magnetization,
or order parameter. The resulting variational free energy is transformed to momentum space. 
Our convention for the Fourier transform employs the position of the local magnetization, 
${\bf R}_i^a$, 
\begin{equation}
\label{FTm}
m_{i}^{a} = \sum_{\bf q} m_{\bf q}^{a} e^{-\imath {\bf q} \cdot {\bf R}_{i}^{a}},
\end{equation}
from which the spin interaction matrix can be expressed in terms of its Fourier components,
\begin{equation}
{\mathcal J}^{a b}(i,j) = \frac{1}{N_{\rm cell}}
\sum_{\bf q} {\mathcal J}^{a b}({\bf q}) e^{\imath {\bf q} \cdot {\bf R}_{ij}^{ab}} \ ,
\label{FTJ}
\end{equation}
where $N_{\rm cell}$ is the number of four atom unit cells, i.e., 
fcc lattice points.
We note that the above convention for the Fourier transform produces 
a symmetric $4\times4$ 
matrix ${\mathcal J}({\bf q})$ at all values of ${\bf q}$. 
An alternate convention
defining the Fourier transform with respect to the Bravais lattice points, 
${\bf R}_{i}$, results
in a complex (Hermitian) ${\mathcal J}({\bf q})$,
as discussed in Ref.~\onlinecite{Enjalran}.
From Eqs.~(\ref{FTm}) and (\ref{FTJ}), we write the 
quadratic part of the MF free energy, 
$F^{(2)}$, 
\begin{equation}
\label{eq-f2}
f^{(2)}(T)= \frac{1}{2} \sum_{\bf q} \sum_{a,b} m_{\bf q}^a 
\left\{ T\delta^{ab}-{\mathcal J}
^{ab}({\bf q})\right\}m_{-\bf q}^b 
\;\;\; ,
\end{equation}
where $f^{(2)}(T)=F^{(2)}(T)/N_{\rm cell}$ and $T$ is the temperature in units of $1/k_{\rm B}$.
Diagonalizing $f^{(2)}(T)$ requires transforming to the normal modes of the 
system,
\begin{equation}
m_{\bf q}^a = \sum_{\alpha=1}^{4}
U^{a,\alpha}({\bf q}) \Phi_{\bf q}^\alpha 	\;\;\; ,
\end{equation}
where the Greek index $\alpha$ labels normal modes, 
$\{ \Phi_{\bf q}^\alpha\}$ are the amplitudes of the 
normal modes, and
$U({\bf q})$ is the unitary matrix that diagonalizes 
${\mathcal J}({\bf q})$
in the sub-lattice space, with eigenvalues $\lambda({\bf q})$,
\begin{equation}
U^{\dag}({\bf  q}){\cal J}({\bf q})U({\bf  q})=\lambda({\bf q})
\;\; .
\label{JqNM}
\end{equation}
In component form, $U^{a,\alpha}({\bf  q})$ represents the
$a$-component of the $\alpha$-eigenvector.
We express $f^{(2)}(T)$ in terms of normal modes as
\begin{equation}
\label{eq-f2q}
f^{(2)}(T) =\frac{1}{2}\sum_{\bf  q} \sum_\alpha 
\Phi_{\bf q}^\alpha 
\left\{ T - \lambda^\alpha({\bf q}) \right \} \Phi_{-\bf q}^\alpha
\;\;\; .
\end{equation} 
In our approach, a minus sign was pulled out in front of the  Hamiltonian in Eq. 6, 
therefore, an ordered state first occurs at the temperature defined 
by the global maximum eigenvalue, 
\begin{equation}
T_c = {\rm max}_{{\bf q}} \{\lambda^{max}({\bf q})\}
\;\;\; ,
\label{Tc}
\end{equation}
where $\lambda^{max}({\bf q})$ is the largest of the four eigenvalues 
($\alpha =1,2,3,4$) at wave vector ${\bf q}$, and 
$\max_{{\bf q}}$ indicates the
global maximum of the spectrum of 
$\lambda^{max}({\bf q})$ for all ${\bf q}$.
The value of ${\bf q}$ for which $\lambda^\alpha({\bf q})$ is maximum is
the ordering wavevector ${\bf q_{\rm ord}}$.

\begin{figure}[ht]
\begin{center}
\includegraphics[height=6.0cm,width=7cm]{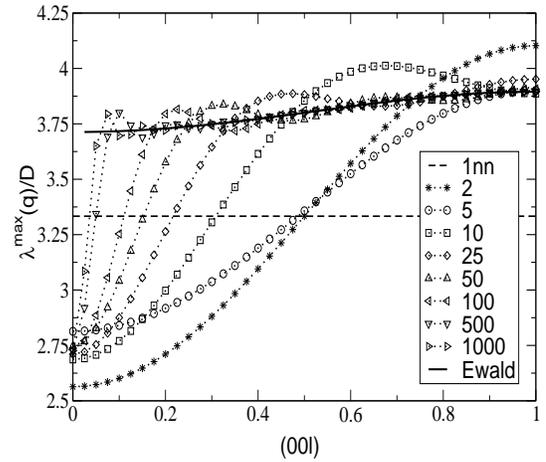}
\end{center}
\caption{$\lambda^{max}({\bf q})/D$ vs.
${\bf q}$ for ${\bf q}$ in the $(00l)$ direction, in units of $2\pi/a$,
for various cut-off distances, $R_c$, and for the infinite
range limit simulated by the Ewald summation technique. Calculations
were made at $J_{\rm nn}/D_{\rm nn}=0$ on the spin ice side of the
phase diagram, Fig.~\ref{PhaseD}.}
\label{ewE_00l}
\end{figure}
In Ref.~\onlinecite{Gingras_cjp}, we evaluated ${\mathcal J}({\bf q})$ in the 
spin ice regime of our model ($J_{\rm nn}/D_{\rm nn}=0$) directly from the 
inverse transform of Eq.~(\ref{FTJ}) for various real-space cut-off distances
of the dipolar term. For the rather dramatic approximation
of nearest-neighbor distances, we found a complete degeneracy (${\bf q}$-independence) of
$\lambda^{\rm max}$ over the whole Brillouin zone. This corresponds to 
the ${\bf q}$-space signature of the 
degenerate nearest-neighbor spin ice model of Harris and Bramwell,
\cite{Harris_jpc} and, equivalently, of the nearest-neighbor (global) Ising antiferromagnet
model of Anderson.\cite{Anderson}  As $R_c$ was increased,
a $R_c$-dependent ${\bf q_{\rm ord}}$ appeared.  In the limit of
infinite range interactions, i.e., $R_c \rightarrow \infty$,
we predicted ${\bf q_{\rm ord}}=(001)$ (in units of $2\pi/a$). 
Results for $\lambda^{\rm max}$ along the 
$(00l)$ direction as a function of $R_c$ are shown in Fig.~\ref{ewE_00l}.

\begin{figure}[ht]
\begin{center}
\includegraphics[width=8cm]{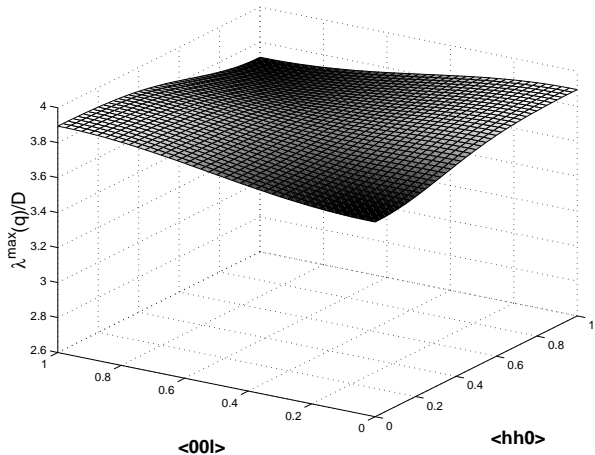}
\end{center}
\caption{The scaled maximum eigenvalues, $\lambda^{max}({\bf q})/D$, in the $(hhl)$ plane 
for the same spin ice model described in Fig.~\ref{ewE_00l}.  
The dipole-dipole interactions are treated with the Ewald approach. 
} 
\label{EVhhlEW}
\end{figure}
Treating dipolar interactions via the Ewald method,
where the true long-range nature of the interactions are respected, 
we observe a completely smooth (without ripples) and 
quasi-degenerate soft mode spectrum 
with a global maximum (critical mode) at $(001)$. 
The spectrum of $\lambda^{\rm max}({\bf q})/D$ in the $(hhl)$ plane of the
pyrochlore lattice for $J_{\rm nn}=0$ is shown in Fig.~\ref{EVhhlEW}, while the
results along the $(00l)$ direction
are included in Fig.~\ref{ewE_00l}.  
Inspection of corresponding eigenvectors of 
the doubly degenerate critical mode  
indicate a two-in two-out spin ice structure, where the spins 
on sub-lattices $a=1,3$ point opposite to those on sub-lattices $a=2,4$ 
in a tetrahedral unit, see Table \ref{tab-evecs}.
Ref.~\onlinecite{Enjalran} discusses how these soft modes can be 
used to reconstruct the long-ranged ordered (equal moment) structure,
where the thermal local magnetization is the same on all sites, and which 
corresponds to the long-range ordered two-in two-out spin ice state of Fig.~4. 

With the implementation of the Ewald method, we are able to study the symmetry 
properties of the long-range dipolar interactions in a controlled manner. 
This is a desirable feature for a problem like spin ice because dipole-dipole 
interactions beyond nearest neighbor distances introduce 
perturbations onto the highly degenerate soft mode spectrum of the parent state, 
the nearest neighbor spin ice model. The form of the dipolar interaction, 
which couples spin and spatial degrees of freedom, permits both positive and
negative contributions to the total dipolar energy as $R_c$ is increased,
so the interactions are self-screening. However, this symmetry is not exact and 
the value for ${\bf q}_{\rm ord}$ in MFT depends crucially but unpredictably
on how far the sum is carried out. The variability 
in ${\bf q}_{\rm ord}$ is readily observed for short cut-off distances,
e.g., $R_c \stackrel{<}{\sim} 100{\rm nn}$.
However, subtle effects are still present even at very large cut-off distances 
and can produce an anomalous ordering wave vector. With the dipolar 
interactions cut-off at $R_c=1000{\rm nn}$, an incommensurate critical 
mode is found at ${\rm q} \sim (0.025,0.025,1.0)$. This mode 
is singly degenerate (unlike the doubly degenerate mode at $(001)$ in the
Ewald limit) and the eigenvectors do not predict a two-in two-out spin ice structure. 
We underscore, again, that this effect is subtle, as is demonstrated by 
the difference in the eigenvalues at $R_c=1000{\rm nn}$, 
$\lambda^{\rm max}(0.025,0.025,1.0) - \lambda^{\rm max}(0,0,1) \approx 
3\times 10^{-4}$. 

In magnetic systems, the development of spin-spin correlations 
can be observed in the ${\bf q}$-dependent susceptibility, $\chi({\bf q})$.
In the Gaussian approximation one has 
$\chi({\bf q}) \propto (T - \lambda^{\alpha}({\bf q}))^{-1}$
(see Appendix \ref{ApHTSE}). Therefore, the critical mode that 
minimizes the free energy, $f^{(2)}(T)$, also controls the
magnetic correlations as $T \rightarrow T_c$, with $T_c$ 
given by Eq.~(\ref{Tc}). In the paramagnetic 
(PM) regime,  $T \gg \theta_{\rm CW}$, one expects all modes to 
contribute to $\chi({\bf q})$. This is also
the regime in which MFT applies, $T > T_c$, and 
is expected to provide quantitatively accurate results.
Therefore, a build up of PM correlations is understood in terms
of an underlying critical mode that marks a transition to an ordered
state at $T_c$.  
This mean-field approach has been used to study the PM elastic neutron scattering 
in the $\langle 111 \rangle$ pyrochlores.\cite{Enjalran}
For the dipolar spin ice model, one obtains results that
are in agreement with the experiments and simulations of Ho$_2$Ti$_2$O$_7$ 
for $R_c > 250{\rm nn}$.\cite{Bramwell_prl} The best results are 
found when dipoles are treated with the Ewald technique. 
If one employs too small a cut-off distance,  
e.g., much less than $R_c \approx 100{\rm nn}$, then one finds that the PM scattering 
is concentrated in regions of the first zone that are inconsistent with experiments.
This is especially true at cut-off distances studied in 
Refs.~\onlinecite{Siddharthan_prl} and \onlinecite{Siddharthan_prb}. 
Again, we believe this is a direct consequence of the failure of a finite 
dipolar sum to much restore the symmetry of the dipolar Hamiltonian, i.e.,
to produce a quasi-degenerate $\lambda^{\rm max}({\bf q})$-spectrum.

\begin{table}
\protect \caption
{The two maximum eigenvalues and corresponding eigenvectors
of ${\mathcal J}({\bf q}_{\rm ord})$ at ${\bf q}_{\rm ord}=(001)$ for spin ice with Ewald
evaluation of the dipolar interactions.
The $U^{a,\alpha}({\bf q}_{\rm ord})$ are normalized
eigenvectors, where the weights indicate a two-in (positive value) two-out
(negative value) spin ice structure. In particular, the spins on sub-lattices
$a=1$ and $a=3$ point out of and spins on sub-lattices $a=2$ and $a=4$
point in to the tetrahedron.}
\label{tab-evecs}
\begin{center}
\begin{tabular}{cp{5mm}cp{5mm}c} 
$\alpha$ && $\lambda^{\alpha}({\bf q}_{\rm ord})/D$ && $U^{a,\alpha}({\bf q}_{\rm ord})$  \\ \hline
$1$ && $3.1575$ && $\frac{1}{\sqrt{2}}(-1,1,0,0)$ \\
$2$ && $3.1575$ && $\frac{1}{\sqrt{2}}(0,0,-1,1)$
\end{tabular}
\end{center}
\end{table}

From MFT, we can also determined the value of 
$J_{\rm nn}/D_{\rm nn}$ at which the ordering changes from an 
``all-in$-$all-out'' ${\bf q}=0$
state (i.e., from large negative antiferromagnetic $J{\rm nn}$) to 
the $(001)$ long-range ordered spin ice state. 
We find that the transition between the two states
occurs at $J_{\rm nn}/D_{\rm nn} = -0.905$. 
This is in full agreement with the value found 
in Monte Carlo simulations results for the transition between
all-in$-$all-out ${\bf q}=0$ ordering and spin ice behavior.\cite{Hertog}

Having obtained strong evidence from MFT that there exists a
well-defined, unique ordering wave vector in the long-range dipolar spin ice 
model at the Gaussian level, we can proceed with our search for a 
dynamically-inhibited transition to long-range order in the model 
by the use of a conventional Metropolis single spin-flip 
MC simulation.

\section{Dynamical Freezing and Loop Moves in Monte Carlo Simulations}

\subsection{Dynamical Freezing in Conventional Single Spin Flip Monte Carlo Simulations}

The mean field theory results presented in the previous section make
it clear that from the point of view of a strictly equilibrium 
(statistical mechanics) magnetic ordering phenomenon,
the dipolar spin ice model of Eq.~(\ref{DSPhamiltonian}) 
should be a rather conventional system with a 
unique and well defined ordering wavevector and staggered magnetization
order parameter. The question then becomes: why
don't Monte Carlo simulations of the dipolar spin ice model, or in fact
the real spin ice materials themselves, develop the 
long-range ordered phase predicted by mean field theory?					
The problem turns out to lie in the local single spin flip
dynamics employed within the Metropolis algorithm and, 
similarly, the local spin 
dynamics at play in the real materials.
As we will show below, Monte Carlo simulations 
of the dipolar spin ice model 
using single spin flips experience a dynamical freezing
at low temperatures. This arises due to the 	
existence of large energy barriers separating distinct quasi-degenerate spin 
ice configurations, and prevents the simulation (and the real materials)
from finding its true energetically-favored long-range ordered ground
state (see Ref.~\onlinecite{Shore} for a related problem).

\begin{figure}[ht]
\begin{center}
\includegraphics[height=6.5cm,width=5cm,angle=90]{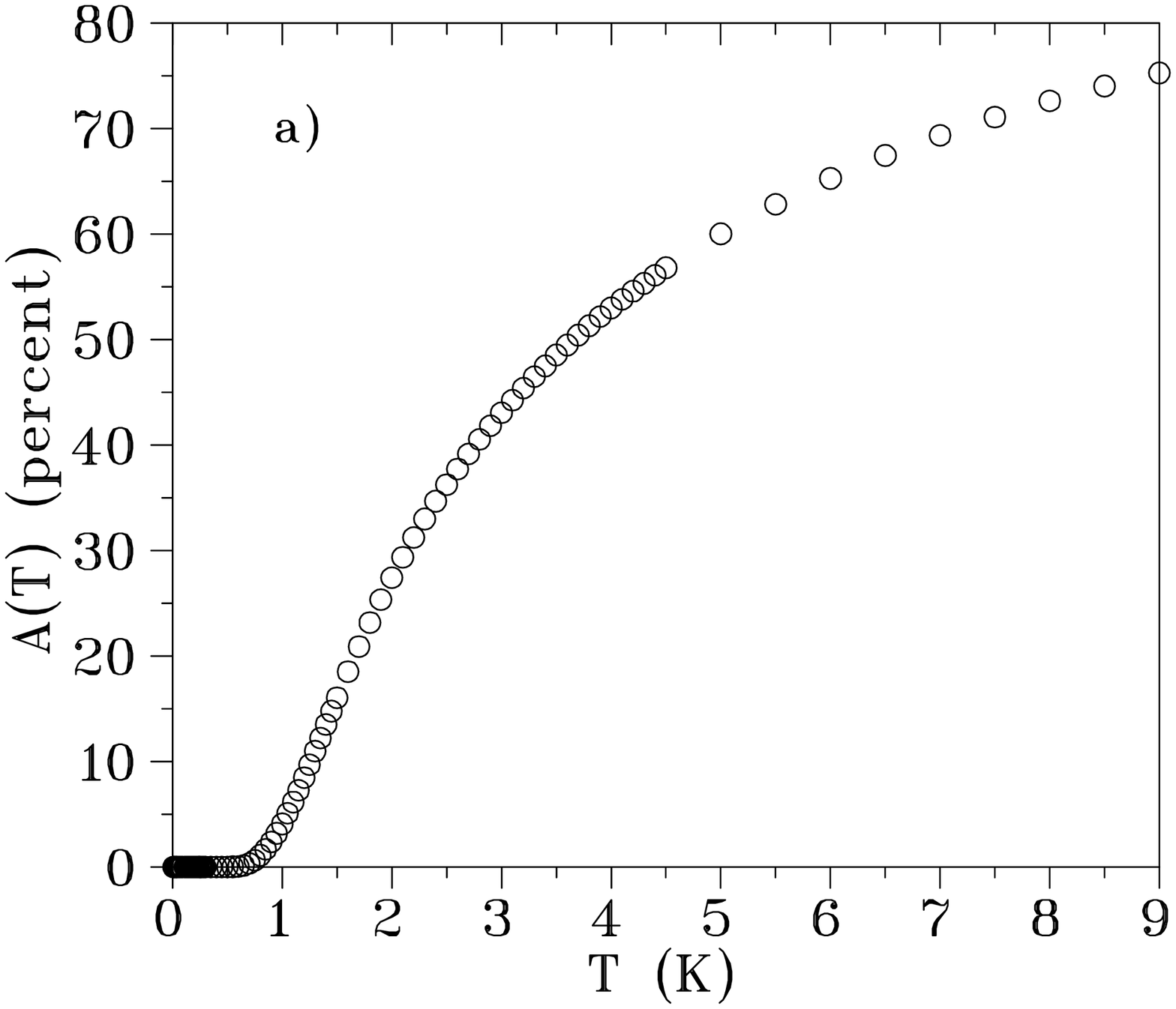}
\includegraphics[height=6.7cm,width=5cm,angle=90]{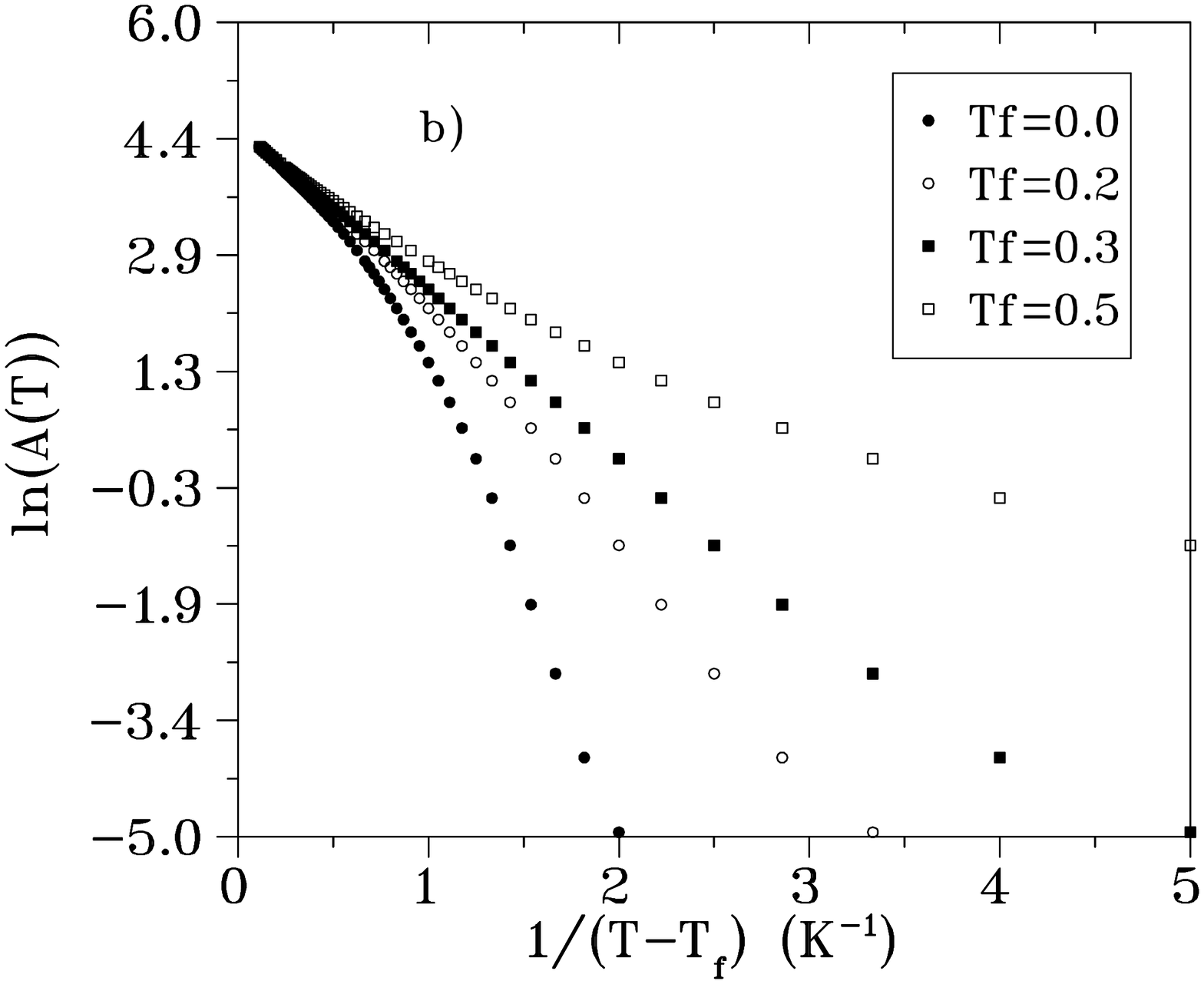}
\end{center}
\caption{(a) Single spin flip Monte Carlo step acceptance rate $A(T)$ for a
simulation of Dy$_{\rm 2}$Ti$_{\rm 2}$O$_{\rm 7}$.  The simulation becomes
frozen when the acceptance rate falls to zero.  (b) The logarithm of the
acceptance rate plotted versus inverse temperature, minus some freezing
temperature $T_{\rm freeze}$, follows a Vogel-Fulcher type law.} 
\label{SSFaccp}
\end{figure}

Observation of the acceptance rate $A(T)$ (percentage of accepted Monte Carlo 
steps) of the dipolar spin ice simulations makes it immediately apparent that  
out-of-equilibrium freezing occurs at low temperatures, 
that is below $T\sim 0.4$ K (as illustrated in Fig.~\ref{SSFaccp}a)
for the $J_{\rm nn}$ and $D_{\rm nn}$ parameters appropriate for 
Dy$_2$Ti$_2$O$_7$. 
Fig.~\ref{SSFaccp}b shows that 
$A(T)$ can be parametrized by a Vogel-Fulcher temperature
dependence as found in numerous freezing phenomena:
$A(T) \propto \text{exp}(\Delta/(T-T_{\rm freeze}))$, where the 
freezing temperature, $T_{\rm freeze}$, is introduced in an {\it ad hoc} 
fashion.  In Fig.~\ref{SSFaccp}b, $\Delta = 1$ K.

It is clear that in order
to investigate the existence of a true energetically-favored
ground state in the dipolar spin ice model, a standard Monte Carlo simulation
employing local single spin flip dynamics is insufficient. Indeed, 
as Fig.~\ref{SSFaccp} shows, these dynamical processes are frozen-out at 
$T$ just slightly below 0.4 K for a model of Dy$_2$Ti$_2$O$_7$.
For $J_{\rm nn}$ and $D_{\rm nn}$ appropriate to describe
Ho$_2$Ti$_2$O$_7$,\cite{Bramwell_prl}
the single spin flip Monte Carlo acceptance rate falls below
10$^{-6}$ at a temperature near 0.6 K. Without ascribing any deep 
significance to it, it is interesting to note that this freezing out
in the simulation at 0.4 K and 0.6 K 
corresponds rather closely to the temperatures 
at which freezing is found in Dy$_2$Ti$_2$O$_7$ \cite{freeze_Dy1} and
Ho$_2$Ti$_2$O$_7$,\cite{freeze_Ho} respectively.
			
This freezing-out occurs due to large free energy barriers 
separating the (almost) degenerate ice-rules states, which develop
rapidly at a temperature of order $T_{\rm peak}$, 
and which are associated with introducing a single 
spin flip to a tetrahedron obeying the ice rules.
As discussed above, and further supported by the mean-field calculation,
the effective (ferromagnetic) nearest neighbor interaction $J_{\rm eff}$ 
favors the ice-rules configuration. As the temperature drops the 
Boltzmann weight $\exp(-4J_{\rm eff}/T)$ suppresses the probability that a 
spin flip will take a given tetrahedron into an intermediate, thermally
activated non ice-rules obeying
configuration.  Thus single spin flip Monte Carlo moves 
are, for all practical purposes, frozen-out and 
dynamically eliminated within the simulation when $T << J_{\rm eff}$.

\subsection{Loop Moves in Monte Carlo Simulations}

In order to explore the low temperature ordering properties 
of dipolar spin ice,
one needs a Monte Carlo algorithm with non-local updates that  
effectively bypass the energy barriers that separate nearly degenerate 
states and allows the simulation to explore the restricted ice-rules
phase space that prevents ordering in the model.
\cite{Melko_prl,Melko_thesis}
In other words, we employ non-local dynamical processes to 
restore ergodicity in the Monte Carlo simulation, and then use
this new algorithm to explore and characterize the long-range ordered 
state that arises out of ice-rules manifold 	
and which is energetically favored by the dipolar spin ice model. 

We first identify the true zero energy modes that can take the near-neighbor spin ice model
 from one ice state to another exactly energetically degenerate ice state.
An example of these zero modes, or loops, is shown in Fig.~1.  
We take as an initial working hypothesis that 
in the dipolar spin ice model, with interactions beyond nearest neighbor, 
the system freezes into an ice-rules obeying state. This is indeed what 
we found: in all of
the tests we performed, systems simulated using conventional single 
spin-flips always froze-out in
an ice-rules obeying state with no ``defects'' 
(by defects we mean violations of the Bernal-Fowler ice rules).  
With interactions beyond nearest-neighbor, these loop moves 
become quasi-zero modes that can take the 
dipolar spin ice model from one ice-rules state to another without introducing 
spin defects into tetrahedra in the lattice.  
This allows all of the quasi-degenerate spin ice states to be sampled 
ergodically, and facilitates the development of a long-range ordered state
by the system at low temperatures.

\begin{figure}[ht]
\begin{center}
\includegraphics[height=5cm]{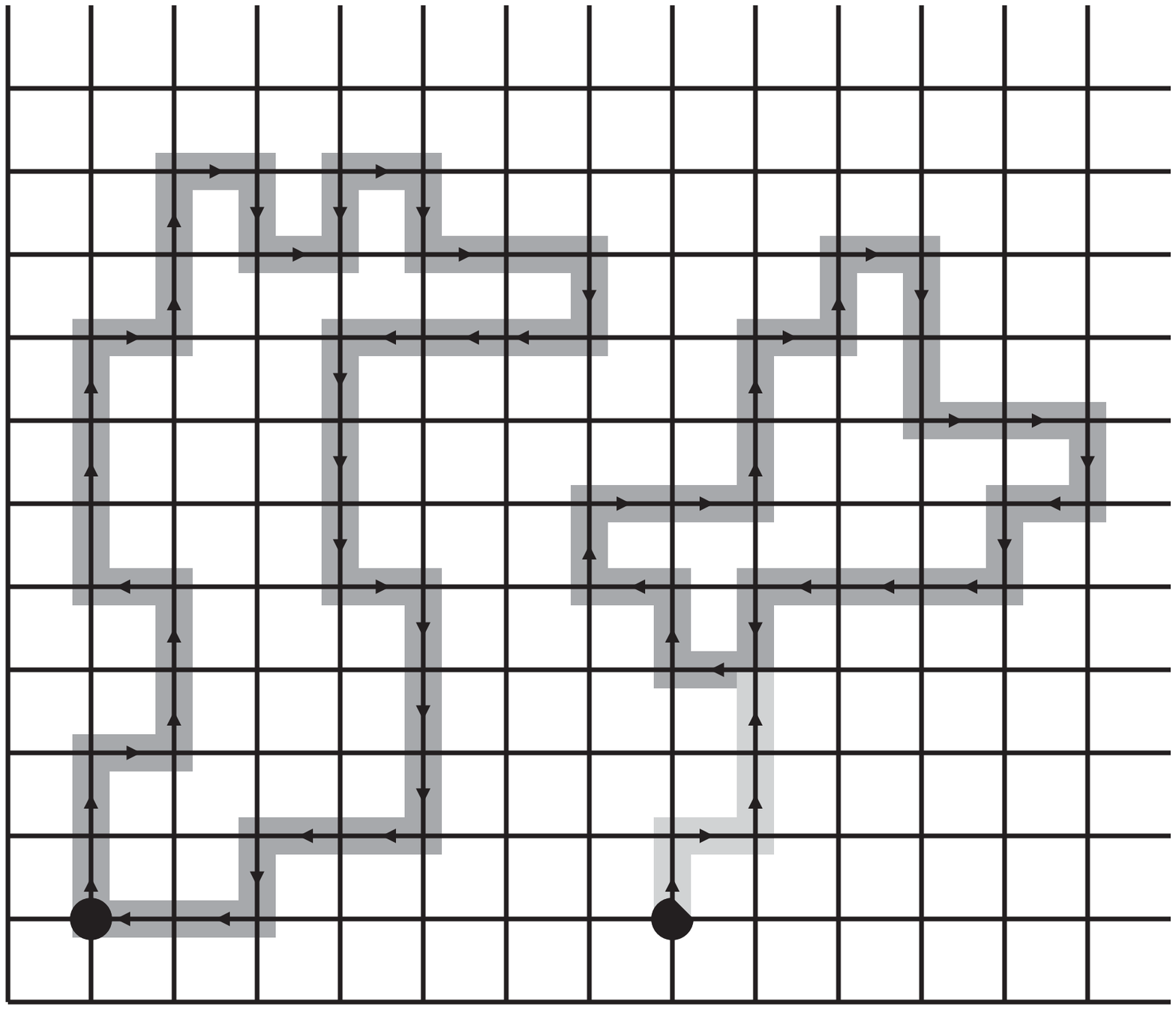}
\end{center}
\caption{Long and short loops formed by the Newmann and 
Barkema algorithm \cite{Barkema,BarkNewBook} on a square ice
lattice.  Vertices are represented by points where 
lattice lines cross.  Each vertex has
two spins pointing in and two spins pointing out, 
however for clarity, only spins which 
are included in the loops are shown.  
Starting vertices are indicated by large black dots.
On the left is an example of a long loop, which is completed 
when it encounters its own
starting vertex.  On the right is a short 
loop, which is complete when it crosses itself
at any point.  Dark gray lines outline completed loops.  
The excluded tail of the short loop is shown in light gray. }
\label{LSloop}
\end{figure}

Within the Monte Carlo simulation, we use the Barkema and 
Newman\cite{Barkema,BarkNewBook}
loop algorithm originally designed for two dimensional square ice models,
and adapt it to work in a similar manner on the three dimensional pyrochlore 
lattice.  In the context of square ice, 
we tested two types of loop algorithms,
the so-called {\it long} and {\it short} loop moves.  
In the square ice model, 
each vertex on a square lattice has four spins associated with
it.  The vertices are analogous to tetrahedron 
centers in the pyrochlore lattice.
The ice rules correspond to ``two spins pointing in, two spins pointing
out'' at
each vertex. In the Newman and Barkema algorithm, a loop is formed by tracing 
a path through ice-rules vertices, alternating between 
spins pointing into and spins pointing
out of the vertices.  A ``long loop'' is completed when the path traced by the 
loop closes upon the same spin from which it started.  A ``short loop'' is 
formed whenever the path traced by the loop encounters any other vertex 
(tetrahedron) already included in the loop -- excluding the dangling tail of 
spins (Fig.~\ref{LSloop}).

We now generalize the Barkema and Newmann loop algorithm
for our study of the three dimensional pyrochlore lattice spin ice problem.
In this system, the 
smallest complete loop that is a zero mode on the pyrochlore lattice 
consists of six spins (see Fig.~\ref{Pyro}). 
Such a loop was previously identified by Bramwell and 		
Harris \cite{Harris_jpc} and also by Anderson\cite{Anderson} 
in the context of the spinel lattice.  
However using the above loop algorithm, much larger loops are possible.
When used with the pyrochlore lattice (Fig.~\ref{Pyro}), such a loop must
pass through two spins on each tetrahedron.  A loop always ``enters'' a 
tetrahedron through an inward pointing spin, 
and ``leaves'' a tetrahedron through 
an outward pointing spin.  The periodic boundary conditions of the lattice may
also be traversed with no ill consequences.  If we form a closed
loop in this manner, and each spin is reversed on it, the entire system stays
in an ice-rules state.  However, small dipole-dipole energy gains or losses may be 
procured due to small energy differences 
between the old and the new ice-rules state.
These small energy changes caused by the loop moves
are evaluated via a Metropolis algorithm within the Monte Carlo.
Specifically, a loop move that takes the system from one ice-rules state to
another one of lower energy is automatically accepted, while a loop move 
that takes the system to a higher energy ice-rule state
(with energy difference $\delta E$ between the two states) is accepted with
$\exp[-\delta E/(k_{\rm B}T)] > {\rm rnd}$, where rnd is a random number taken
from a uniform distribution between 0 and 1.\cite{Barkema,BarkNewBook}
			
Before use in a full-scale Monte Carlo simulation, 
the long and short loop algorithms are subjected to a variety 
of characterizing tests on the three dimensional
pyrochlore lattice.\cite{Melko_thesis}
The first test is a study of the  relative speed (measured by CPU time)
of the algorithms for different sized lattices.  As reported in 
Ref.~\onlinecite{Melko_thesis}, it is found that the small loop algorithm
creates loops that approach a finite size limit as the system
size increases. The long loop algorithm continues creating larger and larger 
loops, that scale approximately linearly with the 
number of spins in the simulation cell.  This forces the algorithm to become 
drastically slower for the larger system sizes considered. 

Second, tests are carried out to investigate 
how the two different loop algorithms 
handle defects that break the ice rules on a tetrahedron.  
As we know, above 
the ``spin ice peak'' in the specific heat of 
the dipolar spin ice model, the ice 
rules (two spins pointing into a tetrahedron, two spin pointing out) are 
generally not obeyed.  However, to retain detailed balance, we want our loop 
algorithm to attempt to form loops at temperatures above the onset of spin ice 
correlations.  The attempt to create a loop is simply aborted in the case 
where the loop path encounters a defect 
(either a three-in one-out vertex, or an all-in or all-out vertex). 
The simulation does not attempt to flip any spins 
on an aborted loop, and the Metropolis algorithm 
is not employed in this case.  
As reported in Ref.~\onlinecite{Melko_thesis},
the result of ice rules defects on loop algorithm performance is significant.  
Within the long loop algorithm, the inclusion of only one defect spin per one 
thousand spins in the system causes almost half of all loops which are 
attempted to be aborted on the grounds that they have encountered the defect,
with efficiency decreasing drastically as more defects are included.
In contrast, the 
short loop algorithm remains 87\%  efficient with the inclusion of one defect 
in one thousand, and retains an efficiency that is much better than
the long loop algorithm as more defects are present in the system.
\cite{Melko_thesis}

\begin{figure}[ht]
\begin{center}
\includegraphics[height=6.5cm,width=5cm,angle=90]{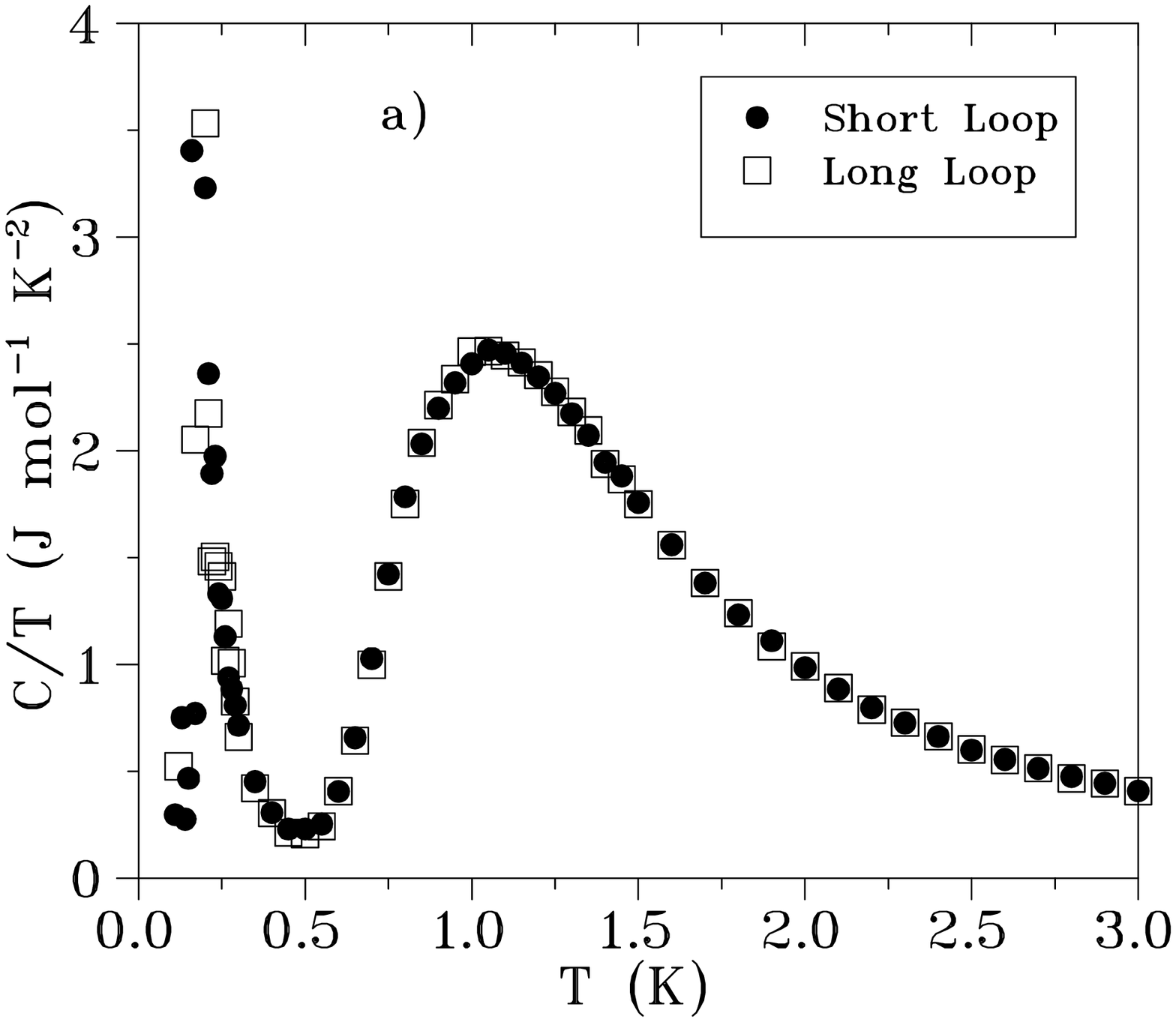}
\includegraphics[height=6.7cm,width=5cm,angle=90]{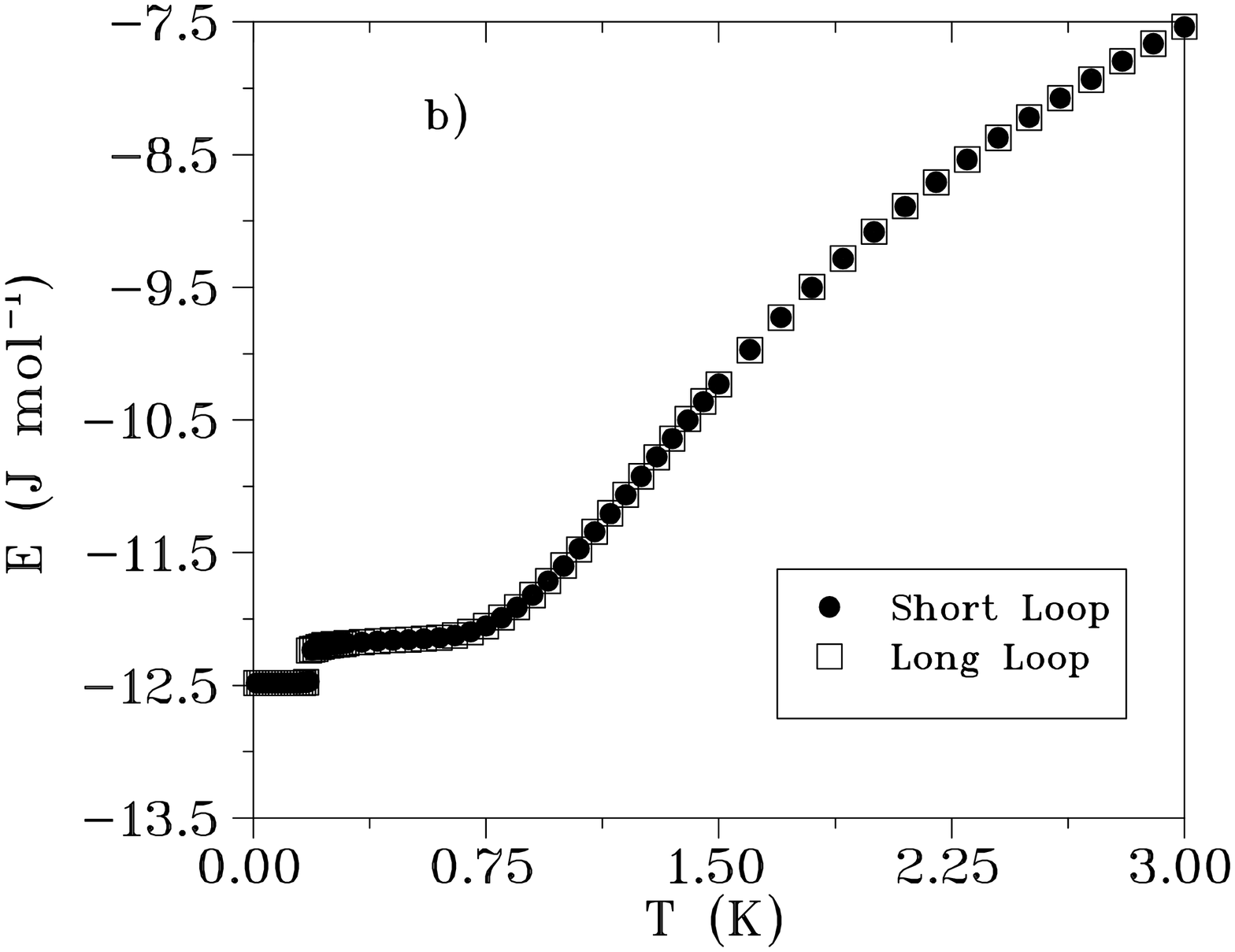}
\caption[Monte Carlo Specific Heat and Energy for the Dipolar Spin Ice Model
 Using Long and Short Loop Moves.]
{Preliminary data for the low temperature magnetic specific heat 
(a) and energy (b) of the dipolar spin ice Monte Carlo, 
system size L=3, with simulation 
parameters set for Dy$_{\rm 2}$Ti$_{\rm 2}$O$_{\rm 7}$.  
The data represent an average taken over approximately $10^5$ 
production Monte Carlo steps.
Closed circles are data from a simulation of the short loop algorithm, 
open squares are data obtained using the long loop algorithm.  
Low temperature features are discussed in the next section.}
\label{ShortLong}
\end{center}
\end{figure}

We use both algorithms to perform a true finite temperature 
Metropolis Monte Carlo simulation of the dipolar spin ice model. 
We code both algorithms separately, and run simulations 
using the regular procedure (cooled 
slowly from a high temperature, equilibrating carefully at every temperature 
step). Our preliminary Monte Carlo results for both the short and long loop 
are given in Fig.~\ref{ShortLong}.
Even though the data in this figure has low statistics (only 
$10^5$ Monte Carlo production steps per spin), it is 
clear that both the short and long loops promote roughly the same thermodynamic
 behavior in the Monte Carlo simulation.  The low temperature features 
(specific heat peak and energy discontinuity at $T\approx 0.2$ K)
of Fig.~\ref{ShortLong}
are induced in the same manner by both algorithms.  These features will be 
discussed in much more detail in Section V. Since both the long and
short loops display equivalent results in the Monte Carlo, 
we are free to choose 
between the two based solely on their performance properties measured above. 
As alluded to above and detailed in Ref.~\onlinecite{Melko_thesis},
the short loop algorithm works more efficiently within the requirements
of our simulation.
However, the disadvantage with using the short loop 
algorithm is that each loop does not cover as large of a percentage of spins 
within the lattice. It is not clear to us, without investigating 
the computational performance (i.e. autocorrelation times) of both
algorithms in much more quantitative detail,
whether a small number of long loops is 
better at bringing the system to equilibrium than a larger number of 
short loops for a fixed CPU time.  However, with the additional observation 
that the long loop algorithm can pass over itself a number of times during its 
creation,
effectively losing additional efficiency in this manner, we ultimately
choose to perform the majority of the simulations on the dipolar spin ice 
model using the short loop algorithm.

\begin{figure}[ht]
\begin{center}
\includegraphics[height=6.5cm,width=5cm,angle=90]{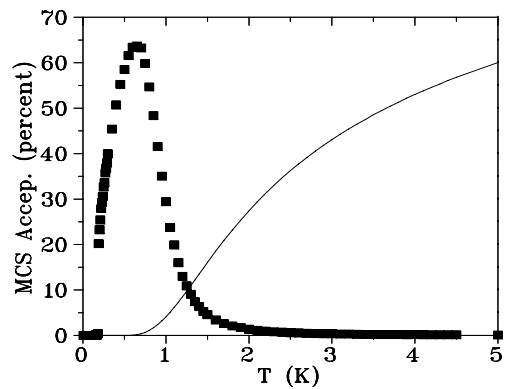}
\caption
{The Monte Carlo acceptance rate for the short loop algorithm (circles)
as compared to the single spin flip algorithm (line), for a simulation using
Dy$_{\rm 2}$Ti$_{\rm 2}$O$_{\rm 7}$ parameters.  Acceptance rates were 
calculated as percentages of attempted Monte Carlo steps (MCS).}
\label{SLaccp}
\end{center}
\end{figure}

With the short loop algorithm chosen for the simulations, 
we re-investigate the Monte Carlo 
Metropolis algorithm acceptance rate. Since each loop successfully 
created (i.e., not aborted by encountering an ice defect) 
by the short loop algorithm is 
still subject to rejection by the Metropolis condition 
on the basis of its change in system energy, 
we expect the maximum acceptance rate to be somewhat 
less than the maximum efficiency of the algorithm given above. 
Results for the loop acceptance rate are shown in Fig.~\ref{SLaccp}.
Clearly, the loop algorithm becomes effective in the temperature range where 
the single spin flip algorithm looses its ability to explore all possible 
configurations of the system.  Above about 1 K, 
the number of accepted loops is 
very low, due to the fact that the system is not 
entirely in an ice-rules configuration.  
As the simulation is slowly cooled, ice-rules constraints 
begin to develop, and the loop algorithm begins to 
work efficiently, moving the 
system between different ice-rules states.  In Fig.~\ref{SLaccp}, 
a sharp drop is observed in the loop			
acceptance rate at approximately 0.18 K.  
As discussed in the next section, this corresponds to 
the temperature where a phase transition develops in the system, 
which locks the system into a long-range ordered
state and eliminates Monte Carlo dynamics once and for all.
			
\section{Loop Monte Carlo Investigation of the Transition to Long Range Order}

\begin{figure}[ht]
\begin{center}
\includegraphics[height=6.5cm,width=5cm,angle=90]{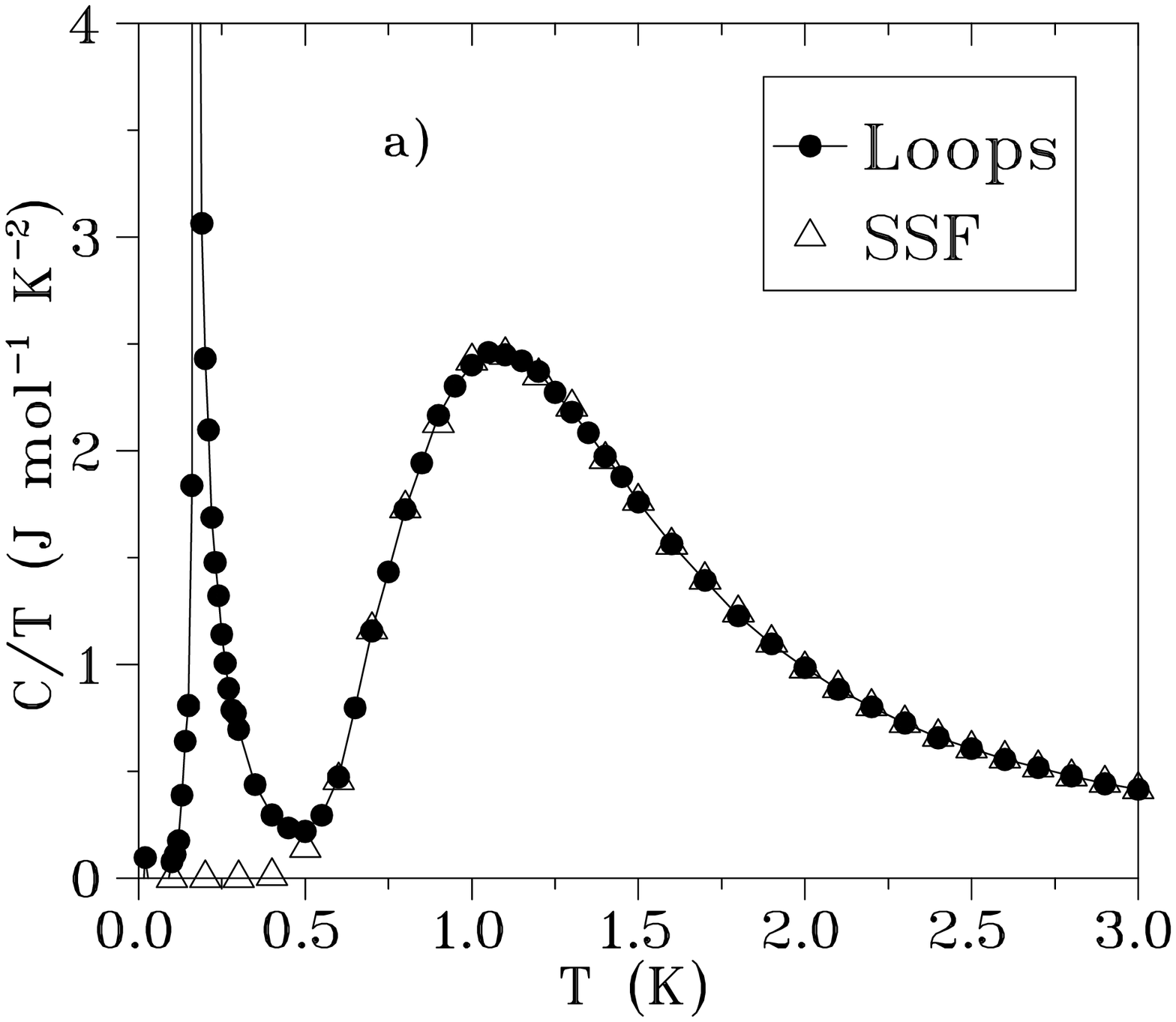}
\includegraphics[height=6.7cm,width=5cm,angle=90]{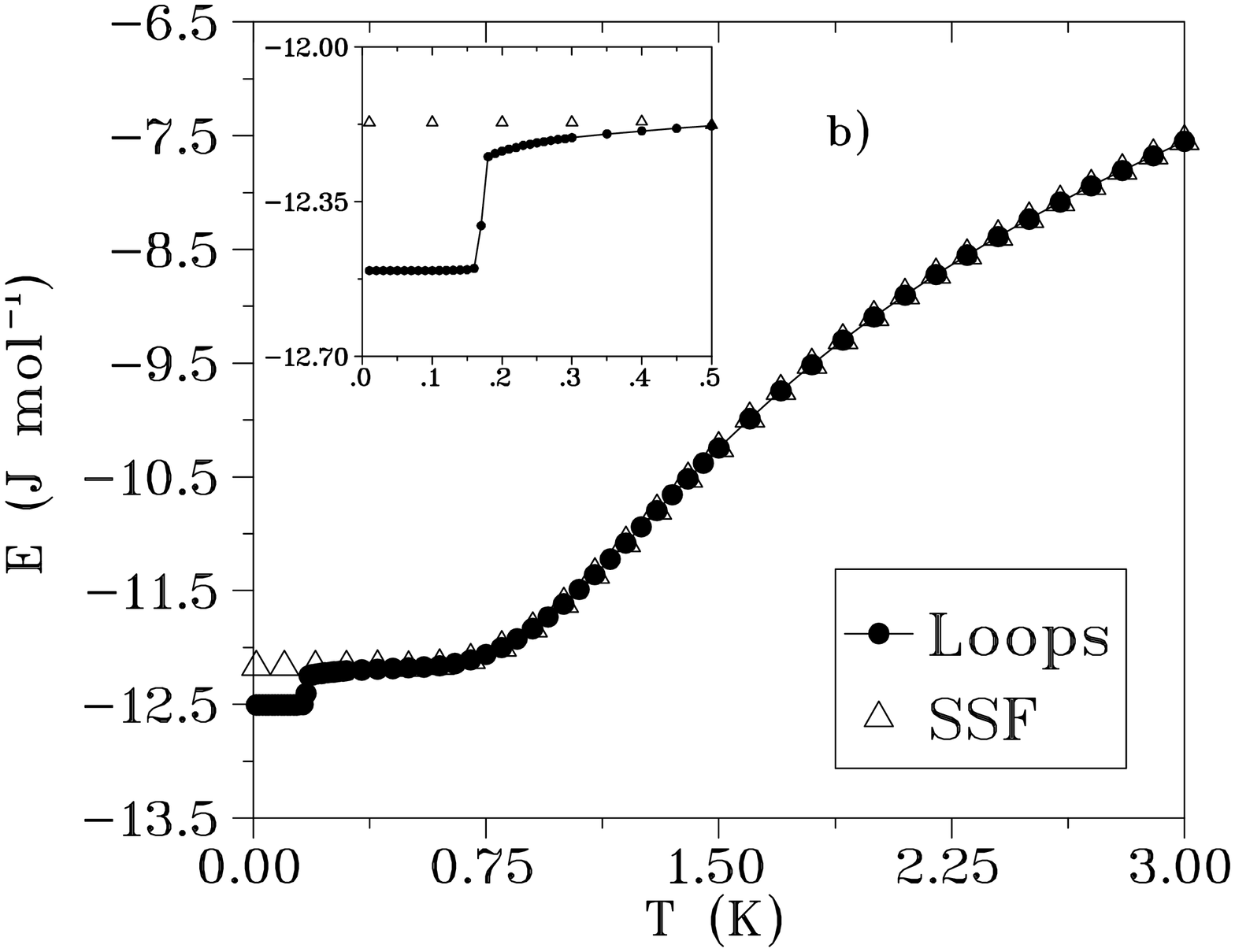}
\caption
{The low temperature magnetic specific heat (a) and energy (b) of the dipolar 
spin ice Monte Carlo, system size L=4, with simulation parameters set for 
Dy$_{\rm 2}$Ti$_{\rm 2}$O$_{\rm 7}$. Closed circles are simulation data run
with the short loop algorithm, open triangles are data 
obtained using the single spin
flip Metropolis algorithm.  In the inset of (b), the energy shows an apparent 
discontinuity at a critical temperature $T_{c} \approx 0.18$ K.  The broad 
feature in the specific heat at $T\approx 1$ K indicates the rapid development 
of the spin ice rule obeying states.
The sharp feature at $T_{c}$ is the appearance of a phase transition 
to a ground state being made (dynamically) accessible via the non-local
loop dynamics. Note that these results are of 
higher statistics than those for Fig.~\ref{ShortLong},
specifically, $1\times 10^5$ equilibriation and and $3\times 10^5$ 
production Monte
Carlo Steps were used.  In addition, the system size is increased to
$L=4$ as opposed to $L=3$. 
Also, note that the location of the specific heat peak is
at roughly the same temperature and is narrower than for $L=3$, indicating
a finite-size effect  on the singular behavior of $C(T)$.
}
\label{SloopCv}
\end{center}
\end{figure}

As suggested by the results above, the short loop algorithm is successful 
in restoring ergodicity in the simulation.  As a consequence of this, 
we observe a low temperature
phase transition in the model.  The most 
familiar and robust indicator of a thermodynamic 
phase transition (as opposed to dynamical freezing) 
is a finite size remnant of a singularity (divergence or discontinuity) in 
the specific heat at the transition temperature $T_{c}$.  
In our simulation of  
Dy$_{\rm 2}$Ti$_{\rm 2}$O$_{\rm 7}$, a sharp cusp in the specific heat is 
observed at a temperature below the spin ice peak (See Fig.~\ref{SloopCv}a).
The feature in the specific heat and the abrupt drop in energy at the same 
temperature gives good preliminary evidence that the loop algorithm is successful in allowing a 
phase transition to occur at a temperature of $T_{c}=0.18$ K.  This is
the same temperature where the loop algorithm acceptance rate goes to zero 
in Fig.~\ref{SLaccp}. 
The energy curve shows a discontinuous drop at $T_{c}$  
(e.g., latent heat) for large lattice sizes, suggesting 
a first order phase transition.  In the remainder of this section, we 
attempt to characterize this ordered state, and the phase transition that leads
to it.  The first step is to identify the order parameter 
associated with the low temperature ordered state.

Direct inspection of the spin directions at $T<0.18$ K
reveals that the ordered state is a long-range ice-rules obeying state with
zero magnetic moment per unit cell and commensurate with the pyrochlore cubic 
unit cell (see Fig.~\ref{GSsmall}). This state corresponds to the critical
mode found above in the mean-field calculation.  There 
are twelve symmetrically equivalent spin configurations 
for the ground state as explained below, 
two for each cubic axis direction and their spin reversed states.
The ordering wavevector {\bf q} lies parallel to one of the cubic axis directions, 
specifically ${\bf q}=(0,0,2 \pi /a)$ or one of its symmetrically equivalent
directions.  To construct the ordered state, first consider a starting 
tetrahedron with its six possible ice-rules states.  For a given ordering 
wavevector {\bf q}, this tetrahedron selects one of the four possible spin
configurations (two independent configurations and their spin-reversals, 
${\bf S}_{i}^{a}\rightarrow -{\bf S}_{i}^{a}$) with a total magnetic 
moment for the tetrahedron perpendicular to {\bf q}.  The entire ordered state 
may then be described by planes (perpendicular to {\bf q}) of such tetrahedra. 
The wavelength defined by this {\bf q} physically corresponds to 
antiferromagnetically stacked planes of tetrahedra, which means that a given 
plane has tetrahedra of reverse spin configuration 
to the plane above and below it.

We construct the multi-component order parameter
\begin{equation}
{\bf \Psi}_{\alpha}^{m}=\frac{1}{N} \left|{ \sum_{j=1}^{N/4} \sum_{a=1}^{4}
 \sigma_{j}^{a} e^{(i \phi_{a}^{m} + i {\bf q}_{\alpha} \cdot {\bf R}_{j})}
 } \right|.
\label{OrderParam}
\end{equation}
This type of labeling is natural given that the 
pyrochlore lattice can be viewed
as an FCC lattice with a ``downward'' tetrahedral basis (Fig.~\ref{Pyro}). Thus 
$j$ labels the FCC lattice points of the pyrochlore lattice, and the index $a$ 
sums over the four spins comprising the basis connected to each $j$.  
The index $\alpha$ labels the three possible symmetry related {\bf q} ordering 
wavevectors. For a given ${\bf q}_{\alpha}$, as described above, there are two 
ice-rules configurations and their reversals which can each form 
a ground state. Thus $m=1,2$ labels these possibilities with the phase factors 
$\{ \phi_{a}^{m} \}$, describing the given configurations $m$.  
Each Ising variable $\sigma _{j}^{a}$ has a value +1 or -1 
when a spin points into or out of its downward tetrahedron $j$, respectively.

As written in Eq.~(\ref{OrderParam}), ${\bf \Psi}_{\alpha}^{m}$ has six 
degenerative components, each of which can take on a value between 0 and 1.  
Upon cooling through the transition, the system selects a unique ordered 
configuration, causing the corresponding component 
of ${\bf \Psi}_{\alpha}^{m}$ to rise to unity and the other 
five to fall to zero  
(provided the finite size system is simulated over a time scale less
than the ergodic time scale where full spin symmetry is restored).
The component which rises to unity is 
equally likely to be any one of the six, selected by random through 
spontaneous symmetry breaking.

\begin{figure}[ht]
\begin{center}
\includegraphics[height=6cm,angle=90]{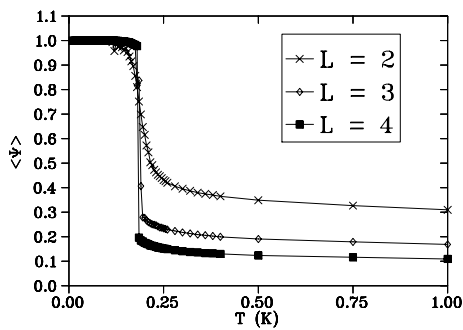}
\caption[The ${\bf q}=(0,0,2 \pi /a)$ Order Parameter.]
{The ${\bf q}=(0,0,2 \pi /a)$ order parameter.  Curves are shown for system
size L = 2, L = 3, and L= 4.}
\label{loopOP}
\end{center}
\end{figure}

Fig.~\ref{loopOP} is a plot of $\left<\Psi \right>$ for three system sizes, 
where
\begin{equation}
\left < \Psi \right>=\sqrt{\sum_{m=1}^{2}\sum_{\alpha =1}^{3}\left( 
{\bf \Psi}_{\alpha}^{m} \right)^{2}}
\label{OPmag}
\end{equation}
is the magnitude of the multi-component order parameter.  These three curves 
illustrate important finite size effects for $\left<\Psi \right>$.  For 
$T<T_{c}$ the different lattice sizes produce identical order parameters. By 
contrast, $\left<\Psi \right>$ for the smaller 
lattice size displays pronounced 
rounding near $T_{c}$ and an increased residual value for large $T$.  
The larger lattice size produces an order parameter with a 
clear discontinuity at $T_{c}$. This discontinuity in the order paramater
combined with the discontinuity of the total energy in Fig. 15b can be 
viewed as strong preliminary evidence for a first order transition.

We now argue the need to study this phase transition with greater numerical
accuracy. This is necessary partially to confirm rigorously its first-order 
nature. More importantly, once this is done, we want to use the data
to confirm the full recovery of Paulings entropy through an estimation of the 
latent heat released by the transition.  To begin, we note that
there are a number of criteria at one's disposal to
demonstrate the occurence of a first-order transition in 
a Monte Carlo simulation. In particular:
\begin{enumerate}		
\item{
 The order parameter $\langle \Psi \rangle$ should have a clear discontinuity 
 at $T_{c}$.}
\item{
 The energy probability histogram, $H(E)$, should have a double peak at  
 $T_{c}$, which identifies the coexistence of two distinct phases at $T_{c}$.}
\item{
 There should be a latent heat at the transition, identifiable  
 by a discontinuity in the internal energy for large system sizes.}
\item{
 In the Monte Carlo, the height (maximum) of the specific heat, 
$C_{\rm peak}$, and the magnetic susceptibility, 
$\chi_{\rm peak}$, should be proportional to the simulation volume:
\begin{equation}
 C_{peak},\chi_{peak} \; \propto \; a+bL^{d}
\label{FSSspecsusc}
\end{equation}
 where $a$ and $b$ are constants and $d$ is the 
dimension of the lattice, in our case equal to three ($d=3$).}
\item{
 The minimum of the fourth order energy cumulant,\cite{Binder}
\begin{equation}
V=1-\frac{ \left<{E^{4}}\right>}{3 \left<{E^{2}}\right>^{2}}
\label{Bind4NRGcum}
\end{equation}
 should vary as
\begin{equation}
V_{\rm min}=V_{0}+cL^{-d}
\label{Bind4NRGcumVar}
\end{equation}
 where $V_{0} \ne 2/3$.}
\item{
 The temperature $T_{\rm peak}(L)$ at which $C_{\rm peak}$ or $\chi_{\rm peak}$
 have a maximum  should vary with the simulation volume as:
\begin{equation}
T_{\rm peak}(L)=T_{c} + cL^{-d}
\label{FSSspsusT}
\end{equation}
where $c$ is a constant, and $T_{c}=T_{\rm peak} (L \rightarrow \infty)$.}
\end{enumerate}

\begin{figure}[ht]		
\begin{center}
\includegraphics[height=6.5cm,width=5cm,angle=90]{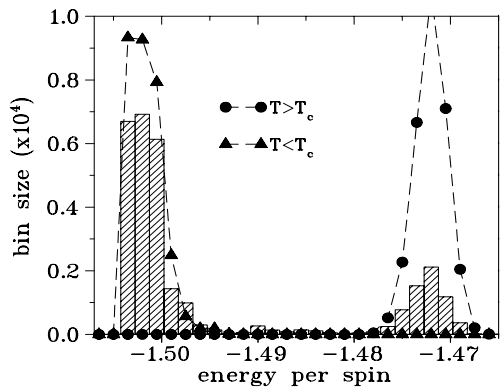}
\caption{
 The energy probability distribution histogram for three temperatures:
 $T=0.178$ K, $T=0.180$ K ($T_{c}$), and $T=0.182$ K.
 For $T > T_{c}$ (filled circles), a single
 peaked Gaussian histogram is present.  At the transition temperature (hashed 
 rectangles), a
 second peak appears which has a lower mean energy.  As the temperature falls
 below $T_{c}$ (filled triangles), the peak with the higher mean energy disappears, and the 
 system energy eventually gathers in the lowest bin.}
\label{PETc}
\end{center}
\end{figure}

We have already confirmed the first condition of our list. 
To check for the second	condition  
the energy probability histogram was calculated by binning the 
simulation energy values for every Monte Carlo 
step as the system passes through 
the transition from higher to lower temperatures (Fig.~\ref{PETc}).
Above $T_{c}$, we observe a single peak Gaussian-like distribution of energies.
At $T_{c}$, the energy probability distribution shows a double peak, 
characteristic  of the coexistence of two phases found at a first order phase 
transition.  Below $T_{c}$ we would normally
 expect to see a Gaussian peak. However, in our 
case,  the histograms below $T_c$ are 
distorted by the accumulation of energies into the lowest bins due to
the proximity of the transition to the ground state.  
At zero temperature, we expect all of the 
energies to lie in the bin for the lowest energy.

\begin{figure}[ht]
\begin{center}
\includegraphics[height=7cm,width=5cm,angle=90]{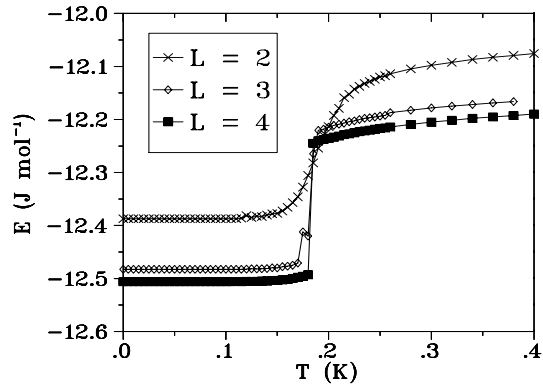}
\caption[Monte Carlo Energy Near the Transition.]
{Details of the simulation energy near the transition for different
 system sizes.}
\label{LatHeat}
\end{center}
\end{figure}

The next condition on our list is the observation of a latent heat at 
the transition.  Fig.~\ref{LatHeat} shows the energy near the transition for 
three different system sizes.   
A clear discontinuity develops as we increase the system size.
The energy discontinuity, $\Delta E$, as read off of this graph for
$L=4$, is 
\begin{equation}			
\Delta E \approx 0.248 \text{ J mol}^{-1}.
\label{LH1}
\end{equation}											
This behavior is also consistent with the transition being first order.  
Below, we use this $\Delta E$ value in calculating 
the entropy recovered at the transition (entropy jump). 
Another calculation
of the latent heat at the transition comes from the finite size scaling
of condition 4 above:\cite{Binder}
\begin{equation}
C_{\rm peak} \approx \frac{(\Delta E)^{2}}{4 k_{\text{B}} T_{c}^{2}}L^{d} + a
\label{FSSlatheat}
\end{equation}
where $L^{d}$ is the system volume as before, and $a$ is the intercept of the 
graph of $C_{\rm peak}(L)$ vs. $L$ to test for condition 4,
which we now discuss.

When attempting to quantify the relationships in conditions 4, 5 and 6 on our 
list, we notice a problem.  The extremely sharp nature of the 
transition makes accurate estimates for 
these quantities almost impossible using 
a traditional temperature cooled MC simulation of the
hybrid single spin flip$-$loop algorithm.  The reason is that the
transition temperature region is so narrow, and first-order metastability
effects are so strong,
obtaining accurate 
data for quantities such
as $C_{\rm peak}$ or $V_{\rm min}$ very near to 
$T_{c}$ is extremely difficult. 
As shown in Fig.~\ref{PETc}, the energy probability 
histogram near a first order transition displays a double hump.  The energies 
that occur between these humps correspond to system configurations that are 
strongly suppressed by the Boltzmann probability distribution near
the transition.  We call these  
``interface configurations''.\cite{Berg}
Traditional Monte Carlo simulations try to ``avoid'' these 
interface configurations as 
the system is cooled through the transition, 
because of their suppression by the 
Boltzmann factor which is the basis of the Metropolis condition.  Therefore, 
the simulation often behaves poorly in this region, moving quickly through 
interface configurations to find more favorable configurations nearby in 
configuration space.  This can lead to erratic behavior and poor statistics 
in thermodynamic quantities of interest near the phase 
transition, thereby reducing the numerical accuracy of the quantities used in
finite-size scaling.

To overcome this problem, Berg and Neuhaus \cite{Berg} proposed the 
{\it multicanonical} method, which is designed to enhance configurations that 
have energies which occur between the double hump of the probability 
distribution.  If these interface configurations are 
artificially enhanced, the 
simulation does not avoid this energy range as strongly
and better statistics can be obtained.  The version of the 
multicanonical Monte Carlo algorithm that we use 
is that proposed by Hansmann and Okamoto,\cite{Hansmann} originally developed 
to be used in the context of protein folding simulations.  The core of the
method is:

{\it Perform Monte Carlo simulations in a multicanonical ensemble instead of 
the usual canonical ensemble.  Then, obtain the relevant canonical distribution 
by using the histogram reweighting techniques of Ferrenberg and
Swendsen.\cite{Ferrenberg}
From this, calculate the thermodynamic quantities of interest.}

In the multicanonical ensemble, we define the probability distribution by
\begin{equation}
p_{\rm mu}(E) = \frac{g(E)w_{\rm mu}(E)}{Z_{\rm  mu}}=\text{constant}
\label{MultiCanonProb}
\end{equation}						
where $g(E)$ is the density of states, 
$w_{\rm mu}(E)$ is the multicanonical weight
factor (not temperature dependent), and $Z_{\rm mu}$ 
is the associated partition 
function. The distribution is constant, meaning that all energies have 
equal weight, which sometimes leads to the name ``flat histogram''
method.
This flatness is important because it ensures that 
configurations in the interface region of the 
transition are not suppressed.			

Unlike for the canonical ensemble, the multicanonical weight 
factor $w_{\rm mu}$ is not {\it a priori} known.  
This turns out to be the crucial step of this scheme:
finding an accurate estimator of $w_{\rm mu}$ that makes the distribution 
$p_{\rm mu}(E)$ flat over the energy range of interest.  The details of how 
to do  this are  somewhat involved, and will not be explicitly outlined here.  
The reader is referred to the relevant technical references for 
details.\cite{Melko_thesis,Hansmann}
Our procedure follows that of Ref.~\onlinecite{Hansmann} very closely.
						
Assuming that we can find a good estimator of $w_{\rm mu}$, 
our method proceeds as follows:
\begin{enumerate}
\item
We find an accurate estimator of the multicanonical weight factor so that 
$p_{mu}(\rm E)$ is reasonably flat over an energy range that includes the 
transition interface.
\item
With this weight factor we perform a multicanonical simulation at one given 
temperature $T$ slightly higher than $T_{c}$.
\item
During this simulation run, we gather statistics for the physical variables of 
choice (for example, the energy $E$). These variables are weighted 
according to the multicanonical distribution.
\item
From this single simulation, we then obtain the Boltzmann-distributed 
variables at any 
temperature for a wide range of temperatures using a reweighting technique.
\end{enumerate}

We use the reweighting technique proposed by Ferrenberg and Swendsen, 
\cite{Ferrenberg} which allows us to 
transform, or reweight, data obtained from another distribution (in our case 
the multicanonical distribution) 
to the relevant Boltzmann distribution, at some inverse temperature $\beta$.  
We use this to obtain an estimate for a given physical quantity 
in the canonical distribution.  

\begin{figure}[ht]
\begin{center}
\includegraphics[height=6.5cm,width=5cm,angle=90]{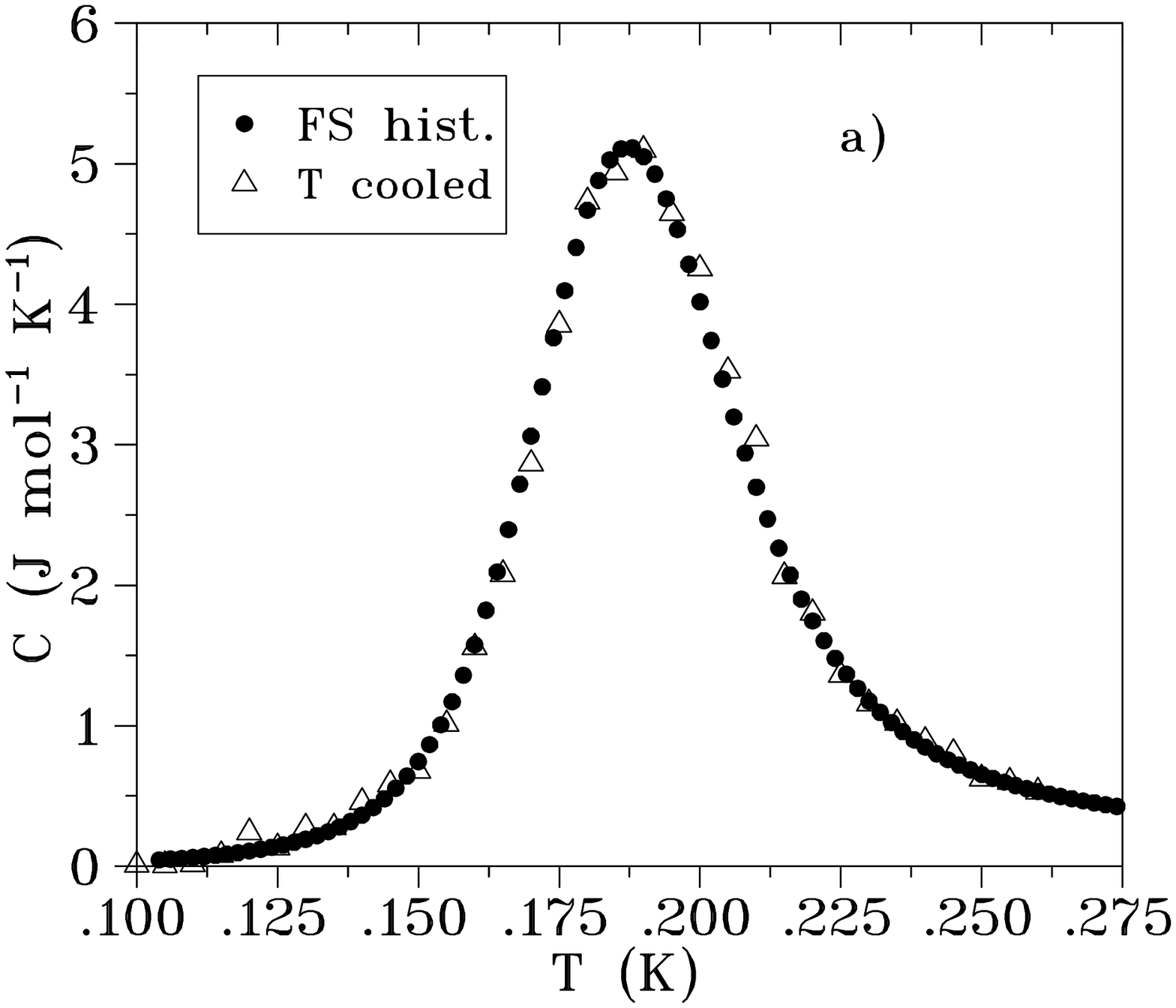}
\includegraphics[height=6.5cm,width=5cm,angle=90]{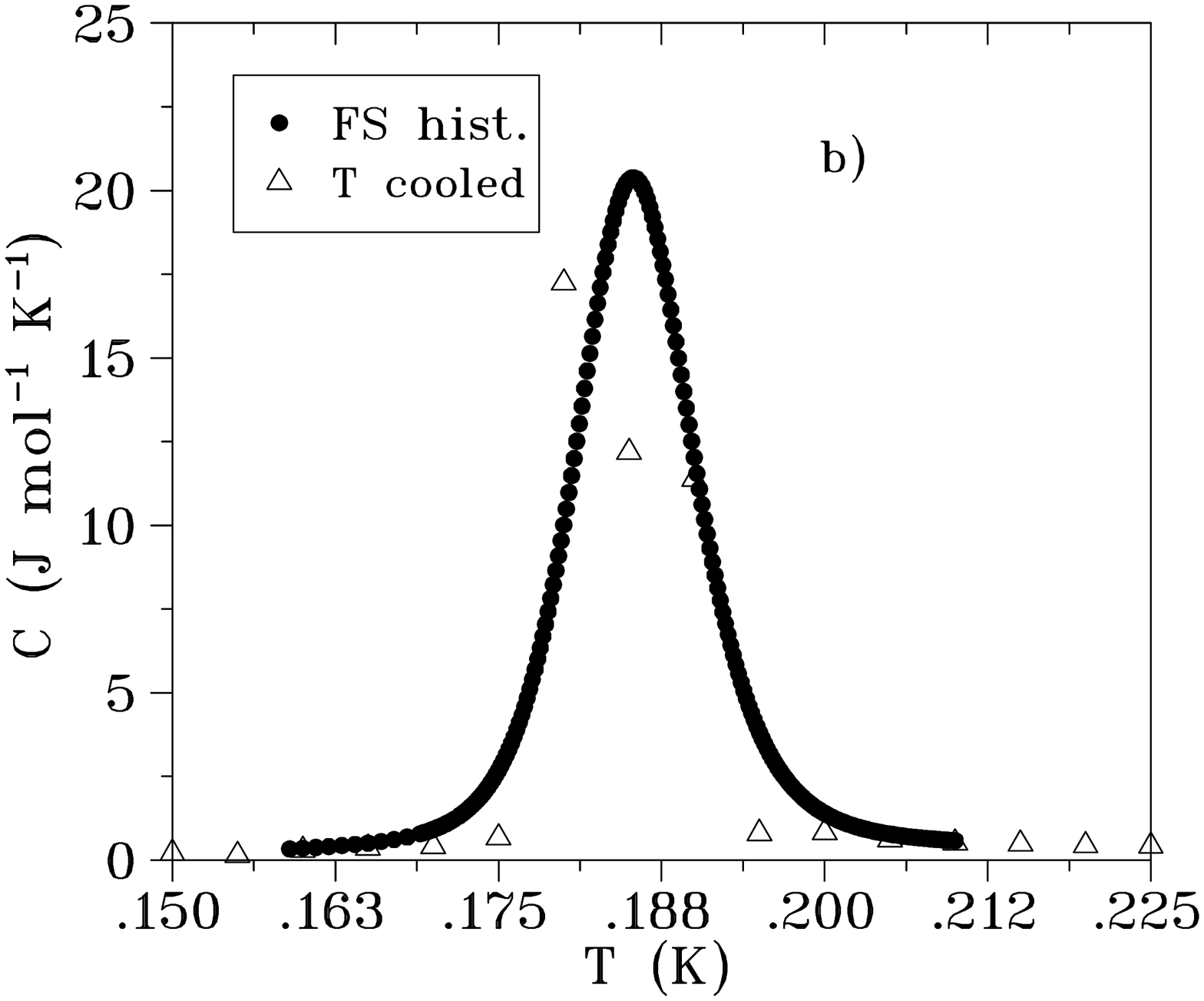}
\caption
{Specific heat curves over the transition temperature, for L = 2 (a) and
 L = 3 (b) system sizes.  Closed circles represent data obtained
with Ferrenberg
and Swendsen's  histogram reweighting technique.
Open triangles represent data taken using a traditional
temperature cooled Monte Carlo simulation.}
\label{Cvcomp}
\end{center}
\end{figure}
We collect data within the multicanonical distribution
and use it to calculate the specific heat 
for the dipolar spin ice model.  For the smallest system size considered, we 
accurately reproduce the specific heat over the transition using the 
flat histogram method.
Fig.~\ref{Cvcomp}a shows a comparison between 
the specific heat of an $L=2$ system obtained using the 
histogram method at one temperature, and the traditional 
Monte Carlo procedure with $8 \times 10^{5}$ 
equilibriation steps and $2 \times 10^{6}$ data production steps for every 
temperature point.  The CPU time that it took to get the histogram data 
was a small fraction of the time it took to obtain the 
regular Monte Carlo data.
Fig.~\ref{Cvcomp}b is a similar result for the next 
highest system size, L = 3.  
The traditional Monte Carlo data was taken with 
$5 \times 10^{5}$ equilibriation 
steps and $1 \times 10^{6}$ data production steps. 
The histogram data was obtained in the same amount of time as for the 
$L=2$ data, and it was only slightly more difficult to 
find a good estimate for $w_{\rm mu}(E)$.  
The poor quality of the traditional Monte Carlo Metropolis data for 
$L=3$ stands in stark contrast to the smooth data obtained using the 
multicanonical simulation.

\begin{figure}[ht]
\begin{center}
\includegraphics[height=6.5cm,width=5cm,angle=90]{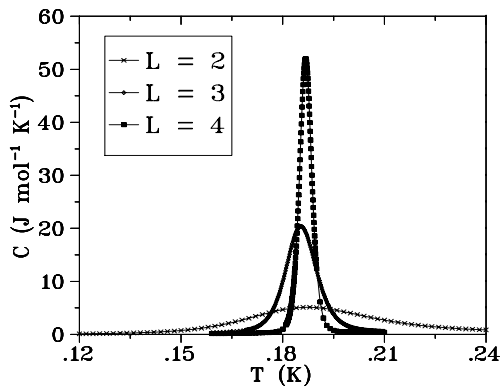}
\caption[Reweighted Monte Carlo Specific Heat Data Through the
Transition.]
{Specific heat of the transition to long-range order, for system sizes
$L=$ 2, 3, and 4. It was found that peak heights for these system sizes
did
not vary much with the flatness of $p_{\rm mu}(E)$.
Critical temperatures $T_{c}$ were more sensitive to the
flatness of the multicanonical distribution, and
hence were harder to estimate than $C_{\rm peak}$.}
\label{CvgraphFSS}
\end{center}
\end{figure}
Unfortunately, one difficulty with the multicanonical algorithm used
here is that, in general, as the system size is increased, 
it becomes increasingly 
difficult to obtain a good estimate for a $w_{\rm mu}(E)$ 
that would give a flat 
$p_{\rm mu}(E)$.  The critical temperature, $T_{c}$, 
of the transition seems to be 
the quantity most sensitive to variations in the 
flatness of $p_{\rm mu}(E)$.
In contrast, the height of specific heat peak is fairly accurately determined
for simulation sizes $L=$ 2, 3, 4, and 5 (Fig.~\ref{CvgraphFSS}), 
showing only a weak sensitivity to the flatness of $p_{\rm mu}(E)$.

\begin{figure}
\begin{center}
\includegraphics[height=6.5cm,width=5cm,angle=90]{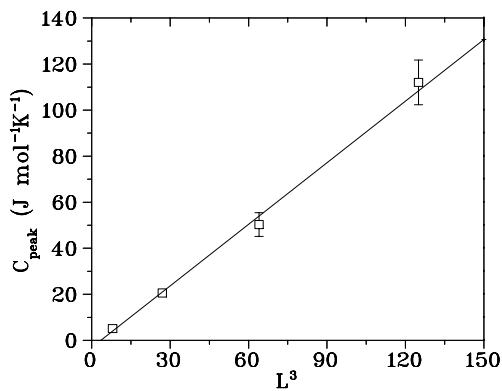}
\caption
{Finite size scaling fit for the specific peak heights of the ordering
 transition.  Data points represent the mean $C_{\rm peak}$ 
value for a given $L$. 
Error bars show one standard deviation.}		
\label{Cvfss}
\end{center}
\end{figure}

The aforementioned error associated with $T_{c}$ for the $L=4$ peak, 
as determined from simulations, is of the order of 0.04 K,  
and becomes increasingly more drastic for the larger system sizes.  
The variation in the height of the specific heat was  found to be much 
less. Nevertheless,  to combat any minor variation in peak 
height and obtain an accurate finite size
scaling results, a statistical average was done on several ($\sim 10$)
 multicanonical  weighting factors to obtain values for $C_{\rm peak}$.  
These results are plotted in Fig.~\ref{Cvfss}.  
A straight line fit to the data using linear regression gives
\begin{equation}
C_{\rm peak} = 0.8924 L^{3} - 3.149.
\label{RegLine}
\end{equation}
The $L^3$ dependence of $C_{\rm peak}(L)$ shows that the finite size 
scaling is consistent with that expected for a first order transition.  
Also, as a second estimator of the latent heat, 
we can use the slope of this line and 
Eq.~(\ref{FSSlatheat}) to extract $\Delta E$.  
Doing so we deduce a latent heat of 
\begin{equation}
\Delta E = 0.245 \text{ J mol}^{-1},
\label{LH2}
\end{equation}
consistent to within 1\% with the value obtained in Fig.~\ref{LatHeat}
(for the $L=4$ system) from reading directly off of the energy graph 
(see Eq.~(\ref{LH1})).

This completes our study of the nature of the ordering transition in dipolar 
spin ice.  As we have shown, the discontinuity in the order parameter, the 
release of latent heat, the double peaked energy probability distribution, and 
the finite size scaling of the specific heat peak all give consistent 
and compelling evidence for the 
transition being first order.  As the technical details concerning this 
transition are understood, we can proceed to study where it, 
and the long-range 
ordered state which results from it, stand in our broader 
picture of ground state entropy found in experiments and 
in standard single spin flip simulations of 
the dipolar spin ice model.									
Since we have confirmed the first order nature of the transition, the 
configuration of the ordered state, calculated the latent heat, and have 
reliable data for the specific heat through the transition, 
we are in a position to re-calculate the total entropy 
that the dipolar spin ice model releases as it 
is cooled to low temperatures.  This calculation must be done carefully.  
We know that in an infinite system, a first order transition is characterized 
by a cusp in the specific heat.  If the transition is 
temperature driven, as in our
case, this first order singularity is the latent heat.  
For an infinite system going through
a first order transition, thermodynamics gives
\begin{equation}
\Delta S = \int_{0}    ^  {T_c^-} \frac{C_<}{T} dT + 
           \int_{T_c^+}^  {\infty    } \frac{C_>}{T} dT + 
 		\frac{\Delta E}{T},
\label{1stOENT}
\end{equation}
where $\Delta E /T$ is the latent heat contribution to the entropy 
(see Fig.~\ref{LatHeat}), and 
$T_c^-$ and and $T_c ^+$ are the temperature limits asymptotically 
close to $T_c$, below and above 
$T_c$, respectively (see Fig.~\ref{LatHeat}). 

\begin{figure}[ht]
\begin{center}
\includegraphics[height=6.5cm,width=5cm,angle=90]{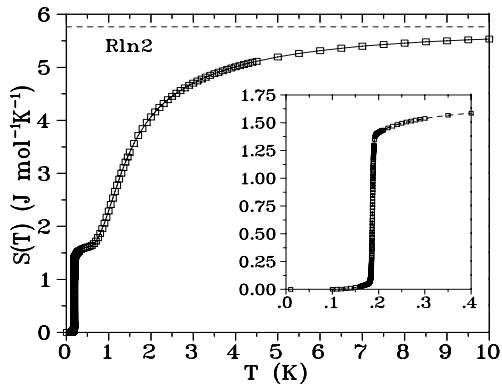}
\caption
{The entropy calculated from integrating 
the simulated specific heat, as explained in the 
text.  The entire value of entropy for the system ($R\ln 2$) 
is recovered in the
high temperature limit.  The inset shows the details of the 
entropy recovered by
the transition to long-range order.}
\label{loopENT}
\end{center}
\end{figure}

To estimate a value for the entropy, we consider the system size 
$L=4$ which has good statistical data for the widest temperature range.  We 
integrate the low temperature data for the specific heat in 
Fig.~\ref{CvgraphFSS} divided by 
temperature obtained from the histogram reweighting technique
 (up to $T\approx 0.21$).  For $T>0.21$ K we use
our regular temperature cooled Monte Carlo data 
(canonical loop $+$ single spin flip) for the integration above this 
point, and up to 10 K, giving $S(T=10)-S(T\approx 0) = 5.530$J mol$^{-1}$
K$^{-1}$.
To integrate up to $T=\infty$, we follow the same high 
temperature extrapolation procedure described  in Section II,
giving $S(T=\infty)-S(T=10) = 0.145$J mol$^{-1}$ K$^{-1}$.
Doing this simple calculation, we find a total 
recovered entropy of 
\begin{equation}
S(T=\infty)-S(T\approx 0) = 5.675 \text{ J mol}^{-1} \text{K}^{-1},
\label{simENT}
\end{equation}
less than 2\% below our expected value of $R \ln2 = 5.764$.
The inset of Fig.~\ref{loopENT} 
clearly shows the entropy recovered by the low temperature transition.  

By considering the entropy recovered by the integration of the 
finite size system specific heat over the transition 
(inset, Fig.~\ref{loopENT}),
we confirm that it is approximately equal to the value of the entropy we would 
expect to recover from the latent heat of an infinite system. Using our latent 
heat calculations above (Eqs.~(\ref{LH1}) and (\ref{LH2})), this value is 
approximately $\Delta S_{T_c}  = \Delta E/T_c \approx 0.2465 / 0.180 
= 1.37 \text{ J}/\text{mol K}$, in good 
agreement with the jump in $S(T)$ in the inset of Fig.~\ref{loopENT}. 

Taking all indicators together, we have demonstrated here that the transition 
to long range order at 180 mK  recovers all residual Pauling 
entropy of the dipolar spin ice model.  Thus we can assert that the degeneracy 
associated with the spin ice model, and the corresponding value of zero point 
entropy, is lifted due to perturbations beyond nearest neighbor dipole-dipole 
interactions, if equilibrium can be maintained at sufficiently low temperatures.

We comment here on a detail
we neglected to discuss above
regarding the relationship between
the mean-field theory and the results from the loop Monte Carlo
simulations.
In the Gaussian mean-field theory presented above the calculation that
was
performed was in effect an identification of the soft-mode against which
the paramagnetic
phase becomes locally unstable upon cooling.
The Monte Carlo simulation finds, however,
that the thermodynamic transition to that ordered state is actually
first order and occurs before the
supercooled critical temperature is reached.

To summarize our results for this section, we refer the reader to the dipolar 
spin ice Monte Carlo phase diagram, Fig.~\ref{PhaseD}.
As illustrated there, the transition between the spin ice phase 
(which retains Pauling's 
entropy) and the ${\bf q}=(0,0,2 \pi /a)$ ordered phase is independent of the
strength of $J_{\rm nn}$.  This is consistent with 
our understanding that the long 
range order results from perturbative interactions beyond nearest neighbor, 
caused by the long-range dipolar interaction. This is also what mean-field
theory finds in the spin ice regime $(J_{\rm nn}/D_{\rm nn} > -0.905)$.
We find that this first order line also slightly 
runs up the boundary between the antiferromagnetic ordered 
phase and the higher temperature paramagnetic phase, and that a tricritical 
point separates these two regions of the line, and occurs 
near the value $J_{\rm nn}/D_{\rm nn} \sim -1.1$. 

Due to the near-vertical nature of the phase boundaries in this region, 
simulations run at a finite $T$ and varying $J_{\rm nn}$ help better 
map out the low temperature phase lines of interest.  However, 
using this method, we observed that the simulations could easily get
``stuck'' in the previous spin configuration (either spin ice disordered, 
${\bf q}=(0,0,2 \pi /a)$ or AF ${\bf q}=0$) when crossing the vertical
phase  boundary.  This history dependence is illustrated 
in the phase diagram as 
hysteresis at low temperatures, mainly between the long-range ordered 
${\bf q}=(0,0,2\pi/a)$ and antiferromagnetic ${\bf q}=0$ phases.
Regardless of this difficulty, we have confirmed 
from direct Ewald energy calculations at zero temperature 
that the true zero-temperature phase boundary between 
the ${\bf q}=(0,0,2 \pi /a)$ and the AF ${\bf q}=0$ phases lies at 
$J_{\rm nn}/D_{\rm nn} = -0.905$, in agreement 
with the result found above in the mean field calculations.

\section{Dipolar Spin Ice in Magnetic Field}
\label{DSImag}

A very interesting problem that pertains to dipolar spin ice materials is their 
behavior in an external magnetic field, {\bf h}.  A number of recent 
experiments \cite{Fukazawa,Higashinaka,hiroi,kagomeMAT,Dy162}
have shown a rich variety of new behavior when spin ice materials are
subjected to such a field, which warrants some 
theoretical investigation. \cite{kagome-ice,Moessner111}
Although not all of the relevant experiments can be described in this short 
section, we briefly outline some of the most important, referring the 
reader to the bibliography for further details on their methods and results.

The first experiments on spin ice materials in an applied magnetic field
were performed by Harris {\it et al.}. \cite{Harris_prl1}  
In a neutron scattering experiment, they applied a 
magnetic field of strength 2 T along the [110] direction of a single crystal of 
Ho$_{\rm 2}$Ti$_{\rm 2}$O$_{\rm 7}$, and looked for signs of ordering. They 
found scattering intensity features which suggest evidence of two ordered 
magnetic structures, the so-called ${\bf q}=0$ and ${\bf q}=X$ phases (Fig.~\ref{q0x}).  
As we will see below, these ordered structures are of fundamental
importance in our study of the ground states of the dipolar spin ice
model.
\begin{figure}[ht]
\begin{center}
\includegraphics[height=8cm,]{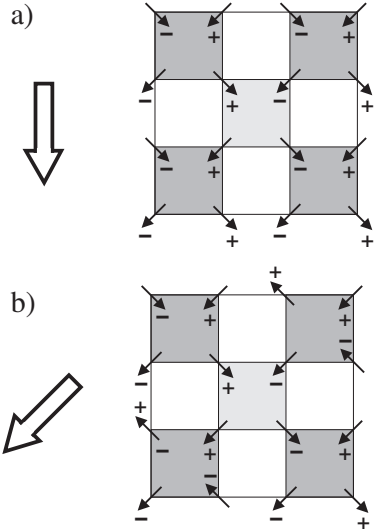}
\caption[The Ordered $q=0$ and $q=X$ Ground States.]
{The pyrochlore unit cell projected down the z-axis.  The symbols
+ and - represent the z component of the spin ``head''.  Configurations
are (a) ${\bf q}=0$ and (b) ${\bf q}=X$.
The large arrows point in the (a) [100] and (b) 
[110] directions, and are included to aid in the discussion of ground states 
in the next section. Note that ${\bf q}=X$, although similar, is 
a spin ice state distinct from the ${\bf q}=(0,0,2\pi/a)$ ordered state shown 
in Fig.~\ref{GSsmall}.
Specifically, the chains of spins parallel to the [100] direction are
staggered antiferromagnetically in the zero field
 ${\bf q}=(0,0,2\pi/a)$ ground state (Fig. 4), while they
are ferromagnetically correlated and parallel to the field
in the ${\bf q}=X$ state of (b) above. }
\label{q0x}
\end{center}
\end{figure}

A quite interesting set of experiments was performed by Ramirez {\it et al.} 
\cite{Ramirez} and Higashinaka {\it et al.},\cite{Maeno2} who subjected 
polycrystalline samples of Dy$_{\rm 2}$Ti$_{\rm 2}$O$_{\rm 7}$ to a variety of 
different field strengths.  Ramirez presented evidence of field-dependent 
phase transitions in a powder sample, manifested as sharp features in the specific 
heat at
    \begin{enumerate}
     \item 0.34 K for $h> 1$T
     \item 0.47 K for $1$T $\stackrel{<}{\sim} h \stackrel{<}{\sim} 3$T
     \item 1.12 K for $h \neq 0$
     \end{enumerate}
where we have used $h$ to represent the magnitude of the applied field {\bf h}. 
Higashinaka and co-workers reproduced the basic features of Ramirez' results
down to $T\cong 0.38 $ K,\cite{Maeno2} confirming the existence of the
two higher-temperature peaks only.
The search for a microscopic explanation of these three peaks has
been a driving force behind much of the experimental and theoretical
work in this field over the past few years, and we discuss it further
in the work that follows below.

Another significant experimental study is the measurement of the single
crystal magnetization curves ($M$ vs. ${\bf h}$) for the spin ice materials.
Fukazawa {\it et al.} performed a number of experiments on single crystals of 
Dy$_{\rm 2}$Ti$_{\rm 2}$O$_{\rm 7}$, obtaining magnetization curves for the 
different applied field directions and a range of temperatures. \cite{Fukazawa}
They showed that magnetization data at 2 K was consistent with the
behavior predicted by the spin ice model, in particular the limiting
field (large {\bf h}) values of the magnetic ``anisotropy'' (which we
illustrate below).
Very recently, measurements \cite{Higashinaka,kagomeMAT} of the magnetization 
curves for {\bf h}//[111] (read ``{\bf h} parallel to the [111] crystal 
direction'') have uncovered a 
novel macroscopically degenerate state corresponding to ice-like behavior on
the kagome planes in the pyrochlore lattice.
\cite{kagome-ice,Moessner111}

We take into account the applied magnetic field {\bf h} in the dipolar spin 
ice model with a simple term added to the Hamiltonian
(Eq.~(\ref{DSPhamiltonian})), 
\begin{equation}
H'=-\sum_{i} {\bf h} \cdot {\bf S}_{i}^{a}
=-\sum_{i} ({\bf h} \cdot {\hat z}^{a})\sigma_{i}^{a} \ .
\label{MagHamil}
\end{equation}
We work strictly with a classical Ising model and neglect any 
transverse field effects and perturbative changes to the moments arising in a 
strong field.  In this classical approximation, the field {\bf h} couples to 
the spins through the simple scalar product Eq.~(\ref{MagHamil}).  
That is, we neglect small corrections to the energy coming from the 
very small, though finite, local susceptibility perpendicular to the 
$\left<{111}\right>$ direction.
As well, we neglect quantum mechanical transverse-field effect that would arise
from admixing the doublet ground state wavefunctions with that of the excited crystal
field levels. For Ho$_{\rm 2}$Ti$_{\rm 2}$O$_{\rm 7}$ 
and Dy$_{\rm 2}$Ti$_{\rm 2}$O$_{\rm 7}$, the first excitation gap is
(very roughly) $\Delta \sim 300$ K.\cite{Rosenkranz} For the magnetic moments
of approximately 10 $\mu_{\rm B}$ for both
Ho$^{3+}$ and Dy$^{3+}$, this means a ground state Zeeman energy splitting
of 12.8 K/Tesla. One can therefore safely neglect magnetic field,
exchange and dipole-dipole induced admixing for fields less than 10 Teslas,
assuming the worse case scenario where the excited doublet also split by 
about 10 K/Tesla.

\begin{figure}[ht]			
\begin{center}
\includegraphics[height=4cm]{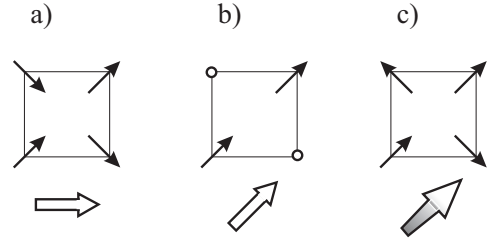}
\end{center}
\caption[Geometric Field Coupling to a Tetrahedral Plaquette.]
{A single tetrahedron projected down the $z$-axis.  Field directions
are (a) ${\bf h}//[100]$, (b) ${\bf h}//[110]$, (c) ${\bf h}//[111]$, depicted 
by the large arrow outline.  Small arrows represent dipole moments coupled to 
the field.  Empty circles represent decoupled spins.}
\label{Ztetra}
\end{figure}

To gain a theoretical understanding of the experimental behavior mentioned 
above, several insightful calculations are possible, using only this simple 
classical Hamiltonian and a knowledge of the possible ground states of 
Fig.~\ref{q0x}.  First, a geometrical understanding of how the magnetic field 
couples to classical 
spins on the pyrochlore lattice is desirable.  We expect that application of a 
magnetic field along the three principle symmetry axes of the crystal will 
result in different spin-field coupling behavior.  To explore this, we begin by 
considering the non-interacting limit 
($h \to \infty$ or $J_{\rm nn}, D_{\rm nn} \rightarrow 0$).  In this 
case, the only constraints on the spins is the local
$\left<{111}\right>$  anisotropy and the 
coupling with the magnetic field.  We can gain more insight by viewing a 
projection of a tetrahedron down the cubic $z$-axis as in Fig.~\ref{Ztetra}.

For ${\bf h}//[100]$, all four spins on a given tetrahedron are coupled with the
field (i.e. all four have a non-zero dot product in Eq.~(\ref{MagHamil})).  The 
expected lowest energy configuration in the absence of spin-spin interaction is 
the one where all spins have their [100] components aligned with the field.  
Knowing this, we can calculate the $h \to \infty$ value for the $M$ vs. {\bf h} 
curves by considering
the average moment $M$ (in units of Bohr magneton per
rare earth ion, $\mu_{B}/ \text{R}^{3+}$), in the direction of the field 
${\bf h}//[100]$:
\begin{eqnarray}
\nonumber
&M&({\bf h} \rightarrow \infty) \\ \nonumber
&=& \frac{1}{4\sqrt{3}} \left({\left[{111}\right] 
+ \left[{1\overline{11}}\right] 
+ \left[{1\overline{1}1}\right] + 
\left[{11\overline{1}}\right]}\right) \cdot \left[{100}\right] \cdot \mu \\
&=& \frac{1}{\sqrt{3}} \cdot \mu \cong 0.5774 \cdot \mu.
\label{M100}
\end{eqnarray}
Also, to calculate $M$ in the appropriate units ($\mu_{B}/
\text{R}^{3+}$, measured in experiments) one must include the factor
$\mu$, which is the magnetic moment of the appropriate rare earth ion. 
If R is Dy$^{3+}$ or Ho$^{3+}$, $\mu \approx 10 \mu_{B}$.
Note that, from Fig.~\ref{Ztetra}a, 
this lowest energy spin-field coupled state is compatible with the ice rules.  
If we decorate the entire lattice with tetrahedra such as this, we recover the 
${\bf q}=0$ state of Fig.~\ref{q0x}a.  This suggests that this ordered state 
should be one of the ground states for the interacting dipolar spin ice model, 
with a sufficiently strong external field ${\bf h}//[100]$.
Indeed, this order has been observed experimentally\cite{Dy162} on samples of
Dy$_{\rm 2}$Ti$_{\rm 2}$O$_{\rm 7}$.

For ${\bf h}//[110]$, only two of the four spins on a tetrahedron couple to the 
field. One expects that, with precise alignment of the sample, these other two 
spins would remain decoupled even in the application of high magnetic fields.  
These decoupled spins are thus free to choose an ordering 
pattern that satisfies their dipolar interaction.  
Because of the complexity of the dipolar interaction, the ground state 
spin configuration is not immediately obvious from studying the geometry.  
However, one expects any zero-temperature phase to be consistent with the 
ice-rules (see Fig.~\ref{Ztetra}b).  In the limit of very high applied field and perfect sample 
alignment, one expects the magnetization to approach 
\begin{eqnarray}
\nonumber
M({\bf h} \rightarrow \infty)&=& \frac{1}{2\sqrt{3}} 
\left({ \left[{111}\right] + \left[{11\overline{1}}\right]}\right) 
\cdot\frac{1}{\sqrt{2}}\left[{110}\right] \cdot \mu \\ 
&=& \frac{1}{\sqrt{6}} \cdot \mu \cong 0.4082 \cdot \mu.
\label{M110}
\end{eqnarray}

Finally, for ${\bf h}//[111]$, all four spins on a tetrahedron are coupled to 
the field.  An interesting complication arises in this case due to crystal 
geometry; any high-field phase of the material will be inconsistent with the 
ice-rules, and the spins will form a {\it three-in one-out} (or its spin 
reverse) tetrahedral configuration (Fig.~\ref{Ztetra}c).  
For zero temperature, both the long-range 
ordered ice-rules state and the three-in one-out state will exist for different 
field strengths.  For low magnitudes 
of {\bf h}, we expect a competition between the exchange, dipolar and magnetic
field 
parts of the Hamiltonian.  At low enough temperatures, one predicts 
\cite{Melko_thesis,Higashinaka,kagome-ice,kagomeMAT,Moessner111} that a plateau 
will develop in the magnetization curve due to the tendency of each tetrahedron 
to stay in the ice rules up to a critical field.  If we couple three of the 
spins to 
the magnetic field, and leave one to oppose the field but obey the ice rules, 
we find a magnetization of
\begin{eqnarray}
\nonumber
&M&({\bf h} = \text{``small''}) \\ \nonumber
&=& \frac{1}{4\sqrt{3}} \left({
\left[{111}\right] + \left[{\overline{1}11}\right] +
\left[{1\overline{1}1}\right] + \left[{\overline{11}1}\right]}\right) 
\cdot \frac{1}{\sqrt{3}} \left[{111}\right] \cdot \mu \\
&=& \frac{1}{3} \cdot \mu \cong 0.3333 \cdot \mu .
\label{M111lowT}
\end{eqnarray}

In the limit of very high applied field, we expect the spin that is coupled 
anti-parallel to the field (in the case above, the last spin vector 
$[\overline{11}1]$) to break 
the ice rules, in favor of minimizing its energy with respect to the field.  
In this case the high field magnetization is
\begin{eqnarray}
\nonumber
&M&({\bf h} \rightarrow \infty) \\ \nonumber
&=& \frac{1}{4\sqrt{3}} \left({
\left[{111}\right] + \left[{\overline{1}11}\right] +
\left[{1\overline{1}1}\right] + \left[{11\overline{1}}\right]}\right) 
\cdot \frac{1}{\sqrt{3}} \left[{111}\right] \cdot \mu \\
&=& \frac{1}{2} \cdot \mu = 0.5 \cdot \mu .
\label{M111}
\end{eqnarray}

We find that our Monte Carlo is successful in reproducing the high field 
limiting values of the experimental $M$ vs. ${\bf h}$ curves.
\cite{Fukazawa} 
In addition, we find that the Monte Carlo also 
reproduces the plateau expected for ${\bf h}//[111]$ and low temperatures.  
\cite{Melko_thesis}
However, because these large ${\bf h}$ results are easily obtainable for 
a nearest-neighbor spin ice model ($D_{\rm nn} \to 0$), we won't discuss
them further in this work.  The reader is referred to 
Ref.~\onlinecite{Melko_thesis} 
and Ref.~\onlinecite{Fukazawa} for the detailed results of this study.
				
\begin{figure}[ht]
\begin{center}
\includegraphics[height=8cm]{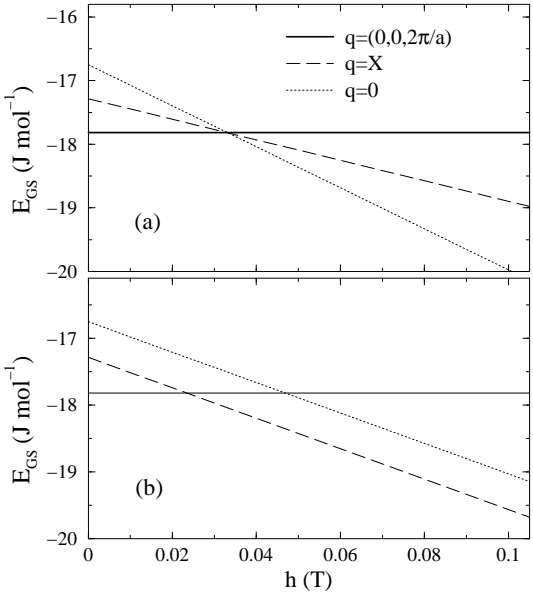}
\caption[Ground State Energies for ${\bf h}$//(100).]
{The $T$=0 energies per spin of the three ice-rules ordered states of the
dipolar spin ice model, as a function of applied internal field 
$h=|{\bf h}|$ along the (a) [100] and (b) [110] directions for 
$J_{\rm nn}=-0.52$ K
and $D_{\rm nn}=2.35$ K (i.e. Ho$_{\rm 2}$Ti$_{\rm 2}$O$_{\rm 7}$ parameters).}
\label{HiceR}
\end{center}
\end{figure}

\begin{figure}[ht]
\begin{center}
\includegraphics[height=4.5cm]{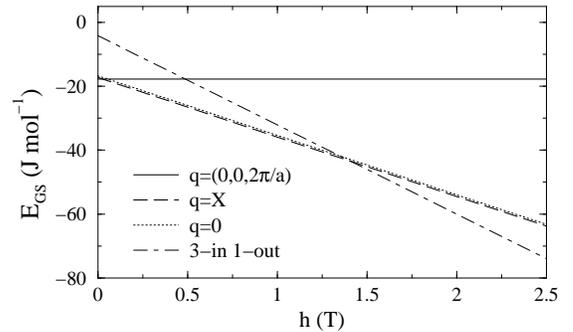}
\caption[Ground State Energies for ${\bf h}$//(111).]
{The $T$=0 energies per spin of the three ice-rules ordered states and
the
three-in one-out spin state of the dipolar spin ice model as a function
of
applied internal field  ${\bf h}$ along the [111] direction.
The ${\bf q}=X$ state becomes the ground state at 0.029 T.
The three-in
one-out configuration becomes the ground state at around 1.4 T, breaking
the
ice rules for each tetrahedron.
The ${\bf q}=X$ line is always 0.534 J mol$^{-1}$ below the ${\bf q}=0$
line at any given field.
}
\label{H3io}
\end{center}
\end{figure}

A numerical calculation of interest that is easily performed is the 
Ewald energies of the various ground state configurations that we have 
encountered so far in the dipolar spin ice model.  The spin ice configurations 
that we consider are both the ${\bf q}=0$ and ${\bf q}=X$ phases 
identified by Harris,\cite{Harris_prl1} and the 
${\bf q}=(0,0,2\pi/a)$ ground state identified previously in this work
(Fig.~\ref{GSsmall}).  In addition, we expect a ``three-in one-out'' state 
to become the lowest energy state for some critical field along the 
${\bf h}//[111]$ direction. 
Figs.~\ref{HiceR} and \ref{H3io} are the results of these ground state energy
calculations for a system size L=2 and parameters appropriate for
Ho$_{\rm 2}$Ti$_{\rm 2}$O$_{\rm 7}$ ($J_{\rm nn} = -0.52 {\rm K}$, 
$D_{\rm nn} = 2.35 \rm{ K}$).  As expected, we find the same qualitative 
behavior for calculations involving Dy$_{\rm 2}$Ti$_{\rm 2}$O$_{\rm 7}$
parameters.  In addition, because all of the ground state configurations 
considered are commensurate with the unit cell of the pyrochlore lattice,
these calculations scale trivially for sizes ${\rm L}>2$.

Fig.~\ref{HiceR}a confirms that the ${\bf q}=0$ configuration becomes
the lowest energy state for large field strength ($h>0.034$ T) 
for ${\bf h}$//[100], as expected from simple geometrical considerations.
Recall that in the ${\bf h}//[110]$ case, there exist two decoupled spins 
per tetrahedron, and subsequently no lowest-energy configuration is obvious 
from the geometric field coupling Eq.~(\ref{MagHamil}).
However, one may anticipate that for ${\bf h}//[110]$, the
decoupled spins (Fig.~\ref{Ztetra}b) order in ``chains'' 
perpendicular to the [110] direction, arranged in such a way as to 
partially satisfy the dipolar interaction.  
We find that this is precisely the ${\bf q}=X$ state, 
which Fig.~\ref{HiceR}b shows
to be the lowest energy state at for $h>0.023$ T. 

Consequently, Fig.~\ref{HiceR}b has direct relevance to experiments by 
Fennel {\it et al.} and Hiroi {\it et al.} that were performed on 
Dy$_{\rm 2}$Ti$_{\rm 2}$O$_{\rm 7}$ with a magnetic field ${\bf h}//[110]$.
\cite{hiroi,Dy162}  
Fennel {\it et al.} observed neutron diffraction patterns that showed
Bragg scattering at ${\bf q}=0$ ``points'' and diffuse scattering at
${\bf q}=X$ ``points'', but no true ${\bf q}=X$ long-range order.  
They suggested that this behavior would arise
from long-range ferromagnetic order occurring in field-coupled spin
chains (called $\alpha$ chains by Hiroi\cite{hiroi}), 
and short-range ``antiferromagnetic'' order occurring in the
field-decoupled spin chains ($\beta$ chains).\cite{Dy162} 
In this argument, the true
ground state is a ${\bf q}=X$ structure that is dynamically inhibited
from being accessed on experimental timescales.

Specific heat measurements by Hiroi {\it et al.} were used to extract
the specific heat contributions of both the $\alpha$ chains
and the $\beta$ chains.\cite{hiroi}
They suggest that the specific heat due to the $\beta$
chains resembles that which one would expect for a low-dimensional
spin system without long-range order.  They also argue for
the presence of
geometrical frustration in the triangular sub-lattice that contains the
$\beta$ chains.  If such a frustration exists, it might be expected to 
destabilize the ``antiferromagnetic'' correlations between these chains that 
would otherwise lead to
${\bf q}=X$ order.  Therefore, Hiroi {\it et al.} argued against 
true long-range order for the system, rather that the $\beta$
chains become effectively isolated and behave as ``pure''
one-dimensional ferromagnetic systems without long-range order in the ground 
state.

At this point, Fig.~\ref{HiceR}b is consistent with the idea of long-range 
${\bf q}=X$ order for the dipolar spin ice model with a magnetic field
${\bf h}//[110]$.  As we will discuss below, finite-temperature Monte
Carlo calculations on the dipolar spin ice model also support the idea
that, similar to the development of ${\bf q}=(0,0,2\pi/a)$  order in the
zero-field case, the development of ${\bf q}=X$ order for ${\bf
h}//[110]$ may in some cases be dynamically inhibited in experimental
systems with local spin dynamics.
The failure of the frustration of the $\beta$ chain sub-lattice invoked by 
Hiroi {\it et al.}\cite{hiroi} to
destroy the long-range ${\bf q}=X$ order may be another example of the small
energy scale left over by the infinite-range dipole-dipole energy
(Eq.~(\ref{DSPhamiltonian})), and why such interactions must be handled
carefully using techniques such as the Ewald method.

It should also be noted here that the ${\bf q}=0$ and ${\bf q}=X$
lines are parallel in Fig.~\ref{HiceR}b {\em only} for samples
that are perfectly aligned with {\bf h} along the [110] crystal axis.
This is an important phenomenon one must consider when comparing theory and 
experiment, as only a small crystal misalignment will partially couple
spins on the $\beta$ chains to the field.  Because {\it precise} 
alignment of a crystal is often very difficult, the possibility of misalignment 
of the order of a degree must be taken into consideration when studying single 
crystal data with ${\bf h}//[110]$.  Repeating our ground state energy 
calculation for misalignment of one degree along the $[100]$ direction, one 
finds a crossing of the ${\bf q}=X$ and ${\bf q}=0$ lines in 
Fig.~\ref{HiceR}b at about 1.3 
T, the ${\bf q}=0$ configuration being of lowest energy above this field
strength.\cite{Fukazawa}

Also, as Fig.~\ref{H3io} confirms, the three-in one-out spin configuration
become the lowest energy state for large ${\bf h}$//[111].
Interestingly, the ${\bf q}=X$ state is the ground state for
$ 0.029 < h < 1.4$T.  This is consistent with the idea that that the z component
of the field for the [111] direction gives a zero net Zeeman contribution to
the unit cell for both the ${\bf q}=0$ and ${\bf q}=X$ spin configurations.  
Therefore, the only energy scale difference left over between the two
states comes from the [110] component of the field, meaning that 
${\bf q}=X$ will be slightly lower in energy than ${\bf q}=0$ (as in
Fig.~25b) for all ${\bf h}$//[111].
Most other features of the ${\bf h}$//[111]
field (such as the intermediate-field plateau and the high-field 
breaking of the ice-rules) are readily explainable in a nearest-neighbor 
spin ice model without dipolar interaction, hence we will discuss them
no further in this work.

We now turn to a set of preliminary results of finite-temperature Monte
Carlo simulations for the dipolar spin ice model in a magnetic field.
As mentioned previously, the desire to explain the three
field-independent specific heat peaks\cite{Ramirez,Maeno2} of polycrystalline 
Dy$_{\rm 2}$Ti$_{\rm 2}$O$_{\rm 7}$ has driven much of the experimental and
theoretical interest in this field over the past few years.
Ramirez {\it et al.} were the first to suggest\cite{Ramirez} that some
of these peaks can be
attributed to a fraction of crystallites whose $[110]$ axes happen to
align (closely) with the applied magnetic field.
More generally, we can interpret this argument as saying that magnetic
moments which are not strongly coupled to the magnetic field through
Eq.~(\ref{MagHamil}) are free to contribute to a dipole-induced phase
transition (and therefore sharp peaks in the specific heat) at low
temperatures.  Historically, the ${\bf h}$//[110] coupling of 
Fig.~\ref{Ztetra}b was considered to be the most likely scenario to
provide these field-decoupled spins in a finite number of crystallites in
the polycrystalline sample.\cite{Ramirez}
However, one may in fact argue that
crystallites with only {\em one} field-decoupled spin 
would occur in much greater number in a real polycrystalline
sample.  This is due to the fact that, for a given crystallite orientation, 
there are an infinite number of applied magnetic field directions for which 
a given sub-lattice is decoupled
(one of these being ${\bf h}$//[112]\cite{Higashinaka}),
corresponding to a rotation degree of
freedom in the choice of ${\bf h}$ that does not exist in the two
spin field-decoupled case (${\bf h}$//[110]).

\begin{figure}
\begin{center}
\includegraphics[height=6.5cm,width=5cm,angle=90]{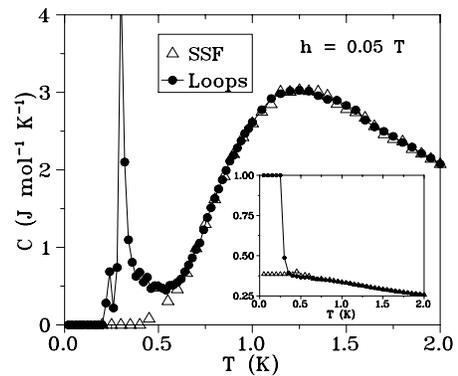}
\caption
{The specific heat of Dy$_{\rm 2}$Ti$_{\rm 2}$O$_{\rm 7}$ for ${\bf h}//[110]$, 
obtained using simulations with single spin flips and loop moves.  Inset: the 
${\bf q}=X$ order parameter calculated by the simulations.  The single spin 
flips (SSF) are unable to find the true ground state of the system, while the 
loops moves (Loops) allow the system to order into the ${\bf q}=X$ structure.
The large value of the high-temperature tail of the order parameter is a
finte-size effect.
}
\label{qXH0p05}
\end{center}
\end{figure}

Hence, we carry out finite-temperature Monte Carlo simulations on the
dipolar spin ice model for various field directions 
to look for signs of an ordering transition in the specific heat.  
In the case of ${\bf h}$//[100] and [110], the ground state is known 
and hence an order parameter can be constructed, facilitating the 
identification of any possible phase transitions.  We perform
simulations on the spin ice model with the added term
Eq.~(\ref{MagHamil}), employing both single spin flips and loop moves 
(and ignoring the demagnetization effects discussed in Appendix C).
For fields parallel to the [110] crystal axis, of magnitude
large enough to favor the ${\bf q}=X$ ground state, but still relatively 
small, we found that the simulations were able to find this fully
ordered state 
only when the loop moves were employed (see Fig.~\ref{qXH0p05}).  
This can be understood by studying the structure
of the ${\bf q}=X$ state and its coupling to this field direction, as
shown in Figs.~\ref{q0x}b and \ref{Ztetra}b.  
As discussed above, the field-decoupled spins occur in 
long chains ($\beta$ chains) in the bulk material.  
Within a single $\beta$ chain,
spins tend to point ferromagnetically along one direction as dictated by the 
exchange and magnetic field energies.  In addition, the dipolar term weakly 
couples nearby chains, and in order to minimize this coupling energy, 
neighboring chains can be expected to seek out a unique ordering pattern.
However, to explore different
chain-orientations at lower temperatures, whole chains must be flipped
at once in the simulation (imagine going from the ${\bf q}=0$ state to
the ${\bf q}=X$ state in Fig.~\ref{q0x}).  Thus, if the system is
attempting to settle into the ${\bf q}=X$ state at a temperature well below
the onset of ice-rules correlations, these global chain-flip moves must be 
employed in the simulation.  Luckily, such chain-flip moves are a subset of the
generic loops created in the Monte Carlo loop algorithm (due to periodic
boundary conditions), and thus no major modification of the simulation
procedure is needed.

\begin{figure}
\begin{center}
\includegraphics[height=6.5cm,width=5cm,angle=90]{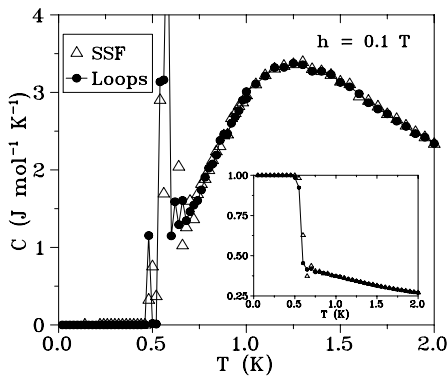}
\caption
{The specific heat of Dy$_{\rm 2}$Ti$_{\rm 2}$O$_{\rm 7}$ for ${\bf h}//[110]$, 
obtained using simulations with single spin flips and loop moves.  Inset: the 
${\bf q}=X$ order parameter calculated by the simulations.
Both the single spin flips and the loop moves are able to find the true
${\bf q}=X$ ground state of the system.}
\label{qXH0p1}
\end{center}
\end{figure}

As we see in Fig.~\ref{qXH0p05}, there is evidence of a feature 
in the specific heat that corresponds to the temperature where the 
${\bf q}=X$ order parameter jumps to essentially the saturation value of one.
The feature in the specific heat and the corresponding jump in the ${\bf q}=X$ 
order parameter are at approximately 0.3 K,  show that this is not 
the same transition as the ${\bf q}=(0,0,2\pi/a)$ transition of the previous 
section.  For small applied field, the transition temperature depends
on the strength of the applied magnetic field.  For example, at 0.075 T the 
transition temperature moves up to 0.4 K, and requires the loop moves to be 
manifest.  
For fields of the order of 0.01 T and larger, the transition has risen to high 
enough temperatures that single spin flip dynamics are still sufficiently 
present to promote development of the ground state, without help 
from the loop moves.
Details of this are illustrated in Fig.~\ref{qXH0p1}.  
Note that the phase transition illustrated in these two figures appears
to be strongly first order, and hence obtaining good error control is
difficult with the current statistics ($8\times 10^4$ equilibration and 
$8\times 10^4$ production MC steps). 
A more detailed study employing multicanonical techniques is currently in 
progress to look more closely at the exact temperature and field dependence of 
this new phase transition.

This sharp specific heat feature persists to a magnetic field value
of $h\sim 0.15$ T.  At higher field strengths, the singularity broadens
and gets absorbed into the spin ice peak of the specific heat at
$T\sim 1.2$ K.  However, for all $h > 0.5$ T, the order parameter is observed to
jump to unity at around 0.9 K, which appears to be the saturation value for 
very strong fields.  Hence, although the nature of the transition to
long-range order changes at some field $h\sim 0.15$ T, the fact remains that
the simulations can always find the ${\bf q}=X$ ground state up to infinite
field values.

For larger field values ($h>0.2T$) and ${\bf h}//[110]$, we reproduce the
experimental\cite{hiroi} features of the specific heat for  
Dy$_{\rm 2}$Ti$_{\rm 2}$O$_{\rm 7}$ nicely.  Fig.~\ref{BC1},
which shows our results for $0<T<8$ K for several field values, is
relegated to Appendix \ref{ApDemag} where we discuss demagnetization
effects.  Comparing this figure to Fig.~2, Ref.~\onlinecite{hiroi}, shows that
our model quantitatively reproduces the development of two peaks as $h$
is increased above 0.5 Tesla.  The $T\sim 7$ K peak, clearly visible 
for $h=1.0$ T (Fig.~\ref{BC1}d), is a Schottky-type peak which
corresponds to the freezing of field-coupled ($\alpha$) spin
chains that are along [110].\cite{hiroi}
The lower peak at $T\sim 1$ K is attributed to the development of
long-range correlations between $\beta$-chains, however its exact
relationship to the ordering of the model as observed by the jump in the
${\bf q}=X$ order parameter is still under study.

Finally, we performed finite-temperature Monte Carlo simulations on the
dipolar spin ice model for fields parallel to the [100], [111] and [112]
crystal directions.  In particular, we used field strengths larger than
those which would favor ${\bf q}=(0,0,2\pi/a)$ order 
(see Figs.~\ref{HiceR} and \ref{H3io}).
In the case of ${\bf h}//[100]$, we see the gradual
development of ${\bf q}=0$ long-range order as the simulation cools down
through the spin ice peak, and no sharp singularities in the specific heat
which may signal a phase transition.  Simulations for ${\bf h}//[111]$
also shows no unexpected anomalies in the specific heat.  Interestingly,
work on the ${\bf h}//[112]$ case has not yet produced any signs of an ordering
transition in the field-decoupled spin sub-lattice, as one might expect
from our discussion above.  Work addressing this problem is ongoing.

To summarize this section, we have performed several calculations of the
properties of the dipolar spin ice model in an external magnetic field.
Ewald energy calculations of various spin configurations reveals the
preferred $T=0$ ordering for various field directions.  In particular,
for ${\bf h}//[100]$ the ${\bf q}=0$ structure is the ground state,
while for ${\bf h}//[110]$ we find the ${\bf q}=X$ state becomes energetically
favored.  For finite-temperature Monte Carlo simulations of the dipolar
spin ice model, we find that the system settles into these ground states
for the respective field directions, although non-local dynamics are
needed to find the ${\bf q}=X$ state for ${\bf h}//[110]$ at low fields.  We
also predict a sharp first-order phase transition to the ${\bf q}=X$
state and a corresponding specific heat spike for $0.02 < h < 0.15$ T  
in Dy$_{\rm 2}$Ti$_{\rm 2}$O$_{\rm 7}$.
Interestingly, unlike the transition to the ${\bf q}=(0,0,2\pi/a)$ state in
zero field, the transition to the ${\bf q}=X$ state {\em can} be found
with single spin flips in the Monte Carlo for fields $0.10 < h < 0.15$ T,
although it is unknown whether local dynamics will be manifest in the
corresponding range in a real experiment.
Finally, for $h>0.15$ T and parallel to [110], 
the simulation exhibits the broad peaks in
the specific heat observed in experiments.\cite{Higashinaka,hiroi}
It is likely that the $T=1.1$ K specific heat feature corresponds to the 
$T=1.1$ K feature found in polycrystalline samples of 
Dy$_{\rm 2}$Ti$_{\rm 2}$O$_{\rm 7}$.\cite{Ramirez,Maeno2}

\section{Conclusion}

We have reviewed much of the early experimental and theoretical 
work on the static magnetic properties of spin ice. We have also 
clarified our principle point of view that long-range dipolar 
interactions are consistent with and responsible for the physics 
observed in spin ice materials based on
Dy$^{3+}$ and Ho$^{3+}$ rare earth ions. Support for our perspective 
resides in the detailed Monte Carlo and mean-field calculations 
presented in this paper.

Monte Carlo simulations were performed on the dipolar spin ice 
model with the long-range dipole-dipole interactions treated 
via the Ewald method. Using a single spin flip Monte Carlo method, 
we were able to study the development of the spin ice manifold. 
We found that spins freeze out at temperatures $O(1{\rm K})$ 
with a macroscopic degeneracy or residual Pauling entropy.  
We also found that single spin flip dynamics are not 
effective at equilibrating the system, thus making it impossible
to determine the ordered state of spin ice by this technique. 

Mean-field theory (Gaussian approximation) was applied to the same 
dipolar spin ice Hamiltonian with the dipolar interactions treated
via the Ewald method in ${\bf q}$-space. 
There, we showed that an ordering wave vector may be selected and 
that a proper treatment of the long-range dipoles is crucial to 
achieving a consistent picture with the experiments. 
A key point is that the symmetry of self-screening is not exact 
for the dipolar Hamiltonian. In the end, we found
a quasi-degenerate spectrum emerges with a commensurate 
critical mode (${\bf q}=(0,0,2\pi/a)$) and a two-in two-out 
spin ice structure. 

In order to find the ordered state of spin ice in a Monte Carlo simulation,
we developed a non-local algorithm that employs loop moves (or updates)
when inside the spin ice manifold. These loop moves represent the 
``nearly'' zero energy collective dynamics that allow our
model to sample the highly degenerate phase space of spin ice.  
Application of this method at temperatures within the 
spin ice manifold, i.e., $T\stackrel{<}{\sim} 1$K, 
leads to the selection of a single spin ice ground state configuration 
with ${\bf q}=(0,0,2\pi/a)$.
The loop Monte Carlo and mean-field results agree. 
In addition, we find a first order transition to the
ground state at $T_c \approx 180$ mK, which recovers all of the
residual Pauling entropy of the spin ice manifold.
Our physical understanding of spin ice is aided by the picture that any 
collective dynamics in real spin ice materials are inhibited by a 
freezing process as the system enters the temperature range where the
ice-rules fulfilling manifold develops,
i.e., $T_{\rm freeze} \approx 0.4$ K for Dy$_2$Ti$_2$O$_7$ and 
$T_{\rm freeze} \approx 0.6$ K for Ho$_2$Ti$_2$O$_7$ 
compared to $T_c \approx 180$ mK.

On the strength of the experimental evidence and the success of the dipolar
spin ice model, we assert that both Ho$_2$Ti$_2$O$_7$ and Dy$_2$Ti$_2$O$_7$ 
are spin ice materials. 

Finally, we have reflected on the application of a magnetic field to the
spin ice materials as means of exploring the possible structures of 
the spin ice manifold and to further characterize the interactions 
present in these intriguing systems.   We find excellent agreement
between the dipolar spin ice model and many experimental
studies to date.  In addition, we find evidence for a low-temperature ordering 
transition to a ${\bf q}=X$ ground state for small magnetic fields parallel to 
the [110] crystal direction, that has not yet been observed.
Some of the results presented here regarding the behavior of spin ice
are intriguing.  This argues for more theoretical, numerical and
experimental work, to resolve all the perplexing issues at stake.

\section{Acknowledgments}

We thank S. Bramwell, J. Gardner, H. Molavian, and A. del Maestro
for useful discussions.  We are grateful to  A. Ramirez for making
available his specific heat data. 
We also thank R. Higashinaka, H. Fukazawa and Y. Maeno for their ongoing
and very fruitful collaboration.
R.G.M. acknowledges support from the US Department of Energy grant\# 
DOE DE-FG02-03ER-46048 for a portion of this work
carried out at UCSB.
M.G. acknowledges financial support from 
NSERC of Canada, the Canada Research Chair Program,
Research Corporation, and the Province of Ontario.

\appendix

\section{Ewald}
\label{ApEwald}

We give only a brief overview of the Ewald \cite{Ewald_orig} 
technique as it applies to 
dipole-dipole interactions in Monte Carlo simulations and at 
the mean-field level. A more detailed discussion of the method can
be found in Refs.~\onlinecite{Ewald} and \onlinecite{DeLeeuw}. The mean-field 
case as it applies to moments on the 
pyrochlore lattice is treated in depth in  
Ref.~\onlinecite{Enjalran}.
 
The dipole-dipole interaction is an infinite sum that falls off as
the inverse cube of the separation distance between dipoles,   
$1/|{\bf R}_{ij}^{ab}|^3$. Hence it is a conditionally convergent series. 
The point of the Ewald method is to convert this  
slowly converging lattice sum into of two absolutely (rapidly) converging
series, one in real space and the other in Fourier space. 
The general lattice sum for $\langle 111 \rangle$ Ising dipoles
on the pyrochlore lattice is
\begin{eqnarray}
{\mathcal A} 
&=& \sum_{i,j}\sum_{a,b} \left ( \frac{{\hat z}^{a} \cdot {\hat z}^{b}}{|{\bf R}_{ij}^{ab}|^3}
- \frac{3({\hat z}^{a} \cdot {\bf R}_{ij}^{ab})
({\hat z}^{b} \cdot {\bf R}_{ij}^{ab})}{|{\bf R}_{ij}^{ab}|^5} \right ), \nonumber \\
&=& -({\hat z}^a \cdot \nabla_{\rm x})({\hat z}^b \cdot \nabla_{\rm x}) 
\left \{
\sum_{i,j}\sum_{a,b} \frac{1}{|{\bf R}_{ij}^{ab} - {\bf x}|} \right \}_{{\bf x}=0} 
\label{dip_latsum}
\end{eqnarray}
where the spin variables $\sigma_i^a$ have been dropped for 
notational convenience. 
The dipole sum excludes terms with $R_{ij}^{ab}=0$. 
Absolute convergence is forced on the sum inside the curly 
brackets of Eq.~(\ref{dip_latsum}) 
by use of a convergence factor. The form of this convergence factor 
differs depending on 
whether the dipolar sum is performed on $N$ particles 
in real space (e.g., Monte Carlo and 
molecular dynamic simulations) or in the thermodynamic limit 
in momentum space (mean-field theory).

In our work, MC simulations are performed on 3 dimensional lattices of 
$L\times L\times L$ cubic cells of the pyrochlore lattice 
under periodic boundary conditions, thus there are 
$N=16\times L\times L\times L$
spins in the simulation cell. The separation of moments 
within a simulation cell is given by 
${\bf R}_{ij}^{ab}$. The dipolar energy for any pair-wise interaction 
is calculated within the minimum image convention 
by summing replicas of the $N$-site simulation cell over 
spherical shells of radii 
${\bf n}=L(n_x,n_y,n_z)$ ($n_x,n_y,n_z$ are integers) with the inclusion 
of a spherical convergence factor $e^{-s|{\bf n}|^2}$.  
The effect of the convergence factor is removed from the 
final form of the Ewald
equations by imposing the limit $s \rightarrow 0$. 
Therefore, the starting point for the Ewald method is 
the dipole-dipole pair interaction, 
\begin{equation}
{\mathcal A}_{ij}^{ab}(s) = -({\hat z}^a \cdot \nabla_{\rm x})
({\hat z}^b \cdot \nabla_{\rm x}) \left \{
\sum_{{\bf n}\prime} \frac{e^{-s|{\bf n}|^2}}{|{\bf n}+{\bf R}_{ij}^{ab} - {\bf x}|} \right \}_{{\bf x}=0},
\label{EWMC}
\end{equation}
where $\sum_{{\bf n}\prime}$ means that ${\bf n}=0$ 
is omitted whenever ${\bf R}_{ij}^{ab}=0$.  
The point charge distribution, 
$1/|{\bf n}+{\bf R}_{ij}^{ab} - {\bf x}|$, is rewritten with
the aid of the $\Gamma$-function identities, 
\begin{eqnarray}
\label{eq-G1}
\frac{1}{|{\bf X}|} &=& \frac{1}{\sqrt{\pi}} \int_{0}^{\infty} t^{-1/2} e^{-t|{\bf X}|^2} dt  \\
&=&  \frac{2}{\sqrt{\pi}} \int_{0}^{\infty} e^{-t^2|{\bf X}|^2} dt \; .
\label{eq-G2}  
\end{eqnarray}
Using Eq.~(\ref{eq-G1}), the pair-wise interaction becomes
\begin{eqnarray}
&{\mathcal A}_{ij}^{ab}(s)& = -({\hat z}^a \cdot \nabla_{\rm x}) 
({\hat z}^b \cdot \nabla_{\rm x}) \frac{1}{\sqrt{\pi}} \int_{0}^{\infty} dt \\ 
&\times& \left \{ \sum_{{\bf n}\prime}t^{-1/2}
e^{-t|{\bf n}+{\bf R}_{ij}^{ab} - {\bf x}|^2 - s|{\bf n}|^2} \right \}_{{\bf x}=0} \nonumber
\; ,
\label{EWMC-2}
\end{eqnarray}
and the remainder of the Ewald for calculation for  
${\mathcal A}_{ij}^{ab}$ 
follows arguments outlined in Ref.~\onlinecite{DeLeeuw}. The
Ewald equations for a Monte Carlo simulation can also be 
found in Appendix A of Ref.~\onlinecite{Gingras_prb}. 
${\mathcal A}_{ij}^{ab}$ is calculated for each pair-wise interaction,
$\{(i,a),(j,b)\}$, in the simulation cell, but this need be done only once
because the spins are fixed to the lattice points. These pair interactions
are stored in a look-up table and used in the stochastic sampling and 
measurement procedures of a Monte Carlo simulation.  

In mean-field theory, one considers the Fourier 
transform of the dipole-dipole interaction;
therefore, the term $e^{-\imath {\bf q} \cdot {\bf R}_{ij}^{ab}}$ 
is included in Eq.~(\ref{dip_latsum}) and plays the role of a 
convergence factor as discussed in Ref.~\onlinecite{DeLeeuw}. 
With a convergence factor that is periodic as opposed to 
spherical in $R$, the derivation of the $q$-dependent Ewald equations 
follows the method of long waves introduced by Born and Huang, 
Ref.~\onlinecite{Ewald}. As noted in Section \ref{sect-mft}, 
the momentum dependent dipole-dipole interaction 
is determined between sub-lattice sites and is a component of 
${\mathcal J}^{ab}({\bf q})$, where ${\mathcal J}({\bf q})$ 
is $4\times 4$ matrix. Using Eq.~(\ref{eq-G2}), we can write the 
dipolar contribution 
to ${\mathcal J}^{ab}({\bf q})$ as 
\begin{eqnarray}
&{\mathcal A}^{a,b}({\bf q})& = -({\hat z}^a \cdot \nabla_{\rm x})
({\hat z}^b \cdot \nabla_{\rm x}) \frac{2}{\sqrt{\pi}} \int_{0}^{\infty} dt\\ 
&\times&
\left \{ 
\sum_{(i,j)\prime} e^{-t^2 |{\bf R}_{ij}^{ab} - {\bf x}|^2 
- \imath {\bf q} \cdot {\bf R}_{ij}^{ab}}
 \right \}_{{\bf x}=0}  \; \nonumber ,
\end{eqnarray}
where $\sum_{(i,j)\prime}$ means the $i=j$ term is excluded when $a=b$.
The remaining steps in the derivation of  
the ${\bf q}$-dependent Ewald equations can be found in 
Ref.~\onlinecite{Enjalran}. 
In this work, ${\mathcal A}^{a,b}({\bf q})$ was determined in at 
each ${\bf q}$-point in the first zone of the $(hhl)$ plane of the 
pyrochlore lattice and used in the formation of ${\mathcal J}({\bf q})$
in Eq.~(\ref{eq-f2}). 
We note that the value of
${\mathcal A}^{a,b}({\bf q})$ in the limit 
${\bf q} \rightarrow 0$ dependents on 
direction. The value of ${\mathcal A}^{a,b}(0)$
can be related to the demagnetization factor. 
\cite{CohenKeffer,AharonyFisher}

\section{$\chi({\bf q})$ from a high temperature series expansion}
\label{ApHTSE}

We demonstrate that the mean field description of the 
static susceptibility,\cite{Reimers}
$\chi(\bf{q})$, can also be derived from a high temperature
series expansion (HTSE) to lowest order in
$\beta$. In our HTSE, where there is no reliance on a mean-field approximation,
we show that the same functional dependence on the eigenvalues that leads to
a minimum in the free energy also defines the maximum in the response function
of the local magnetization, which we interpret as a transition to 
a long-range ordered state at the Gaussian level. 
We use the model for local $\langle 111 \rangle$ Ising spins on
the pyrochlore lattice, Eq.~(\ref{HIsing}),
but the final results are applicable, with minor modifications for the
spin components, to a general
Heisenberg Hamiltonian. The Ising
$\langle 111 \rangle$ Hamiltonian is
\begin{equation}
H = - \frac{1}{2}\sum_{i,j} \sum_{a,b}
{\mathcal J}^{a b}(i,j) \sigma_{i}^{a} \sigma_{j}^{b}
\label{eq-Ham111}
\end{equation}
where $\sigma_{i}^{a}=\pm 1$ are Ising variables and
${\mathcal J}^{a b}(i,j)$ is given by Eq.~(\ref{J111}). 

The ${\bf q}$-dependent susceptibility is defined as the Fourier
transform of the two-point correlation function,
\begin{equation}
\chi({\bf q}) = \frac{\beta}{N_{\rm cell}} \sum_{a,b} \sum_{i,j} 
(\hat{z}^{a}\cdot\hat{z}^{b}) \langle \sigma_{i}^{a} \cdot
\sigma_{j}^{b}\rangle e^{\imath {\bf q} \cdot {\bf R}_{ij}^{ab}}, 
\label{eq-chiq}
\end{equation}
where $\beta = 1/T$ with temperature in units of $k_B$.
We note that at this point $\chi({\bf q})$ is a 4$\times$4 matrix
for Ising spins on the pyrochlore lattice.  A HTSE 
for $\langle \sigma_{i}^{a} \sigma_{j}^{b} \rangle$
is expressed most clearly as an expansion of cummulants,\cite{dombgreen}
\begin{equation}
\langle \sigma_{i}^{a} \sigma_{j}^{b} \rangle = \sum_{m=0}^{\infty}
\frac{(-\beta)^{m}}{m!} \langle \sigma_{i}^{a} \sigma_{j}^{b}
H^{m}\rangle_{c}.
\label{eq-htse1}
\end{equation}
On the left hand side of Eq.~(\ref{eq-htse1}), $\langle ... \rangle$ represents
a thermal average and is thus a trace over states at finite $T$. The
cummulants, $\langle ... \rangle_c$, are expressed as traces over the $T=0$ states
(i.e., $\langle ... \rangle_o = Tr\{...\}/2^N$).
For our purposes, we need consider only the first two terms in the expansion,
\[
\langle \sigma_{i}^{a} \sigma_{j}^{b} \rangle \approx
\langle \sigma_{i}^{a} \sigma_{j}^{b} \rangle_c
- \beta \langle \sigma_{i}^{a} \sigma_{j}^{b} H\rangle_c.
\]

In the case of Ising spins, a non-zero trace has an even number of spin
variables at each site. The zeroth order cumulant is determined trivially,
\[
\langle \sigma_{i}^{a} \sigma_{j}^{b} \rangle_c =
\langle \sigma_{i}^{a} \sigma_{j}^{b} \rangle_o = \delta_{ij}\delta^{ab}.
\]
The first order contribution involves two terms, 
\[
\langle \sigma_{i}^{a} \sigma_{j}^{b} H \rangle_c =
\langle \sigma_{i}^{a} \sigma_{j}^{b} H \rangle_o -
\langle \sigma_{i}^{a} \sigma_{j}^{b} \rangle_o \langle H \rangle_o ,
\]
but the second does not
contribute because $\langle H \rangle_o \propto
\langle \sigma_{l}^{c} \sigma_{m}^{d} \rangle_o = 0$.
Therefore, one has
\begin{eqnarray}
\langle \sigma_{i}^{a} \sigma_{j}^{b} H \rangle_o &=& -\frac{1}{2} \sum_{lm}
\sum_{cd} {\mathcal J}^{cd}(l,m) \langle \sigma_{i}^{a} \sigma_{j}^{b}
\sigma_{m}^{c} \sigma_{l}^{d} \rangle_o \nonumber \\
&=& - {\mathcal J}^{ab}(i,j). \nonumber
\end{eqnarray}
The $\bf{q}$-dependent susceptibility in the high temperature limit reads,
\begin{equation}
\chi({\bf q}) \approx \frac{\beta}{N_{\rm cell}} 
\sum_{a,b} \sum_{i,j} ({\hat z}^{a} \cdot {\hat z}^{b})
(\delta_{ij} \delta^{ab} + \beta {\mathcal J}^{ab}(i,j))
e^{\imath {\bf q} \cdot {\bf R}_{ij}^{ab}}.
\label{eq-chiq3}
\end{equation}
Performing the Fourier transform of ${\mathcal J}^{ab}(i,j)$, 
the inverse of Eq.~(\ref{FTJ}), we obtain
\begin{equation}
\chi({\bf q}) = \beta \sum_{a,b} ({\hat z}^{a} \cdot {\hat z}^{b})
(\delta^{ab} + \beta {\mathcal J}^{ab}({\bf q})) , 
\label{eq-chiq4}
\end{equation}
where ${\mathcal J}(-{\bf q})={\mathcal J}({\bf q})$ for a symmetric
interaction matrix. 
The spin-spin interaction matrix and thus the susceptibility are 
diagonalized via the normal mode transformation given by 
Eq.~(\ref{JqNM}). The unitary matrix, $U({\bf q})$, 
that diagonalizes ${\mathcal J}({\bf q})$ contains 
the orthonormalized eigenvectors of
${\mathcal J}({\bf q})$, i.e., 
$\sum_{\alpha} U^{a,\alpha}({\bf q}) U^{b,\alpha}(-{\bf q})= {\bf I}$, 
where ${\bf I}$ is the $4\times4$ unit matrix. 
We use this to rewrite the two terms in Eq.~(\ref{eq-chiq4}),
\begin{equation}
\delta^{ab} = \sum_{\alpha} \sum_{a,b} U^{a,\alpha}({\bf q}) 
U^{b,\alpha}({\bf -q})\ , \nonumber
\end{equation}
and
\begin{equation}
{\mathcal J}^{ab}({\bf q}) = 
\sum_{\alpha}\lambda^{\alpha}({\bf q}) 
U^{a,\alpha}({\bf q}) U^{b,\alpha}(-{\bf q})\ . \nonumber
\end{equation}
The expression for $\chi({\bf q})$ now becomes,
\begin{equation}
\chi({\bf q}) = \beta \sum_{\alpha}\sum_{a,b} 
(\hat{z}^{a} \cdot \hat{z}^{b})
(1 + \beta \lambda^{\alpha}({\bf q}))U^{a,\alpha}({\bf q})U^{b,\alpha}({\bf -q}).
\label{eq-chiq5}
\end{equation}
In the high temperature limit, $\beta \rightarrow 0$, and 
$1 + \beta \lambda^{\alpha}({\bf q}) \approx 1/(1 - \beta \lambda^{\alpha}({\bf q}))$;
therefore, the static, ${\bf q}$-dependent susceptibility now reads,
\begin{eqnarray}
\chi({\bf q}) &=& \beta \sum_{\alpha} \sum_{a,b}
\frac{({\hat z}^{a} \cdot {\hat z}^{b})U^{a,\alpha}({\bf q}) 
U^{b,\alpha}({\bf -q})}
{(1 - \beta \lambda^{\alpha}({\bf q}))} \\
&=& \beta \sum_{\alpha}
\frac{|\sum_{a} \hat{z}^{a} U^{a,\alpha}({\bf q})|^2}{(1 - \beta \lambda^{\alpha}({\bf q}))}.
\nonumber
\label{eq-chiq6}
\end{eqnarray}
Hence, as $T$ approaches the ordering temperature defined by Eq.~(\ref{Tc}),
$T_c=\lambda^{\rm max}({\bf q}_{\rm ord})$, the susceptibility diverges and 
signals a transition to a long-range ordered state. 

\section{Demagnetization effects}
\label{ApDemag}

When doing finite temperature Monte Carlo 
simulations on magnetic materials in an applied magnetic field,
the effect of the boundary of the simulation cell must 
be carefully considered. 
For systems of interest, the dipolar spin ice Hamiltonian 
is augmented with a field dependent term, Eq.~(\ref{MagHamil}).  
The inclusion of this term in our Monte Carlo 
simulations leads to subtle effects. 
In a microscopic Hamiltonian, the field {\bf h} referred to 
in Eq.~(\ref{MagHamil}) is the sample {\it internal} field, 
i.e., the magnetic field that directly couples to each magnetic 
dipole moment.  However in real materials, 
bulk demagnetization effects alter the magnitude of the 
internal field in a complicated manner that depends on sample size, 
shape, alignment, and surrounding medium.  In general, 
experimentalists define three separate quantities
(the magnetic flux density {\bf B}, the magnetic field strength {\bf H},
and the magnetization {\bf M}) to account for these effects.  
In a macroscopic material, these quantities are related by
\begin{equation}
{\bf B}=\mu_{0}({\bf H} + {\bf M}) \ ,
\label{BulkMag}
\end{equation}
where {\bf B} is the independent quantity controlled in the experiment,
but {\bf H} is the field strength that couples to the spins (through
which the bulk susceptibility is defined).
In order to benchmark an experiment to a theory such as ours, 
an attempt must be made to relate the {\it external} 
experimental (applied) {\bf B} controlled by external sources of current
to the internal {\bf h} of our Hamiltonian.  From an experimental side, 
this amounts to knowing the internal {\bf M} associated with 
the specific sample being measured.  This {\bf M} is in general not easily 
deduced; although, for certain sample shapes 
(e.g., ellipsoids of revolution) it is at least uniform, 
and fairly accurate estimates can be made.  
The procedure of correcting for {\bf M} to obtain 
{\bf H} is called making a demagnetization correction.

Theoretically, demagnetization effects are incorporated 
into a Monte Carlo simulation by imposing certain boundary 
conditions on the microscopic Hamiltonian in question. 
As described earlier, we use the Ewald summation 
method to calculate the long-range dipolar interactions of our model.  
We follow the standard approach in which the pair 
wise interactions are evaluated by summing over periodic 
copies of the $N$ site simulation cell until convergence is
obtained,\cite{DeLeeuw} effectively simulating the
infinite range nature of the dipoles. 
A consequence of this technique is that the 
finite size nature of the simulation cell is 
suppressed. We are, therefore, faced with the 
question of how to interpret an ``infinite boundary''.
If one wishes to simulate materials with no net magnetic moment, or materials 
with no internal demagnetizing field, no correction due to sample boundary is 
needed, and the simple Ewald sum results may be used.  
This is equivalent to simulating a long  thin ``needle'' of 
the bulk material.  However, if one wishes to simulate a 
material in which the unit cell has a net magnetic moment or internal 
demagnetizing fields, then we must modify the 
Ewald sum to take into account the necessary boundary effects. 
This is especially important in our simulation because the 
long-range nature of the dipole-dipole interactions 
greatly accentuates these effects.

The approach described by de Leeuw {\it{et al.}} \cite{DeLeeuw} is to include 
a {\it boundary term} in the Ewald sum of the form
\begin{equation}
\left({ \frac{4\pi}{2\mu'+1} }\right)
\frac{ {\bf \mu}_{i} \cdot {\bf \mu}_{j} }{ L^{3}}
\label{BndryTerm}
\end{equation}
where ${\bf \mu}_{i}$ is the magnetic dipole moment of a spin, and $L$ is the 
system linear dimension. Physically, the inclusion of this term corresponds to 
the consideration of a region external to the spherical Ewald boundary (see 
discussion in Ref.~\onlinecite{DeLeeuw}).  This external region is a continuum 
with magnetic permeability constant ranging from $\mu'=1$ to infinity.
Because of the nature of our Ewald sum, 
the $\mu'=1$ case will in effect simulate the bulk of a spherical sample 
surrounded by a vacuum.  The $\mu' =\infty$ case
corresponds to simulating a bulk sample which is needle-like 
and parallel to an applied {\bf B}, and hence 
contains no internal demagnetizing field.

In summary, to make a meaningful comparison between simulation and experiment 
within the dipolar spin ice model in an applied magnetic field, one of two 
scenarios must happen:
\begin{enumerate}
\item
  Simulations are performed using the regular Ewald summation method, and
  bulk demagnetization effects are accounted for by experimentalists.
\item
  Simulations are performed with the inclusion of a boundary term
  Eq.~(\ref{BndryTerm}).  Experimentalists are restricted to measurements on
  spherical samples to make quantitative comparisons.  However, measurements on
  other sample shapes (with approximately constant internal fields) may allow
  some qualitative comparison.
\end{enumerate}

\begin{figure}
\begin{center}
\includegraphics[height=7cm]{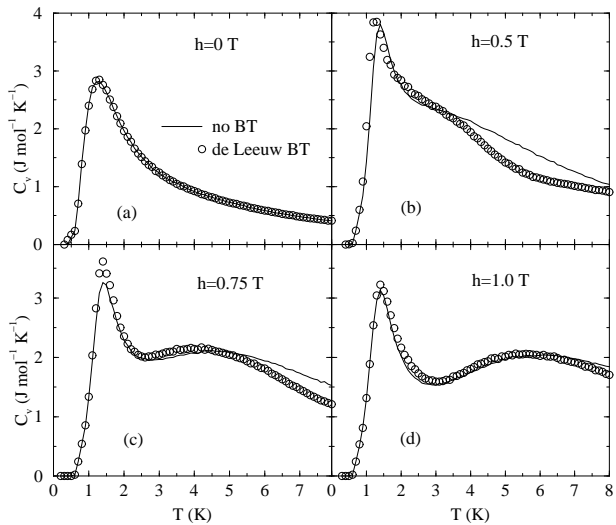}
\caption[Monte Carlo Specific Heat and the de Leeuw Boundary Term.]
{Specific heat curves for Dy$_{\rm 2}$Ti$_{\rm 2}$O$_{\rm 7}$ in an
applied
field ${\bf h}//[110]$, in simulations without (line) and with (circles)
the
boundary term (BT) in the Ewald energy summation.  These simulations
employed
single spin flip dynamics in the Monte Carlo.  The significance of this
data in the context of experimental measurements on
Dy$_{\rm 2}$Ti$_{\rm 2}$O$_{\rm 7}$ is discussed in detail in Section
\ref{DSImag}.}
\label{BC1}
\end{center}
\end{figure}

The inclusion of the boundary term is a non-trivial matter in 
many simulations. 
For example, it will promote effects such as domain formation 
in simulations of 
global Ising ferromagnets.  It is therefore always 
important to check ground state configurations of system 
where the term Eq.~(\ref{BndryTerm}) is absent against those 
where is has been included, in order not to miss any 
important secondary effects.  

As a means of addressing some of these issues, we present some
preliminary results on Monte Carlo simulations of the dipolar spin ice
model.
Fig.~\ref{BC1} shows specific heat curves for 
Dy$_{\rm 2}$Ti$_{\rm 2}$O$_{\rm 7}$ in 
an applied magnetic field, with and without the inclusion 
of the boundary term, Eq.~(\ref{BndryTerm}) with $\mu'=1$.
The results presented in Fig.~\ref{BC1} were performed using
single spin flip dynamics on a system of size $L=3$.
As we see, the boundary term does not effect the $h=0$ specific heat curve.  
This is consistent with the understanding that the boundary term is only 
necessary in Monte Carlo programs where the simulation cell has a net magnetic 
moment (which is not true in general for the spin ice manifold).  For moderate 
fields, we see that the boundary term significantly alters the shape of the 
specific heat curve, as expected, since the simulation cell is acquiring a net 
magnetic moment.  For very large fields, the boundary term begins to lose its 
effect, as the field term (Eq.~(\ref{MagHamil})) becomes dominant in the 
Hamiltonian.

\end{document}